%% file: thesis.tex
\documentclass[11pt, oneside, a4paper, oldfontcommands]{memoir}

\usepackage[T1]{fontenc}
\usepackage{lmodern}
\usepackage[a4paper,width=142mm,top=35mm,bottom=35mm,bindingoffset=10mm]{geometry}
\usepackage{amsmath,amssymb,mathrsfs,empheq}
\usepackage[utf8]{inputenc}
\usepackage{multirow}
\usepackage{array}
\usepackage[UKenglish]{babel}
\usepackage{xcolor}
\usepackage{graphicx,amstext,amsmath}
\usepackage{subcaption}
\usepackage{mathtools}
\usepackage{tensor}
\usepackage{bm}
\usepackage{comment}
\usepackage{esvect}
\usepackage{hyperref}
\hypersetup{
    colorlinks=true,
    linkcolor=blue,
    filecolor=magenta,      
    urlcolor=red,
    citecolor=red
}
\usepackage[sort&compress,merge,numbers]{natbib}
\usepackage{empheq}
\usepackage{float}
\usepackage{rotating}
\usepackage{acronym}

\def\CC{{\cal C}}

\def\IR{{\mathbb R}}

\def\IC{{\mathbb C}}

\def\bS{{\boldsymbol S}}
\def\T{{\boldsymbol T}}

\newcommand{\be}{\begin{equation}}
\newcommand{\ee}{\end{equation}}
\newcommand{\p}{\partial}

\nonzeroparskip

\newsubfloat{figure}

\newcommand{\thesisTitle}{An effective theory for higher-dimensional black holes and applications to metastable antibranes}

\newcommand{\thesisName}{Nam Huy Hoai Nguyen}

\newcommand{\thesisFirstSupervisor}{Vasileios Niarchos}

\chapterstyle{madsen}

\DoubleSpacing

\setsecnumdepth{subsection}

\makepagestyle{myruled}
\makeheadrule {myruled}{\textwidth}{\normalrulethickness}
\makeevenhead {myruled}{}{}{\itshape\leftmark}
\makeoddhead {myruled}{}{}{\itshape\rightmark}
\makeevenfoot {myruled}{}{\thepage}{}
\makeoddfoot {myruled}{}{\thepage}{}

\makepsmarks {myruled}{
\nouppercaseheads
\createmark {chapter} {both} {shownumber}{}{. \ }
\createmark {section} {both}{shownumber}{} {. \ }
\createmark {subsection} {both}{shownumber}{} {. \ }
\createmark {subsubsection}{both}{shownumber}{} {. \ }
\createplainmark {toc} {both} {\contentsname}
\createplainmark {lof} {both} {\listfigurename}
\createplainmark {lot} {both} {\listtablename}
\createplainmark {bib} {both} {\bibname}
\createplainmark {index} {both} {\indexname}
\createplainmark {glossary} {both} {\glossaryname}
}

\makepagestyle{plain}
\makeevenfoot {plain}{\thepage}{}{}
\makeoddfoot {plain}{}{\thepage}{}

\makepsmarks {plain}{
\nouppercaseheads
\createmark {chapter} {both} {shownumber}{}{. \ }
\createmark {section} {both}{shownumber}{} {. \ }
\createmark {subsection} {both}{shownumber}{} {. \ }
\createmark {subsubsection}{both}{shownumber}{} {. \ }
\createplainmark {toc} {both} {\contentsname}
\createplainmark {lof} {both} {\listfigurename}
\createplainmark {lot} {both} {\listtablename}
\createplainmark {bib} {both} {\bibname}
\createplainmark {index} {both} {\indexname}
\createplainmark {glossary} {both} {\glossaryname}
}

\setsecnumdepth{subsubsection}
\begin{document}
\frontmatter
\include{content/front_matter}

\newpage

\include{content/acronyms}

\mainmatter

\pagestyle{myruled}

\include{./content/chapter1} 

\cleardoublepage

\bibliographystyle{JHEP}
\bibliography{thesis}

\input{content/back_matter}

\end{document}

%% file: content/front_matter.tex
\pagenumbering{roman}           
\pagestyle{empty}               
\input{content/title_page}      
\cleardoublepage

\pagestyle{plain}               
\input{content/abstract}        
\cleardoublepage
%
\input{content/acknowledgement} 
\cleardoublepage
\setcounter{tocdepth}{2}        
\tableofcontents*                
\cleardoublepage
%
\input{content/declaration}
\cleardoublepage
%

%% file: content/title_page.tex

\begin{titlingpage}
    \begin{center}
        \vspace*{1cm}

        \huge
        \textbf{\thesisTitle{}}

        \vspace{0.5cm}

        \vspace{1.5cm}

        \textbf{\thesisName}

        \vfill

        \large
        A thesis presented for the degree of\\
        Doctor of Philosophy

        \vspace{0.8cm}

        \includegraphics[width=0.4\textwidth]{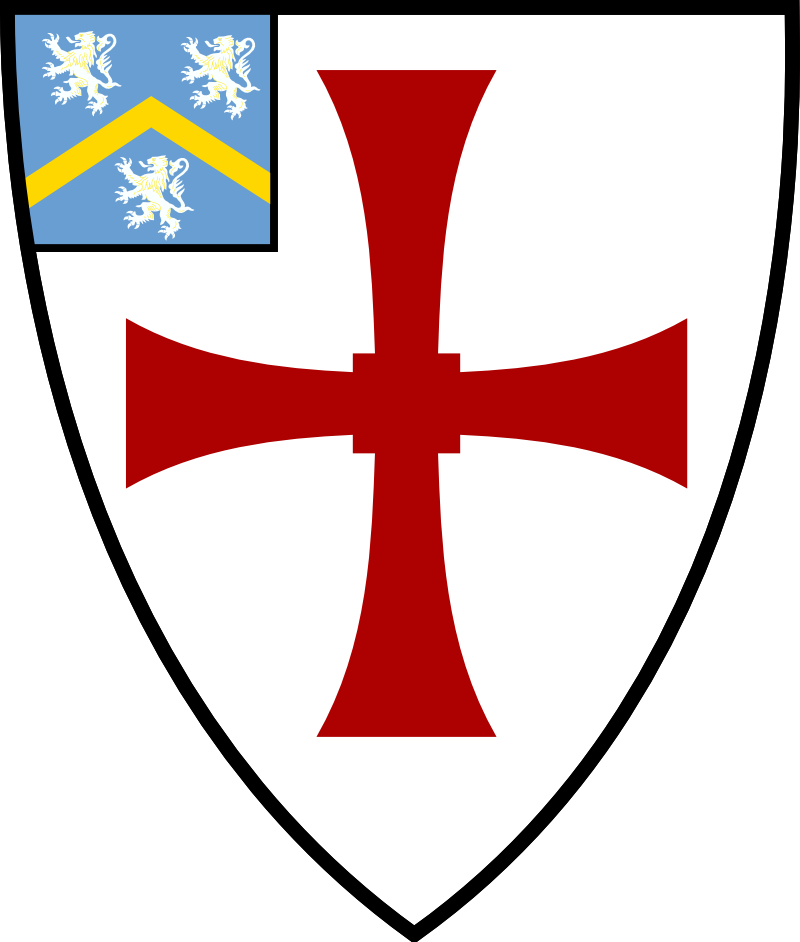}

        Department of Mathematical Sciences\\
        The University of Durham\\
        United Kingdom\\
        June 2021

    \end{center}
\end{titlingpage}

%% file: content/abstract.tex
%
\hfill
\begin{center}
\textbf{{\Large \thesisTitle}}\\
\textbf{{\large \thesisName}}\\
\end{center}

\begin{abstract}
Despite their consequential applications across the subfields of high energy physics, metastable states of antibranes in warped throats are not yet fully understood. In this thesis, we provide new information on various aspects of these metastable antibranes through applications of the blackfold effective theory for higher-dimensional black holes. As concrete examples, we study the conjectured metastable state of polarised anti-D3 branes (namely, wrapped NS5 branes with dissolved D3 brane charge) at the tip of the Klebanov-Strassler (KS) throat in type IIB supergravity and the analogous state of polarised anti-M2 branes (namely, wrapped M5 branes with dissolved M2 brane charge) at the tip of the Cvetic-Gibbons-Lu-Pope (CGLP) throat in eleven-dimensional supergravity.

For anti-D3 branes in KS throat, from a finite-temperature analysis in the wrapped NS5 regime, we provide novel evidence for the existence of the metastable state exactly where no-go theorems are lifted. In particular, in the extremal limit, we recover directly in supergravity the metastable states originally discovered by Kachru, Pearson, and Verlinde (KPV). Away from extremality, we uncover a metastable wrapped black NS5 state (the thermalised version of the KPV state) and observe that such metastability is lost when we heat the wrapped NS5 state sufficiently that its horizon geometry resembles that of a black anti-D3 state.

All claims regarding metastability of antibranes in warped throats only refer to a balance of force and not statements of robustness under perturbations. For their various applications, it is important to determine whether these configurations are truly metastable by probing them with perturbations. Here, we study the classical stability of the KPV state under generic long-wavelength deformations. We observe that, with regards to considered perturbations and regime of parameters, the state is classically stable.

A study of anti-M2 branes in CGLP throat reveals many similarities to that of the anti-D3 branes. We recover directly in supergravity the Klebanov-Pufu (KP) state at extremality, and our finite temperature results fit suggestively well with known, complementary no-go theorems. However, a unique feature of the anti-M2 state is that when considering the effects of non-zero temperature on the KP metastable state, we discover an exotic pattern of thermal transitions different from that of the KPV.

This thesis contains detailed discussions on all the above results as well as a pedagogical introduction to the blackfold formalism, focusing on aspects immediately relevant to applications to metastable antibranes.
\end{abstract}

\vfill
    {\small Supervisor: \thesisFirstSupervisor}

%% file: content/acknowledgement.tex
%
\chapter*{Acknowledgements}
\label{sec:acknowledgement}
\begin{OnehalfSpacing}
I would like to give my deepest thanks to my supervisor, Vasileios Niarchos, for his skillful guidance and continued encouragement throughout my Ph.D. His mentorship shapes my approach to physics, research, and beyond. I would also like to thank my undergraduate supervisor, Aristomenis Donos, for his passion inspired me to pursue a career in theoretical physics.

I want to give my sincere gratitude to my valuable collaborators, Niels Obers, Thomas Van Riet, and Jay Armas, whose knowledge, experience, advice, and suggestions have helped me tremendously. My heartiest thanks go also to all (past and current) members of the Department of Mathematical Sciences at Durham University. It is because of them that my time in Durham is so special.

I would like to thank my family for their constant love and support. All my achievements are only possible because of them. Lastly, I want to say my heartfelt thanks to my wife, Linh. Without her, the universe might not be as colourful.

\end{OnehalfSpacing}

%% file: content/declaration.tex
%

\chapter{Declaration}
\label{sec:declaration}
\begin{OnehalfSpacing}
The work in this thesis is based on research carried out at the Centre for Particle Theory, Department of Mathematical Sciences, University of Durham, United Kingdom. No part of this thesis has been submitted elsewhere for any other degree or qualification, and it is the sole work of the author unless referenced to the contrary in the text.

Some of the work presented in this thesis has been published in journals and conference proceedings - the relevant publications are listed below.

\section*{Publications}
\begin{itemize}
\item J. Armas,\textbf{ N. Nguyen}, V. Niarchos, N. A. Obers, and T. Van Riet, \textit{Metastable
Nonextremal Antibranes}, \textit{Phys. Rev. Lett. } \textbf{122} (2019), no. 18 181601,
 \href{https://arxiv.org/abs/1812.01067}{[arXiv:1812.01067]}.

\item J. Armas, \textbf{N. Nguyen}, V. Niarchos, and N. A. Obers, \textit{Thermal transitions of metastable M-branes}, \textit{JHEP} \textbf{08} (2019) 128, \href{https://arxiv.org/abs/1904.13283}{[arXiv:1904.13283]}.

\item \textbf{N. Nguyen}, \textit{Comments on stability of KPV metastable}, \textit{JHEP} \textbf{11} (2020) 055, \href{https://arxiv.org/abs/1912.04646}{[arXiv:1912.04646]}.

\end{itemize}
%
%
%

\vfill
    \textbf{Copyright \textcopyright~2020 by \thesisName.}

    ``\emph{The copyright of this thesis rests with the author. No quotation from it should be published without the author's prior written consent and information derived from it should be acknowledged}''.

\end{OnehalfSpacing}


%% file: content/acronyms.tex
\chapter{Nomenclature}
\label{sec:acronyms}

\begin{acronym}

\acro{KS}{Klebanov-Strassler}

\acro{CGLP}{Cvetic-Gibbons-Lu-Pope}

\acro{KPV}{Kachru-Pearson-Verlinde}

\acro{KP}{Klebanov-Pufu}

\acro{KKLT}{Kachru-Kallosh-Linde-Trivedi}

\acro{MAE}{Matched asymptotic expansion}

\end{acronym}

%% file: content/chapter1.tex

\setcounter{equation}{0}
\graphicspath{{/Users/matthewtownson/Documents/PhD/Latex/thesis_3/content/lowsnrfigs/}}

\chapter{Introduction}
\label{IN}
\begin{center}
\begin{minipage}{30em}
“Physics is really nothing more than a search for ultimate simplicity, but so far all we have is a kind of elegant messiness.”
\end{minipage}
\end{center}
\begin{flushright}
\textit{Bill Bryson}
\end{flushright}

\section{An invitation}
\label{IN1}
An understanding of controlled supersymmetry (SUSY) breaking in string theory is among the principal goals in string phenomenology for \textit{establishing connections between string theory and our reality}. Even in the case that string theory turns out to be not the quantum gravity of our universe, non-SUSY string configurations are still relevant because they can be used as pseudo experimental data for \textit{constructions of Swampland conjectures on how all quantum gravity theories must behave} and they describe holographically \textit{non-SUSY quantum field theories} (QFT) that have immediate practical applications.  

One of the canonical methods for breaking supersymmetry in string theory involves balancing antibranes in warped throats \cite{Maldacena:2001pb}. Despite their consequential applications in holographic QFT \cite{Kachru:2002gs, Klebanov:2010qs}, string cosmology \cite{Kachru:2003aw, Kachru:2003sx}, and black hole physics \cite{Bena:2012zi}, some aspects of the resulting antibranes configurations are not yet well-understood. Through applications of the \textit{blackfold effective theory for higher-dimensional black holes} \cite{Emparan:2009cs,Emparan:2009at,Armas:2016mes}, we have uncovered novel properties regarding the \textit{existence} \cite{Armas:2018rsy}, \textit{stability} \cite{Nguyen:2019syc}, and \textit{thermal transitions} \cite{M2M5brane} of these metastable antibranes. This thesis is a report on those findings. 

In recent debates of string cosmology, a prominent role is played by the de Sitter Swampland conjecture, which considers the possibility that de Sitter vacua cannot come from a consistent quantum gravity theory \cite{Obied:2018sgi, Palti:2019pca}. The conjecture, which applies to all quantum gravity theories, is proposed based primarily on the observation that we cannot construct a rigorous de Sitter vacuum in string theory \cite{Danielsson:2018ztv}. Among all constructions of string de Sitter vacua, the most promising is perhaps the Kachru-Kallosh-Linde-Trivedi (KKLT) construction \cite{Kachru:2003aw}. Lie at its heart is a metastable configuration of polarised anti-D3 branes at the tip of the Klebanov-Strassler (KS) throat \cite{Klebanov:2000hb}, first discovered by Kachru, Pearson, and Verlinde (KPV) in \cite{Kachru:2002gs}. Many debates on the validity of the KKLT construction are rooted in our incomplete understanding of the KPV configuration. As a result, to determine whether de Sitter vacua live in the Landscape or the Swampland, we begin with a study of the KPV configuration of metastable antibranes.

The initial discovery of the KPV state was based on brane probe analysis. The status of this analysis beyond probe became controversial when unphysical singularities were found in the backreacted supergravity description of localised anti-D3 branes in KS throat \cite{Bena:2009xk, Gautason:2013zw}. The presence of these singularities was viewed by some authors as evidence that backreaction can change dramatically the conclusions of the probe approximation, casting doubt to the very existence of the metastable state. Though such singularities disappear for polarised anti-D3 branes, or equivalently wrapped NS5 branes with dissolved anti-D3 brane charge (wrapped anti-D3-NS5), which is what the KPV state really is \cite{Cohen-Maldonado:2015ssa}, because of technical difficulties, the majority of follow up computations were performed in the localised anti-D3 regime instead of the wrapped NS5 regime. Nevertheless, a proper understanding of the KPV state beyond probe requires information about the backreacted wrapped NS5 state. This motivates our \textbf{first research objective}: 
\begin{center}
\begin{minipage}{\textwidth}
\textit{Investigate the backreacted profile of wrapped anti-D3-NS5 branes at the tip of the KS throat in relation to the existence of the KPV state.}
\end{minipage}
\end{center}
As the KPV state is an exemplar antibranes metastable state, the findings regarding the KPV state are also applicable to other antibranes metastable states, e.g. the Klebanov-Pufu state \cite{Klebanov:2010qs}.

Another important question regarding the KPV state and its application in string cosmology is whether such configuration is robust under perturbations. What is shown in \cite{Kachru:2002gs}, and subsequently in \cite{Armas:2018rsy}, is only that homogeneous, spherical anti-D3-NS5 branes living at the tip of the KS throat can form an equilibrium by balancing ``electromagnetic'' forces pushing them over the $S^3$ of the tip and gravitational forces doing the opposite. Based on these works alone, we cannot say whether the resulting state is short-lived or truly metastable under perturbations. This motivates our \textbf{second research objective}:
\begin{center}
\begin{minipage}{\textwidth}
\textit{Investigate the classical stability of the KPV state with respect to generic perturbations.}
\end{minipage}
\end{center}

Studies on the effects of non-zero temperature on antibranes can reveal much about their physics. Already in the scope of our first research objective, the behaviours of nonextremal antibranes are crucial. However, because of their relevance to further topics such as holographic QFT or black hole physics, it is fruitful to study nonextremal metastable antibranes beyond the scope of existence discussions. For our study, we find it most interesting to take as our specimen the Klebanov-Pufu (KP) \cite{Klebanov:2010qs} metastable state of polarised anti-M2 branes, or equivalently wrapped M5 branes with dissolved anti-M2 brane charge (wrapped anti-M2-M5 branes), at the tip of the Cvetic-Gibbons-Lu-Pope (CGLP) throat \cite{Cvetic:2000db}. This forms our \textbf{third research objective}: 
\begin{center}
\begin{minipage}{\textwidth}
\textit{Study the effects of non-zero temperature on the KP metastable state of wrapped anti-M2-M5 branes at the tip of the CGLP throat.}
\end{minipage}
\end{center}

\paragraph{Outline of thesis}
This thesis consists of 6 chapters and 4 appendices. Chapter \ref{IN} contains an introduction to the research topic and research questions, a summary of the subsequent chapters, and a statement of notations and conventions. Chapter \ref{BF} (summarised in section \ref{IN2}) provides a pedagogical introduction to the blackfold effective theory for higher-dimensional black holes, focusing on aspects immediately relevant for applications to metastable antibranes. Chapters \ref{EX}, \ref{STA}, and \ref{THE} (summarised in sections \ref{IN3}, \ref{IN4}, and \ref{IN5}) are developed from our three research questions, providing discussions on respectively the existence, the classical stability, and the thermal transitions of metastable antibranes. To avoid repetition, for the study of the anti-D3-NS5 state in chapter \ref{EX}, we provide comprehensive explanations on the blackfold construction of the metastable state but only a brief discussion on the thermal effects. On the other hand, for the study of the anti-M2-M5 state in chapter \ref{THE}, while the discussion on the blackfold components is short, the discussion on the effects of nonextremality is thorough. Lastly, in chapter \ref{IN6}, we conclude with a presentation of the outlook.

The 4 appendices are integral to the thesis. They contain a presentation of the preliminaries, crucial components that are shared by different chapters in the main text, as well as important calculations. In particular, appendix \ref{EG} provides the preliminaries on embedding geometry. Appendix \ref{CON} discusses relevant throat geometries and prepares them for our applications. Appendix \ref{BS} presents the brane bound states and extracts from them relevant information. Appendix \ref{secB} holds the derivation of the blackfold perturbation equations for the KPV state. 

With references to the relevant appendices, the chapters of this thesis can be read individually as they are self-contained and each carries a complete story.

\section{Blackfolds: An effective theory for higher-dimensional black holes}
\label{IN2}

The \textit{blackfold} (short for \textit{black-manifold}) formalism \cite{Emparan:2009cs,Emparan:2009at} is an effective theory for higher-dimensional black holes, first developed in the study of neutral higher-dimensional black ring solutions in pure Einstein gravity \cite{Emparan:2007wm}. In this thesis (chapter \ref{BF}), we introduce the blackfold approach in two steps: (i) derive the blackfold equations as effective branes dynamical equations and (ii) demonstrate that the blackfold equations provide information on the backreacted (super)gravity description of the branes configurations.  

For step $(i)$, we begin with a derivation of the \textit{effective dynamical equations} for charged branes in fluxed backgrounds from Dirac action (section \ref{BF1}) \cite{Carter:2000wv,Armas:2017pvj} and from conservation principles (section \ref{BF2}) \cite{Armas:2016mes}. These effective dynamical equations can be written in term of \textit{forced conservation equations}, which for the example of branes in Einstein-Maxwell theory take the form:
\begin{align}
\label{in1}
&\nabla_a T^{a b} = \p^b X_\mu \, \mathcal{F}^\mu ~ , & &T^{ab} K_{ab}^{\,\,\, \,\,\, (i)}  = \mathcal{F}^\mu \, n^{(i)}_\mu ~, \\
\label{in2}
&\nabla_{a_1} (J_{q+1})^{a_1 ... a_{q + 1}} = 0 ~ , & &\nabla_{a_1} (\mathcal{J}_{D - q - 3})^{a_1 ... a_{D - q - 3}} = 0 
\end{align}
where $T^{a b}$, $J_{q+1}$, and $\mathcal{J}_{D - q - 3}$ are respectively the energy-stress tensor, the electric current, and the magnetic current carried by the branes. The force term $\mathcal{F}$ is given by
\begin{multline}
\mathcal{F}^\mu = \frac{1}{(q+1)!} \left(F_{q+2} \right)^{\mu a_1 ... a_{q+1}} (J_{q+1})_{a_1 ... a_{q+1}} \\
+ \frac{1}{(D-q-3)!} (F_{D-q-2})^{\mu a_1 ... a_{D-q-3}} (\mathcal{J}_{D-q-3})_{a_1 ... a_{D-q-3}} ~ .
\end{multline}

To make use of equations \eqref{in1}-\eqref{in2}, we have to write down an expression for the currents carried by the branes: $T^{\mu \nu}$, $J_{q+1}$, and $\mathcal{J}_{D - q - 3}$. Generally, as these currents depend on the background profile, this is not an easy task. However, if we can tune the parameters of  our branes/background configurations to the \textit{blackfold regime} where the configurations possess a large separation of scales: $\mathcal{R}, \mathcal{R}_E \gg r_b$ where $\mathcal{R}$, $\mathcal{R}_E$, and $r_b$ are respectively the characteristic length scale of the background, the curvature radius of the bending in the configuration, and the characteristic near horizon scale of the branes, we can approximate these currents by a set of \textit{equivalent currents}, which can be computed from the branes' uniform flat-space solutions (section \ref{BF3}). 

As $\mathcal{R} \gg r_b$, the leading order effective branes dynamics is probe dynamics, e.g. backreactions of the branes to the background profile can be ignored. By substituting the set of equivalent currents into the forced conservation equations \eqref{in1}-\eqref{in2}, we obtain the \textit{blackfold equations}. In this derivation, it is clear that the blackfold equations describe the effective dynamics of branes in fluxed background. We note that the presented derivation can be straightforwardly generalised to branes in a more general (super)gravity theory \cite{Armas:2016mes}.

For step $(ii)$, we start with a discussion of the \textit{matched asymptotic expansion} (MAE) \cite{Harmark:2003yz,Gorbonos:2004uc}, a procedure for constructing perturbatively the backreacted description of the branes configurations everywhere in spacetime  by matching two asymptotic regions: the near zone ($r \ll \mathcal{R}, \, \mathcal{R}_E$) where the profile can be approximated by a seed solution and the far zone ($r \gg r_b$) where the profile can be approximated by a background solution. Applying this procedure to the example of bending black branes in flat space, we show explicitly that \textit{the blackfold equations provide the necessary conditions for the MAE procedure} (section \ref{BF4}) \cite{Camps:2010br, Camps:2012hw}.  In a similar manner, this statement can also be proven for generic branes/background configurations in various (super)gravity theories \cite{Armas:2016mes}.

In the case of bending black branes, one can show that \textit{the blackfold equations provide also the sufficient conditions for the MAE procedure} \cite{Camps:2012hw}. Naturally, we introduce the \textit{blackfold conjecture} \cite{Niarchos:2015moa}, which is the generalisation of this statement to generic configurations, that there is \textit{a one to one correspondence between a solution of the blackfold equations and a regular solution of the gravitational equations}. This conjecture is almost analogous to the statement in Fluid/Gravity \cite{Bhattacharyya:2008jc} that there is a one to one map between a solution of the fluid equations and a regular solution of the gravitational equations.   

From the perspective of effective branes dynamics, blackfold equations can be easily derived. As one can show that these equations provide the necessary conditions and perhaps also the sufficient conditions for a leading order matched asymptotic description of the backreacted configurations, they can be used to extract information on such description. This is the essence of the blackfold approach. Evidence for the power of such approach can be found in the wealth of examples in the literature, e.g. \cite{Emparan:2009vd, Caldarelli:2010xz, Armas:2012bk, Niarchos:2012pn}. 

\section{On the existence of metastable antibranes}
\label{IN3}
For discussions on the existence of metastable antibranes, our specimen is the Kachru-Pearson-Verlinde (KPV) conjectured metastable state of polarised anti-D3 branes, or equivalently wrapped NS5 branes with dissolved anti-D3 brane charge (wrapped anti-D3-NS5 branes), at the tip of the Klebanov-Strassler (KS) throat \cite{Kachru:2002gs, Klebanov:2000hb}. Originally in \cite{Kachru:2002gs}, it was shown from brane probe analysis that, in the regime of $p/M$\footnote{$p$ denotes the number of the anti-D3 branes and $M$ the strength of the Klebanov-Strassler background flux.} between 0 and $p_{crit}$ with $p_{crit} \approx 0.080488$, the anti-D3-NS5 brane can balance its own ``weight'' with ``electromagnetic'' forces from the fluxes to form a metastable state at the tip of the KS throat.

The existence of this KPV state has been refuted in various works starting with the investigations of \cite{Bena:2009xk}. The problem, found at the time, arises when trying to go beyond the probe limit and investigate what happens once the branes backreact. In particular, \cite{Bena:2009xk} and many subsequent works \cite{Gautason:2013zw} found that the backreacted profile of localised anti-D3 branes had singular 3-form fluxes in such a way that it would cause immediate brane-flux decay \cite{Blaback:2012nf}. It is further noted in \cite{Bena:2014jaa} that there is a conflict in regimes of validity in the original DBI derivation of the KPV state in \cite{Kachru:2002gs}. As a response, \cite{Michel:2014lva} argued that the singularity can be renormalised in such a way that does not affect stability when $p=1$, which is a case that is not amenable to a supergravity analysis. Subsequently, \cite{Cohen-Maldonado:2015ssa, Cohen-Maldonado:2016cjh}  argued that metastability can also be retained when $p\gg 1$ since the observed singularities cannot be proven to exist once one backreacts spherical NS5 branes instead of point-like anti-D3 branes. In fact, all proofs of unphysical singularities rested on an assumption which was in contradiction with KPV from the start, since polarised anti-D3 metastable states are really NS5 states. 

Several studies \cite{Bena:2012ek,Bena:2013hr, Blaback:2014tfa, Hartnett:2015oda} have investigated the effect of adding temperature to the anti-D3 branes. Most of these works were motivated by the would-be singularity in the 3-form fluxes. Whether or not a singularity can be cloaked by a horizon that arises when moving away from extremality is believed to be an important criterium for deciding the fate of singularities \cite{Gubser:2000nd}. Although strong indications were found that one should not worry about singularities at all, \cite{Michel:2014lva, Cohen-Maldonado:2015ssa,Cohen-Maldonado:2016cjh}, it remains an outstanding problem to understand what happens when the antibranes are at finite temperature. 

In this thesis (chapter \ref{EX}), we present a finite-temperature study on the backreacted profile of anti-D3-NS5 branes in KS throat, recovering the KPV state at extremality \cite{Armas:2018rsy}. To study the KPV and its thermal analogue state, we are interested in the type IIB supergravity solution describing wrapped anti-D3-NS5 branes at the tip of the KS throat. Because an exact description of such configuration is technically unfeasible, we turn to approximate descriptions. In particular, a perturbative description of the configuration can be obtained through the technique of \textit{matched asymptotic expansion}, where the solution is approximated in the far zone by the background solution of interest, here the KS throat, and in the near zone by an uniform flat-space p-brane solution, here the D3-NS5 bound state. As the blackfold equations provide the \textit{necessary conditions} and perhaps also the \textit{sufficient conditions} for the leading order matched asymptotic solution, we make use of such equations to learn about the configuration of anti-D3-NS5 branes in KS throat.

After introducing the KPV state in section \ref{extra2}, we derive, in section \ref{EX1}, the blackfold equations for nonextremal anti-D3-NS5 branes at the tip of the KS throat, recovering consistently in the supergravity regime ($g_s p \gg 1$) the KPV \cite{Kachru:2002gs} results at extremality. In section \ref{EX2}, we provide a discussion on the regime of validity of our analysis. Subsequently, in section \ref{EX3}, we describe effective potentials for nonextremal anti-D3-NS5 branes and study the effects of nonextremality on the conjectured KPV metastable state. In particular, we observe that, as soon as the branes become nonextremal, an additional unstable vacuum appears. This is a novel, `fat' unstable NS5-brane state. Increasing the entropy of the solutions leads to a merger of the fat unstable state with the thin metastable state, annihilating both vacua. By plotting the ratio $d \sim (\text{NS5 Schwarzschild radius}/ \text{Radius of the wrapped } S^2)$, we show that the origin of this transition is closely related to the geometric properties of the corresponding black hole horizon.

The picture from our blackfold analysis is suggestively consistent with the exact analysis of \cite{Cohen-Maldonado:2015ssa}. A no go-theorem, based on singularities, implies that the black NS5 solution, if it exists, has a maximum horizon radius until it disappears. A black anti-D3 solution also escapes the no go, but was deemed unphysical in \cite{Cohen-Maldonado:2015ssa}, since it does not persist in the extremal limit. We have now shown that, in all cases where \cite{Cohen-Maldonado:2015ssa} lifted the no-go theorem, our blackfold analysis produce a go with concrete quantitative predictions. Since the prevailing picture regarding the singularities has now been argued for in complementary regimes, we believe the findings constitute strong evidence for the existence of the metastable state.

In the spirit of keeping this summary in appropriate length, we have been brief in our explanations. For detailed discussions on how the anti-D3-NS5 blackfold results together with existing literature on the topic form an argument for the existence of the metastable state, we refer readers to section \ref{EX4}.

\section{On the stability of metastable antibranes}
\label{IN4}
The claims regarding metastability of antibranes in warped throats from DBI analysis in \cite{Kachru:2002gs,Klebanov:2010qs}  and subsequently from the blackfold approach in \cite{Armas:2018rsy,M2M5brane} (presented in chapter \ref{EX} and \ref{THE}) only refer to a balance of force that allows the branes to be in equilibrium in certain directions and not statements of robustness under perturbations. In particular, what is shown for the anti-D3-NS5 branes in \cite{Kachru:2002gs}, and subsequently in \cite{Armas:2018rsy}, is that a homogeneous, spherical NS5 state living at the tip of the KS throat feel a balance of ``electromagnetic'' forces pushing it over the $S^3$ and gravitational forces doing the opposite. Based on these works alone, we cannot say whether the state is short-lived or robust under perturbations. For their numerous applications, particularly cosmological string de Sitter construction \cite{Kachru:2003aw}, it is important to determine whether these branes are truly metastable.

In \cite{Bena:2014jaa}, from the perspective of localised anti-D3 branes, it was argued that there exists a direction along which the branes feel repulsive forces among themselves and destabilise away from the KPV state. This suggests that, in an appropriate regime of parameters, the KPV configuration suffers from fragmentation instability. 

From the complementary perspective of anti-D3-NS5 branes\footnote{For a discussion on the localised D3 perspective versus the spherical NS5 perspective, we refer readers to section \ref{EX4}.}, we study the stability of the KPV state using the blackfold approach. Before summarising our results, let us stress what our analysis does \textit{not} do. As blackfold is based on the idea of matched asymptotic expansion, one need to specify a seed metric as the description of the solution in the near zone. By choosing the stacked anti-D3-NS5 branes solution as the near zone seed, we have effectively ignored all brane splitting and fragmentation deformations. Moreover, as noted in chapter \ref{EX}, the analysis is reliable when $p/M$ is not too close to zero, at which point the size of the NS5 sphere shrinks ($\psi \approx 0$) and the localised anti-D3 perspective becomes the better description. Since the analysis in \cite{Bena:2014jaa} is done from the localised anti-D3 branes perspective and the discovered instabilities are brane splitting instabilities, the blackfold results presented here should be thought of as complimentary and not contradictory to that of \cite{Bena:2014jaa}. Another important caveat is that, as blackfold theory is an effective theory of long-wavelength interactions, our claim of stability is made only with respect to long-wavelength perturbations. 

Let us note that a preliminary study of the stability of the KPV state was done in \cite{Bena:2015kia} where it was argued that the spherical NS5 shell is unstable under perturbations. While keeping in mind that the regime of validity of the D3 perspective analysis done in \cite{Bena:2015kia} and ours are different, as we shall see shortly, our results do not support the picture proposed in \cite{Bena:2015kia}.  

In chapter \ref{STA}, by introducing generic long-wavelength deformations to the blackfold description of the KPV state, we observe that the blackfold equations (constraints on long-wavelength deformations) prohibit the existence of tachyonic modes. It is interesting to mention also that counter-intuitively, the KPV state, a polarised state of anti-D3 branes, can feel an electromagnetic repulsion away from the tip of the KS throat. Nevertheless, this electromagnetic repulsion is ``out-weighted'' by the gravitational pull so the KPV state is still stabilised radially by a net force downward.

Even though this result is evidence supporting the metastability of the KPV state under generic perturbations, since our current approach cannot observe directly\footnote{It is interesting to note that, in certain cases, the leading order blackfold equations can detect the onset of a fragmentation instability. For example, the onset of the Gregory-Laflamme (GL) instability in black strings \cite{Gregory:1993vy, Hovdebo:2006jy} and black rings \cite{Santos:2015iua} whose end point is fragmentation \cite{Lehner:2010pn, Figueras:2015hkb} can be observed by an analogous blackfold stability analysis \cite{Emparan:2009at,Armas:2019iqs}. } fragmentation instabilities (if they persist in the large NS5 sphere regime, $M \gg 1$)\footnote{As the regime of validity of the analysis in \cite{Bena:2014jaa, Bena:2015kia} is that of small/finite size NS5 sphere, it is possible that the instability observed there does not persist in the large sphere regime. An analogous example of this picture can be found in the study of black rings where a fragmentation instability (elastic mode instability) is observed for fat rings but is not observed for very thin rings \cite{Figueras:2015hkb, Armas:2019iqs}.}, we believe that further works needed to be done to reach a conclusive statement.

\section{On the thermal transition of metastable antibranes}
\label{IN5}
For the study of thermal transition of metastable antibranes, our exemplar candidate is the Klebanov-Pufu (KP) metastable state of polarised anti-M2 branes, or equivalently M5 branes with dissolved anti-M2 brane charge (wrapped anti-M2-M5 branes), at the tip of the Cvetic-Gibbons-Lu-Pope (CGLP) throat \cite{Klebanov:2010qs, Cvetic:2000db}. Originally in \cite{Klebanov:2010qs}, it was shown that, in the regime of $p/\tilde M \leq \mathfrak p_{crit} \simeq 0.0538$\footnote{$p$ denotes the number of the anti-M2 branes and $\tilde M$ the strength of the CGLP background flux.}, the anti-M2-M5 brane balances its own ``weight'' with ``electromagnetic'' forces from the fluxes to form a metastable state at the tip of the CGLP throat.

The anti-M2-M5 metastable state is the eleven-dimensional supergravity analogue of the type IIB supergravity anti-D3-NS5 metastable state discussed previously. As such, in our blackfold study of the anti-M2-M5 state, we unsurprisingly observed that many of the properties proven for the anti-D3-NS5 state also hold true for the anti-M2-M5 state. To avoid repetition, here and in the main text, we shall not discuss these common properties but instead focus on exploring the \textit{exotic} pattern of thermal transition of the anti-M2-M5 state, which is different from that of the anti-D3-NS5 state.  

In this thesis (chapter \ref{THE}), through an application of the blackfold formalism, we present a black hole phase diagram that is not only consistent with the no-go theorems of \cite{Cohen-Maldonado:2016cjh} but also reveals new unexpected patters of finite-temperature transitions. At zero temperature, we show that the blackfold equations recover faithfully the abelian DBI equations used by KP in \cite{Klebanov:2010qs} and the same extremal metastable vacuum that they found. At finite, sufficiently small temperature, we uncover (in direct analogy to the case of anti-D3-NS5 branes) three main branches of wrapped M5 black brane solutions: a fat unstable state, a metastable state and a thin unstable state. The terms `fat' and `thin' refer to the relative size of the $S^3$ that the M5 brane wraps and the size of the Schwarzschild radius. The behaviour of these branches at higher temperatures depends on the value of $p/\tilde M$. In particular, we discover three separate regimes of $p/\tilde M$ (inside the window of the metastable state, $p/\tilde M \lesssim 0.054$) that exhibit different patterns of thermal transitions.

There is a low-$p/\tilde M$ regime where the anti-M2 physics in CGLP is very similar to the anti-D3 physics in KS. In this regime, there is a single finite-temperature transition that involves the merger of a fat unstable black M5 with the metastable black M5. Beyond this merger the metastable state is lost. We present non-trivial quantitative evidence that supports the scenario where this merger is driven by properties of the horizon geometry. 

In addition, for the anti-M2 system we find two regimes of $p/\tilde M$ that have no known counterpart in the system of anti-D3 branes in KS. In the large-$p/\tilde M$ regime, there is a single merger between the metastable state and the {\it thin} unstable M5 brane state. In this case, unsurprisingly, there are no indications that the loss of the metastable state is driven by properties of the horizon geometry. In an intermediate regime of $p/\tilde M$, the phase diagram exhibits three (instead of one) transitions: two of them involve mergers of the metastable state with the thin unstable state and one involves a merger of the metastable state with the fat unstable state. These patterns are new, unexpected predictions of the blackfold formalism for the supergravity solution and the dual QFT.

\section{Notations and conventions}
\label{IN7}

\begin{itemize}{}

\item The signature is mostly plus $(- + + + ...)$.

\item Greek letters ($\alpha, \beta, \, ...$) are used for background indices. Latin letters ($a, b, \, ...$) are used for worldvolume indices. 

\item The Hodge star operator of a $p$-form on an $n$-dimensional manifold is defined as 
\be
(* A)_{\mu_1 ... \mu_{n-p}} = \frac{1}{p!} \epsilon_{\nu_1 ... \nu_p \mu_1 .... \mu_{n-p}} A^{\nu_1 ... \nu_p}
\ee
with $\epsilon_{\nu_1 ... \nu_p \mu_1 .... \mu_{n-p}}$ the Levi-Civita tensor.

\item Electric currents appear with a $-$ sign in the sourced Maxwell equations: 
\be
d \star F_{p+2} = - 16 \pi G \ J_{p+1} ~ .
\ee

\item Magnetic currents appear with a $+$ sign in the sourced Maxwell equations: 
\be
d F_{p + 2} = 16 \pi G  \ j_{n - q - 3}  ~ .
\ee

\item The type IIB supergravity action is given by
\begin{multline}
\mathcal{I}_{IIB} = \frac{1}{16 \pi G} \int_{\mathcal{M}_{10}} d^{10} x \Bigg\{ \sqrt{- g} \Big[ e^{- 2 \phi} \left( R + 4 \p_\mu \phi \p^\mu \phi - \frac{1}{2} | H_3|^2 \right) \\- \frac{1}{2} |\tilde{F}_1|^2 - \frac{1}{2} | \tilde{F}_3|^2 - \frac{1}{4} |\tilde{F}_5|^2 \Big] 
- \frac{1}{2} C_4 \wedge H_3 \wedge F_3
\Bigg\}
\end{multline}
where the gauge invariant field strengths are defined as
\be
\tilde{F}_{q+2} = F_{q+2} - H_3 \wedge C_{q-1} 
\ee
with the exception of the self-dual $\tilde{F}_5$ which is defined as
\be
\tilde{F}_5 = F_5 + B_2 \wedge F_3
\ee
where $F_{q+2} \equiv d C_{q+1}$. As we shall be using the type IIB supergravity equations a lot in this thesis, for the convenience of the readers, let us write them explicitly here. In particular, in our conventions, the type IIB supergravity equations are made up of the following equations:

The $\phi$ equation
\be
4 e^{2 \phi} \nabla^\mu \left( e^{-2 \phi} \p_\mu \phi \right) + R + 4 \p_\mu \phi \p^\mu \phi - \frac{1}{12} (H_3)_{\mu_1 \mu_2 \mu_3} (H_3)^{\mu_1 \mu_2 \mu_3}  = 0 ~ .
\ee
The $B_2$ equation
\be
d \left( e^{- 2\phi} \star H_3 - \star \tilde{F}_3 \wedge C_0 - \frac{1}{2} \tilde{F}_5 \wedge C_2 + \frac{1}{2} C_4 \wedge F_3  \right) = 0 ~ .
\ee
The $C_0$ equation
\be
d (\star F_1) + H_3 \wedge \star \tilde{F}_3 = 0 ~ .
\ee
The $C_2$ equation
\be
d \left( \star \tilde{F}_3 \right) + H_3 \wedge \star \tilde{F}_5 = 0 ~ .
\ee
The $C_4$ equation
\be
d \left( \star \tilde{F}_5 \right) - H_3 \wedge F_3 = 0 ~ .
\ee
And finally, the $g_{\mu \nu}$ equation
\be
e^{- 2 \phi} G^{\mu \nu} +  2 \nabla^\mu \left( e^{- 2 \phi} \p^\nu \phi \right) - 2 \nabla_\rho \left( e^{- 2 \phi} \p^\rho \phi \right) g^{\mu \nu} = T^{\mu \nu}_{(\phi)} + T^{\mu \nu}_{(H_3)} + T^{\mu \nu}_{(F_1)} + T^{\mu \nu}_{(F_3)} + T^{\mu \nu}_{(F_5)} 
\ee
with 
\begin{align}
T^{\mu \nu}_{(\phi)} &= 4 e^{- 2 \phi} \left( \p^\mu \phi \p^\nu \phi - \frac{1}{2} g^{\mu \nu} \p_\lambda \phi \p^\lambda \phi \right) \\
T^{\mu \nu}_{(H_3)} &= \frac{e^{-2 \phi}}{4} \left( H_3^{\mu \mu_1 \mu_2} H^{\ \nu}_{3 \ \mu_1 \mu_2}  - \frac{1}{6} g^{\mu \nu} |H_3|^2\right) \\
T^{\mu \nu}_{(F_1)} &= \frac{1}{2} \left( F_1^\mu F_1^\nu - \frac{1}{2} g^{\mu\nu} |F_1|^2 \right) \\
T^{\mu \nu}_{(F_3)} &= \frac{1}{4} \left( \tilde{F}_3^{\mu \mu_1 \mu_2} \tilde{F}^{ \ \nu}_{3 \  \mu_1 \mu_2} - \frac{1}{6} g^{\mu \nu} |\tilde{F}_3|^2 \right) \\
T^{\mu \nu}_{(F_5)} &= \frac{1}{2} \frac{1}{48} \left( \tilde{F}_5^{\mu \mu_1 ... \mu_4} \tilde{F}^{\ \nu}_{5 \ \mu_1 ... \mu_4} - \frac{1}{10} g^{\mu \nu} |\tilde{F}_5|^2 \right) 
\end{align}
where $|F_{p}|^2 = \frac{1}{p!} (F_p)_{\mu_1 ... \mu_p} (F_p)^{\mu_1 ... \mu_p}$.

\item The eleven-dimensional supergravity action is given by
\be
\mathcal{I}_{M} = \frac{1}{16 \pi G} \int_{\mathcal{M}_{11}} \left[ \star R - \frac{1}{2} G_4 \wedge \star G_4 - \frac{1}{6}  A_3 \wedge G_4 \wedge G_4  \right] ~ .
\ee
\end{itemize}

\chapter{Blackfolds: An effective theory for higher-dimensional black holes}\label{BF}
The purpose of this chapter is to introduce an effective theory for higher-dimensional black holes named \textit{blackfolds} \cite{Emparan:2009cs,Emparan:2009at,Armas:2016mes}. We begin with a discussion of brane dynamics, deriving the \textit{branes effective dynamical equations} from Dirac action in section \ref{BF1} and from conservation principles in section \ref{BF2}. Subsequently, in section \ref{BF3}, we introduce the \textit{blackfold equations} and demonstrate how such equations can be used to describe the dynamics of generic branes in generic background. In section \ref{BF4}, we discuss blackfolds in the framework of \textit{matched asymptotic expansion} (MAE). Through the example of bending black branes in flat space, we show that the same blackfold equations provide the \textit{necessary conditions} for a matched asymptotic description of the backreacted configurations. We further discuss the \textit{blackfold conjecture}, which states that the blackfold equations provide also the \textit{sufficient conditions} for a matched asymptotic solution. Lastly, in section \ref{BF5}, we note down the blackfold equations for configurations in type IIB supergravity and eleven-dimensional supergravity as their explicit forms are important for discussions in later chapters. 

\section{Brane dynamics from Dirac action}\label{BF1}
Let us start with the Dirac action for a p-brane embedded in a D-dimensional manifold in a standard Einstein-Maxwell type theory, coupling to a background metric $g_{\mu \nu}$ and a $(p+1)$-form gauge field $A_{\mu_1 ... \mu_{p+1}}$:
\begin{align}
\label{bf1}
\mathcal{I}_{Dirac} &= \int_{\mathcal{W}_{p+1}} d^{p+1} \sigma \sqrt{- \gamma} \,\,\, \mathcal{L} \, \left( A_{\mu_1 ... \mu_{p+1}}, g_{\mu\nu} \right) \\
&= - T \int_{\mathcal{W}_{p+1}} d^{p+1} \, \sigma \sqrt{- \gamma} \, +  \, Q_p \int_{\mathcal{W}_{p+1}} \mathbb{P} [ A_{p+1}] 
\end{align}
with $\mathcal{W}_{p+1}$ the worldvolume of the p-brane, $\sigma$ the worldvolume coordinates, $\gamma$ the determinant of the worldvolume induced metric $\gamma^{ab}$, $\mathcal{L}$ the Lagrangian density, $T$ the tension of the brane, $Q_p$ the charge of the brane under the $A_{p+1}$ gauge field, and $\mathbb{P}[A_{p+1}]$ the pullback of $A_{p+1}$ to the worldvolume. As an example, for a point particle coupling to the background metric and the Maxwell gauge field with mass $m$ and charge $e$ respectively, the action takes the form
\be
\label{bf2}
\mathcal{I}_{\, e} = \int_{\mathcal{W}_{1}} d \tau \sqrt{- \gamma} \,\,\, \left( - m + \frac{e}{\sqrt{- \gamma}} \p_\tau X^\rho A_\rho \right)
\ee
where $\gamma = g_{\mu \nu} \p_\tau X^\mu \p_\tau X^\nu$.

If we vary the background fields $g_{\mu \nu}$ and $A_{\mu_1 ... \mu_{p+1}}$ in \eqref{bf1}, we obtain the variational equation 
\be
\label{bf3}
\delta \mathcal{I}_{Dirac} = \int_{\mathcal{W}_{p+1}} d^{p+1} \sigma \sqrt{- \gamma} \left( \frac{1}{2} {T}^{\mu \nu} \delta g_{\mu \nu}  + \frac{1}{(p+1)!} {J}^{\mu_1 ... \mu_{p+1}} \delta A_{\mu_1 ... \mu_{p+1}} \right)
\ee
where 
\begin{align}
\label{bf10}
&{T}^{\mu \nu} = 2 \frac{\p {\mathcal{L}}}{\p g_{\mu \nu}} + {\mathcal{L}} \,\, h^{\mu \nu} ~ ,    & {J}^{\mu_1 ... \mu_{p+1}} = (p+1)! \frac{\p {\mathcal{L}}}{ \p A_{\mu_1 ... \mu_{p+1}}} ~ .
\end{align}

Considering the gauge field variation $\delta A_{\mu_1 ... \mu_{p+1}} = \nabla_{[\mu_1} \chi_{... \mu_{p+1}]}$, as such variation is a pure gauge transformation, we have that $\delta \mathcal{I}_{Dirac}$ has to vanish. Explicitly, we have
\be
\delta \mathcal{I}_{Dirac} = \int_{\mathcal{W}_{p+1}} d^{p+1} \sigma \sqrt{- \gamma} \left({J}^{\mu_1 ... \mu_{p+1}} \nabla_{\mu_1} \chi_{... \mu_{p+1}} \right) = 0 ~ .
\ee
Assuming that our currents vanish at the boundaries, by integration by parts, we obtain
\be
\label{bf4}
\nabla_{\mu_1} {J}^{\mu_1 ... \mu_{p+1}} = 0 ~ .
\ee
It is important to note that ${J}^{\mu_1 ... \mu_{p+1}}$ is a worldvolume projected tensor\footnote{For further discussions on embedding geometry and projected tensors, see appendix \ref{EG}.}. For example, in the case of the point charge \eqref{bf2}, such current is given by $\frac{e}{\sqrt{- \gamma}} \p_\tau X^\rho$, which is clearly a projected tensor. Therefore, the current conservation equation becomes 
\be
\overline{\nabla}_{\mu_1} J^{\mu_1 ... \mu_{p+1}} = 0
\ee
where $\overline{\nabla}_{\mu} = h^{\nu}_{\mu} \nabla_{\nu}$. As proven in \eqref{a1}, this equation is equivalent to the worldvolume conservation equation:
\be
\label{bf9}
\nabla_{a_1} J^{a_1 ... a_{p+1}} = 0 
\ee
where $\nabla_a$ is a covariant derivative with respect to the induced metric.

Turning our attention to the metric variation $\delta g_{\mu \nu}$, the analogous gauge symmetry is diffeomorphism, which is nothing but the ``active'' version of a coordinate transformation. The difference between diffeomorphism and the $(p+1)$-form gauge symmetry above is that diffeomorphism acts on both the metric $g_{\mu \nu}$ and the $(p+1)$-form gauge field $A_{\mu_1 ... \mu_{p+1}}$. Under an infinitesimal diffeomorphism, the background gauge fields vary as
\begin{align}
\delta A_{\mu_1 ... \mu_{p+1}} &= \pounds_\xi A_{\mu_1 ... \mu_{p+1}} = \xi^\sigma \nabla_\sigma A_{\mu_1 ... \mu_{p+1}} + (p+1)! A_{\sigma [ \mu_1 ....  }  \nabla_{\mu_{p+1]}} \xi^\sigma ~ , \\
\delta g_{\mu \nu} &= \pounds_{\xi} g_{\mu \nu} = 2 \nabla_{( \mu} \xi_{\nu )}
\end{align}
where $\pounds$ denotes Lie derivative and $\xi^\rho$ is a vector specifying the direction of infinitesimal displacement. Substituting these variations into \eqref{bf3}, we have
\begin{multline}
\delta \mathcal{I}_{Dirac} = \int_{\mathcal{W}_{p+1}} d^{p+1} \sigma \sqrt{- \gamma} \Bigg(  {T}^{\mu \nu} \, \nabla_{\mu} \, \xi_{\nu } \\
+ \frac{1}{(p+1)!}  {J}^{\mu_1 ... \mu_{p+1}} \Big( \xi^\nu \nabla_\nu A_{\mu_1 ... \mu_{p+1}} + (p+1)! \, A_{\nu \mu_1 ....  }  \nabla_{\mu_{p+1} } \xi^\nu \Big) \Bigg) ~ .
\end{multline}
Integrating by parts, assuming all our currents vanish at the boundaries, and making use of \eqref{bf4}, we have 
\be
\label{bf5}
\delta \mathcal{I} = \int_{\mathcal{W}_{p+1}} d^{p+1} \sigma \sqrt{- \gamma} \, \xi_\nu \left( \frac{1}{(p+1)!}  F^{\nu \mu_1 ... \mu_{p+1}} {J}_{\mu_1 ... \mu_{p+1}} - {\nabla}_{\mu} \, {T}^{\mu \nu} \right) = 0
\ee
where $F_{p+2} \equiv d A_{p+1}$. Analogous to before, as $T^{\mu \nu}$ is a worldvolume projected tensor, the divergence of ${T}^{\mu \nu}$ should be computed with $\overline{\nabla}$ instead of $\nabla$. As $\xi^\nu$ is arbitrary, from \eqref{bf5},  we have 
\be
\label{bf6}
\overline{\nabla}_\mu T^{\mu \nu} = \mathcal{F}^{\nu}
\ee
with 
\be
\label{bf14}
\mathcal{F}^{\nu} = \frac{1}{(p+1)!} F^{\nu a_1 ... a_{p+1}} {J}_{a_1 ... a_{p+1}} ~ .
\ee
As discussed in \eqref{a2}-\eqref{a3}, we can decompose equation \eqref{bf6} into \textit{intrinsic} and \textit{extrinsic} equations given respectively by
\begin{align}
\label{bf7}
\nabla_a T^{a b} &= \p^b X_\mu \, \mathcal{F}^\mu ~ , \\
\label{bf8}
T^{ab} K_{ab}^{\,\,\, \,\,\, (i)}  &= \mathcal{F}^\mu \, n^{(i)}_\mu
\end{align}
where $K_{\mu \nu}^{\ \ \, (i)} \equiv K_{\mu \nu}^{\ \ \, \rho} n^{(i)}_{\rho}$ with $n_\rho^{(i)}$ being the normal vectors of the introduced branes.

Obtained from the consideration of the Dirac brane action, the current conservation equation \eqref{bf9}, the intrinsic \eqref{bf7}, and extrinsic \eqref{bf8} equation form a set of \textit{effective dynamical equations} for a p-brane coupling to a background metric $g_{\mu \nu}$ and a $(p+1)$-form gauge field $A_{\mu_1 ... \mu_{p+1}}$. As an illustration, let us apply these equations to the point charge \eqref{bf2}. In such case, the currents are computed using \eqref{bf10} to be $T^{\tau \tau} = - m \, \gamma^{\tau \tau}$ and $J = e \, d \tau$. The current conservation equation, the intrinsic, and extrinsic equation become respectively
\begin{align}
\label{bf11}
\p_\tau e &= 0 ~ , \\
\label{bf12}
\p_\tau m &=  0  ~ ,\\
\label{bf13}
m \, \p^2_\tau X^\rho  &= - m \, \Gamma^{(i)}_{\tau \tau} + e \, \p^\tau X_{\mu}  \,   F^{\mu \nu} \, n_{\nu}^{(i)}
\end{align}
where
\be
\Gamma^{(i)}_{\tau \tau} \equiv \Gamma^{\rho}_{\mu \nu} \p_{a} X^{\mu} \p_{b} X^{\nu} n_{\rho}^{(i)} ~ . 
\ee
Equations \eqref{bf11} and \eqref{bf12} are trivial, however, the extrinsic equation \eqref{bf13} has a very interesting interpretation: it is the generalisation of Newton's second law $F= m a$. In particular, we see that the LHS of equation \eqref{bf13} is already in the form $m a$ while the RHS describes the generalisation of the force term with both the gravitational effects (via the term $- m \, \Gamma^{(i)}_{\tau \tau}$) and the electromagnetic effects (via the term $e \, \p^\tau X_{\mu}  \,   F^{\mu \nu} \, n_{\nu}^{(i)}$). Let us note that, even for generic configurations, the extrinsic equation can still be written as 
\be
T^{ab} \p_a \p_b X^\rho = - T^{ab} \, \Gamma^{(i)}_{a b} +  \mathcal{F}^\mu \, n^{(i)}_\mu
\ee
and, thus, interpreted as the generalisation of Newton's $F = m a$. Of course, we can also consider branes that couple simultaneously to many gauge fields. In such case, the force term \eqref{bf14} shall be straightforwardly modified to include those terms. For further discussions on the dynamics of submanifolds from  action variational principles, we refer readers to the review by Carter \cite{Carter:2000wv} and references therein. One might also be interested in reading a modern generalisation of the formulation to surfaces with non-trivial edges in \cite{Armas:2017pvj}.  

\section{Brane dynamics from conservation equations}\label{BF2}
In this section, we demonstrate how one can obtain the effective dynamical equations for branes coupling to non-trivial background fields from conservation principles, taking Einstein-Maxwell theory as an illustrative example. Let us start with the action for the Einstein-Maxwell theory:
\be
\mathcal{I}_{EM} =  \frac{1}{16 \pi G} \int_{\mathcal{M}_{D}} \star R  - \frac{1}{2} F_{q+2} \wedge \star F_{q+2} ~ .
\ee
From this action, we obtain respectively the Einstein and Maxwell equations:
\begin{align}
\label{bf15}
G_{\mu \nu} - 8 \pi G \, T_{ \textbf{M} \mu \nu} &= 0 ~ , \\
\label{bf16}
d \, \star F_{q+2} &= 0  
\end{align}
where $G_{\mu \nu}$ is the standard Einstein tensor, and $T_{ \textbf{M} \mu \nu}$ is the electromagnetic energy-stress tensor:
\begin{align}
 T_{ \textbf{M} \mu \nu} &= \frac{2}{\sqrt{-g}} \frac{\delta \mathcal{I}_{\text{matter}}}{ \delta g_{\mu \nu}}\\
\label{bf21}
&= \frac{1}{16 \pi G \, (q+1)!} \left ( \left( F_{q+2}\right)^{\mu \mu_1 ... \mu_{q+1}}  \left(F_{q+2}\right)^{\nu}_{\ \mu_1 ... \mu_{q+1}}  - \frac{1}{2(q+2)} g^{\mu \nu} \left( F_{q+2} \right)^2  \right ) ~ .
\end{align}
Note that for the complete set of Einstein-Maxwell equations, we also have to include the Bianchi identity for the $A$ gauge field:
\be
\label{bf17}
d F_{q+2} = 0   ~ .
\ee

Let us have a background with metric $g_{\mu \nu}$ and gauge field $A_{\mu_1 ... \mu_q}$, satisfying the Einstein-Maxwell equations \eqref{bf15}, \eqref{bf16}, and \eqref{bf17}. If we introduce to this background branes carrying energy-stress tensor $T_{\mathbf{B}}^{\mu \nu}$, electric current $J_{q+1}$, and magnetic current $\mathcal{J}_{D - q - 3}$, the effects are described by the Einstein equation
\be
\label{bf20}
G_{\mu \nu} - 8 \pi G \, T_{ \textbf{M} \mu \nu} = 8 \pi G \, T_{ \textbf{B} \mu \nu} ~ ,
\ee
the sourced Maxwell equation
\be
\label{bf18}
d \star F_{q +2} = - 16 \pi G \star  J_{q +1} ~ ,
\ee
and the modified Bianchi identity
\be
\label{bf19}
d F_{q + 2} = 16 \pi G  \star \mathcal{J}_{D - q - 3} ~ . 
\ee
For later convenience, note that equations \eqref{bf18}-\eqref{bf19} can also be equivalently expressed in index notation respectively as
\begin{align}
\label{bf22}
\nabla^\mu \left( (F_{q+2})_{ \mu \mu_1 ... \mu_{q+1}} \right) &= - 16 \pi G \, (J_{q+1})_{\mu_1 ... \mu_{q+1} } ~ ,\\ 
\label{bf23}
\p_{[\mu} (F_{q+2})_{\mu_1 ... \mu_{q+2}]} &= \frac{16 \pi G }{(D - q - 3)! \, (q+3)}  \ (\mathcal{J}_{D - q - 3})^{\nu_1 ... \nu_{D - q - 3}} \epsilon_{\nu_1 ... \nu_{D - q - 3} \, \mu ... \mu_{q+2}} 
\end{align}
where $\epsilon_{\nu_1 ... \nu_{D - q - 3} \, \mu ... \mu_{q+2}}$ is the Levi-Civita tensor. Taking the divergence of both sides of \eqref{bf20}, as $\nabla_\mu G^{\mu \nu} = 0$, we have
\begin{align}
\nabla_\mu T_{ \textbf{B}}^{\mu \nu} = &- \nabla_\mu  T_{ \textbf{M}}^{\mu \nu} \\
= &\frac{1}{(q+1)!}  \nabla_\mu \left(F_{q+2} \right)^{\mu }_{\ \mu_1 ... \mu_{q+1}}  (F_{q+2})^{\nu  \mu_1 ... \mu_{q+1}} \\
 &- \frac{1}{(q+2)!} F_{ q+2}^{ \mu \mu_1 ... \mu_{q+1}} g^{\nu \rho}  (q+3) \nabla_{[\rho}  F_{  q+2 \mu  \mu_1 ... \mu_{q+1}]}  \\
= &\frac{1}{(q+1)!} \left(F_{q+2} \right)^{\nu \mu_1 ... \mu_{q+1}} (J_{q+1})_{\mu_1 \mu_2 .... \mu_{q+1}} \\
&+ \frac{1}{(D-q-3)!} (F_{D-q-2})^{\nu \mu_1 ... \mu_{D-q-3}} (\mathcal{J}_{D-q-3})_{\mu_1 \mu_2 ... \mu_{D-q-3}}
\end{align}
where in the second line we have used \eqref{bf21} and in the third line we have used \eqref{bf22}-\eqref{bf23}. Note that we have also defined the field strength $F_{D-q-2}$ as $F_{D-q-2} \equiv \star F_{q+2}$. From equations \eqref{bf18}-\eqref{bf19}, we have the current conservation equations
\begin{align}
\label{extra6}
d \star J_{q+1} &= 0 ~ ,\\
\label{extra7}
d \star \mathcal{J}_{D - q - 3} &= 0
\end{align}
which, in index notation, take the form
\begin{align}
\nabla_{\mu_1} (J_{q+1})^{\mu_1 ... \mu_{q  + 1}} &= 0 ~ , \\
\nabla_{\mu_1} (\mathcal{J}_{D - q - 3})^{\mu_1 ... \mu_{D - q - 3}} &= 0 ~ .
\end{align}
As our introduced energy-stress tensor and currents are carried by branes, using the same arguments as that of the previous section, we can write the energy-momentum conservation equation as 
\begin{align}
\label{bf24}
\nabla_a T^{a b} &= \p^b X_\mu \, \mathcal{F}^\mu ~ , \\
T^{ab} K_{ab}^{\,\,\, \,\,\, (i)}  &= \mathcal{F}^\mu \, n^{(i)}_\mu
\end{align}
and the current conservation equations as
\begin{align}
\nabla_{a_1} (J_{q+1})^{a_1 ... a_{q + 1}} &= 0 ~ , \\
\label{bf25} 
\nabla_{a_1} (\mathcal{J}_{D - q - 3})^{a_1 ... a_{D - q - 3}} &= 0
\end{align}
where we have drop the subscript $\mathbf{B}$ from $T^{a b}$, and defined the force term $\mathcal{F}$ as
\begin{multline}
\mathcal{F}^\mu = \frac{1}{(q+1)!} \left(F_{q+2} \right)^{\mu a_1 ... a_{q+1}} (J_{q+1})_{a_1 ... a_{q+1}} \\
+ \frac{1}{(D-q-3)!} (F_{D-q-2})^{\mu a_1 ... a_{D-q-3}} (\mathcal{J}_{D-q-3})_{a_1 ... a_{D-q-3}} ~ .
\end{multline}
Equations \eqref{bf24}-\eqref{bf25} can be called \textit{forced conservation equations}\footnote{For discussions of analogous conservation equations in the framework of generalised global symmetries \cite{Gaiotto:2014kfa}, we refer readers to \cite{Hofman:2017vwr,Grozdanov:2016tdf}.}. Of course, it is straightforward to write down the analogous equations for branes coupling to more gauge fields in a more complicated gravity theory \cite{Armas:2016mes}.

Though our forced conservation equations may look identical to the branes effective dynamical equations derived in the previous section, there are subtleties that are worth discussing. In our derivation of equations \eqref{bf24}-\eqref{bf25}, we have only used generic properties of the gravitational theory while making no assumptions on the branes or the background\footnote{As opposed to the Dirac action treatment of branes dynamics where the assumptions are that the branes have negligible thickness, curvature and their backreactions are mild.}. As such, these equations should hold even for the fully backreacted solution. In particular, if we plug into \eqref{bf24}-\eqref{bf25} the (perturbative) backreacted expressions for the metric, the gauge field, and the currents, we will be able to extract information beyond the leading order effective dynamics. This possibility will be discussed further at a later stage (page \pageref{ab12}). For now, let us consider how one can use equations \eqref{bf24}-\eqref{bf25} to describe the leading order effective dynamics of branes. 

To make use of equations \eqref{bf24}-\eqref{bf25}, we have to write down an expression for the metric $g_{\mu \nu}$, the gauge field $A_{\mu_1 ... \mu_q}$, and the introduced currents $T^{a b}$,  $J_{q+1}$, and $\mathcal{J}_{D - q - 3}$. Imagine if we are in a regime of parameters where the leading order effective dynamics is the probe dynamics, i.e. backreactions from the introduced branes can be ignored, we can simply use the background metric $g_{\mu \nu}$ and gauge field $A_{\mu_1 ... \mu_q}$ in our computations. If the regime also allows us to approximate the introduced currents $T^{a b}$,  $J_{q+1}$, and $\mathcal{J}_{D - q - 3}$ with known equivalent currents, our task is done. A regime that satisfy both of the above is the \textit{blackfold regime}, which shall be discussed in the next section.

\section{Blackfold equations as dynamical equations}\label{BF3}
A thorough discussion of the blackfold approach usually begins with a discussion of the \textit{matched asymptotic expansion} (MAE), a procedure for constructing approximate solutions in gravity theories. However, for our first contact with the blackfold approach, it is beneficial for our intuition if we first introduce the blackfold equations as effective branes dynamical equations. Don't worry about missing out, though, a discussion of blackfolds in the framework of MAE will be provided in subsequent sections.

\begin{figure}\centering
\includegraphics[width= 0.6 \textwidth]{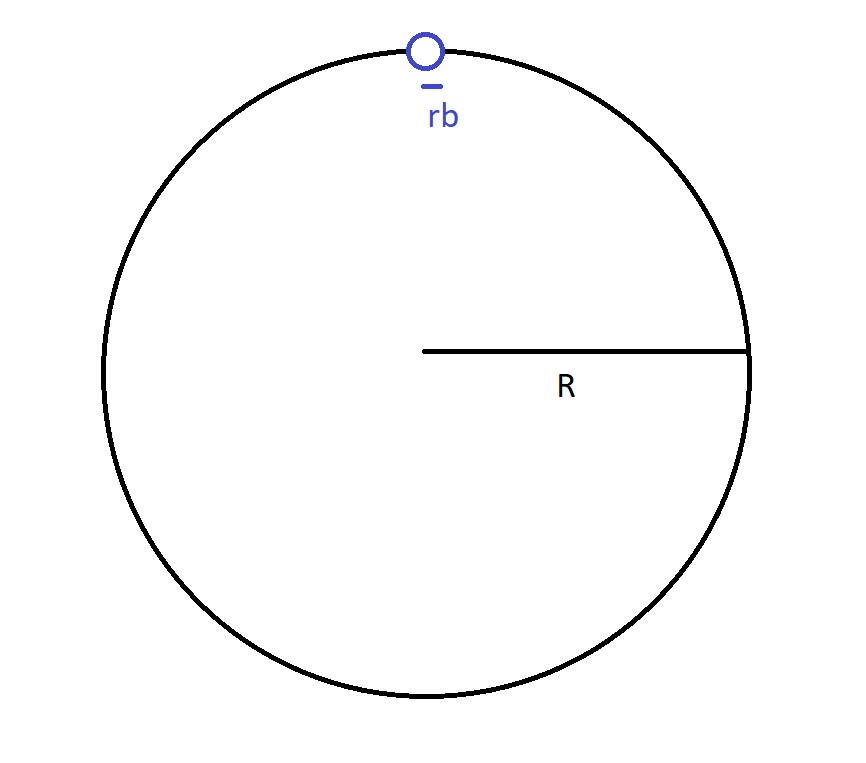}
\caption{\label{bfFig1} An illustrative picture describing a localised black hole with horizon radius $r_b$ living in a background with a spacelike dimension compactified into a circle with curvature radius $\mathcal{R}$.}
\end{figure}

We begin with the definitions of some terminology. Let us have branes with characteristic near horizon scale $r_b$ and background with characteristic length scale $\mathcal{R}$. Then, the \textit{blackfold regime} is defined as the parameters space where we have a large separation of scales $r_b \ll \mathcal{R}$. Consider as an example the configuration of a black hole on a cirle, illustrated by figure \ref{bfFig1}, where the scale $r_b$ is given by the horizon radius of the black hole and the scale $\mathcal{R}$ is given by the curvature radius of the circle. In this example, the blackfold regime is guaranteed when the horizon radius is much smaller than the curvature radius.

Equivalent currents are currents that mimic the far-zone effects of our added objects on the background. The idea of equivalent currents is nothing new. An intuitive example is ``equivalently'' modelling the Sun as a point mass and solving for its far-zone description, which is nothing but the linearised Schwarzschild solution. 

As the blackfold regime requires the scale of the introduced branes $r_b$ to be small (compared to all other scales in the problem), the branes leading order effective dynamics is that of probe branes\footnote{For example, in \cite{Gralla:2008fg}, it is rigorously proven that in General Relativity, the leading order dynamics of small bodies with low size and mass is that of probe particles.} and we can mimic their introduced currents with equivalent currents. As the blackfold regime also requires the scale of the background $\mathcal{R}$ to be large (compared to all other scales in the problem), we can approximate such equivalent currents with the currents carried by the branes in flat space\footnote{It can be helpful to think about the example of a black hole on a circle (figure \ref{bfFig1}). In particular, it is easy to persuade ourselves with this example that when $r_b$ is small, the black hole can be approximated as a point-like object and its leading order dynamics should be that of a probe particle. Moreover, when $\mathcal{R}$ is large, the far-zone effects of the black hole on the circle can be approximated by the analogous effects of the black hole on a straight line.}.  Plugging the expressions for the equivalent currents into equations \eqref{bf24}-\eqref{bf25}, we arrive at a set of equations named the \textit{blackfold equations}, which approximates the dynamics of the branes. 

For concreteness, let us consider the example of a charged particle moving in a non-trivial background. If we tune the parameters of the particle/background configuration in such a way that there is a large separation of scale $\mathcal{R} \gg r_b$, we can approximate the currents carried by the point charge using its description in flat space, which in 4-dimension takes the familiar form of the Reissner-Nordstrom solution\footnote{The description of the Reissner-Nordstrom black hole in higher dimension is known, e.g. \cite{Gath:2013qya}. Nevertheless, we decided to consider the 4-dimension case for familiarity.}: 
\be
\label{bf26}
ds^2 = - \left(1 - \frac{r_0}{r} + \frac{r_Q^2}{r^2} \right) dt^2 + \left(1 - \frac{r_0}{r} + \frac{r_Q^2}{r^2} \right)^{-1} dr^2 + r^2 d \Omega^2_{2} ~ ,
\ee
\be
A = \frac{Q}{r} \, dt
\ee
where $r_Q$ and $Q$ are related by 
\be
r_Q^2 = K \, Q^2 G  
\ee
with $K$ being the Coulomb force constant.

For the construction of equivalent currents, we want to know what spatially point-like sources would reproduce the far-zone description of the Reissner-Nordstrom solution when substituting into the forced Einstein-Maxwell equations \eqref{bf20}-\eqref{bf19}. This question can be easily answered through the notion of the \textit{mass} and \textit{charge} of the seed solution. In particular, as the ADM mass is defined as the localised point-like source that replicates the metric in the far-zone, our energy-stress tensor is $T^{\tau \tau} = m_{\text{ADM}} \, \gamma^{\tau \tau}$ where $m_{\text{ADM}}$ can be read off from \eqref{bf26}:
\be
m_{\text{ADM}} = \frac{r_0}{2 G} ~ .
\ee
Similarly, we have $J = \mathcal{Q} \, d \tau$ where $\mathcal{Q}$ is the electric charge of the solution, computed using Gauss's law as
\be
\mathcal{Q} = - \frac{1}{16 \pi G} \lim_{r \rightarrow \infty} \int \star F_2 = \frac{Q}{4 G} ~ .
\ee
Substituting these currents into \eqref{bf24}-\eqref{bf25}, we get the blackfold equations
\begin{align}
\p_\tau \mathcal{Q} &= 0 ~ , \\
\p_\tau m_{\text{ADM}} &= 0 ~ , \\
m_{\text{ADM}} \, \p^2_\tau X^\rho  &= - m_{\text{ADM}} \, \Gamma^{(i)}_{\tau \tau} + \mathcal{Q} \, \p^\tau X_{\mu}  \,   F^{\mu \nu} \, n_{\nu}^{(i)} ~ .
\end{align}
Comparing these equations to \eqref{bf11}-\eqref{bf13}, we can easily see that the blackfold equations describe the effective dynamics of a particle with mass $\frac{r_0}{2 G}$ and charge $\frac{Q}{4 G}$. 

Our analysis for the point particle in Einstein-Maxwell theory generalises straightforwardly to branes in (super)gravity theories. In such cases, the approximation of the energy-stress tensor $T^{ab}$ shall be obtained through a generalisation of the ADM mass, or equivalently the Brown-York energy-stress tensor \cite{Brown:1992br}. The approximation of the gauge fields/dilaton currents in a supergravity theory might require some care as there can be different, nonequivalent notions of charge \cite{Marolf:2000cb}. In fact, these subtleties do appear in our later discussions, e.g. in the computation of equivalent currents of anti-D3-NS5 bound state in appendix \ref{BS1} or anti-M2-M5 bound state in appendix \ref{C2}. 

In the case of non-trivially embedded branes, there is another scale in the story, the curvature radius of the bending in the configuration, $\mathcal{R}_E$. Even in flat space, there are not many known description of non-trivially embedded branes. As a result, we often further approximate the equivalent currents of the branes configurations by their trivially-embedded description. In such case, the required blackfold regime is  $r_b \ll \mathcal{R}, \mathcal{R}_E$. 

After deriving the dynamical equations from conservation equations, discussing equivalent currents, stating the blackfold regime, and introducing the blackfold equations, we obtain as an example the effective dynamics of a point charge. Naturally, one might say ``We already know the effective dynamics of a point charge from Dirac action. Why bother with all these?''. One reason is that our method for computing branes dynamics can be used in situations where other methods are difficult or even not feasible. Examples of these include the dynamics of nonextremal anti-D3-NS5 branes (chapter \ref{EX}) and anti-M2-M5 branes (chapter \ref{THE}) in warped throats. A more important reason is that, even though we introduce the blackfold equations as effective dynamical equations, they should not be understood solely as such. As argued in details in the next section, the blackfold equations are part of the (super)gravity equations for the first order matched asymptotic description of the backreacted configuration. As such, they give information on the backreacted profile of the branes configuration.

\label{ab12} In describing effective branes dynamics with the blackfold equations, we have made two approximations. Firstly, we have approximated the metric and gauge fields in equations \eqref{bf24}-\eqref{bf25} with their values in the unperturbed background profile, effectively ignoring all effects of backreations from the branes. Secondly, we have approximated the currents carried by the branes by their equivalent currents, which are further approximated by the far-zone currents carried by the flat-space homogeneous description of the introduced branes. Improvements can be made on both of these approximations. In particular, we can improve our first approximation by obtaining the matched asymptotic description of the configuration and use it as the background profile in the computation of the blackfold equations. Improvements of this type are related to the self-force of the branes, i.e. our blackfold equations now include the effects of backreactions from the branes and the obtained dynamics is beyond probe dynamics. We can also improve our second approximation. Specifically, we can improve how well the equivalent currents mimic the actual currents carried by the branes by including in their descriptions finite-thickness and bending effects \cite{Armas:2011uf}. We should also make use of the matched asymptotic description to write down a better approximation for the equivalent currents. Improvements of this type are related to the intrinsic information of the branes, i.e. they allow us to study non-linear effects (viscosity, damping) in the blackfold (fluid) equations. For discussions and applications of the next-to-leading order blackfold equations, we refer readers to \cite{Gath:2013qya, Armas:2013hsa,Armas:2011uf,Armas:2012ac,Armas:2012jg,Armas:2013goa, Armas:2018ibg, Camps:2010br}.

\section{Blackfold equations as constraints of MAE}\label{BF4}
In the previous section, we have introduced the blackfold equations as effective dynamical equations for branes in background satisfying the condition $\mathcal{R}, \mathcal{R}_E \gg r_b$ where $\mathcal{R}$, $\mathcal{R}_E$, and $r_b$ are respectively the characteristic length scale of the background, the curvature radius of the bending in the configuration, and the characteristic near horizon scale of the branes. In this section, we demonstrate that the same blackfold equations provide the necessary conditions for a matched asymptotic (super)gravity description of the backreacted configuration to first order in derivative expansion.

Let us begin with an introduction to the procedure of matched asymptotic expansion (MAE), see e.g.  \cite{Harmark:2003yz,Gorbonos:2004uc} for a discussion of the MAE in the context of caged black holes. The MAE can be understood as a procedure where the (super)gravity profile is analysed in two asymptotic regions: the near zone, $r \ll \mathcal{R}, \mathcal{R}_E$, where the profile can be approximated by a seed solution and the far zone, $r \gg r_b$, where the profile can be approximated by a background solution. Then, given that we can tune our system to have $\mathcal{R}, \mathcal{R}_E  \gg r_b$, information on the full (super)gravity profile can be extracted from the matching of these two asymptotic regions in an overlap zone, $\mathcal{R}, \mathcal{R}_E \gg r  \gg r_b$. The information obtained in the overlap zone can tell us a lot about the system. In particular, we can make use of such information to construct an ansatz for a perturbative description of the configuration everywhere in spacetime. Plugging this ansatz into the (super)gravity equations, solving the resulting differential equations, we will have a leading order matched asymptotic description of the (super)gravity profile. 

To implement the MAE procedure, let us introduce the idea of \textit{derivative (long-wavelength) expansion}, see e.g. \cite{Bredberg:2011jq, Bhattacharyya:2008jc} for discussions of derivative expansion in the context of fluid-gravity correspondence. As an illustrative example, let us consider a family of p-branes solutions in flat space in pure Einstein gravity: 
\be
\label{bf27}
d s^2_{seed} = \left( \eta_{ab} + \frac{r_0^n}{r^n} u_a u_b \right) d \sigma^a d \sigma^b + \frac{d r^2}{1 - \frac{r_0^n}{r^n}} + r^2 d \Omega_{n+1}^2 
\ee
where $\sigma^a$ are the worldvolume coordinates along the branes and $r, \, \Omega_{n+1}$ are the transverse coordinates orthogonal to them. In \eqref{bf27}, $\eta_{ab} = g_{\mu \nu} \p_a X^\mu \p_b X^\nu = \eta_{\mu \nu} \delta_a^\mu \delta_b^\nu$ can be view as the induced metric of Minkowski space on the trivially-embedded branes. We can categorise the parameters of this solution family into \textit{intrinsic} parameters, e.g. the horizon thickness $r_0$ and boost velocity $u^a$, and \textit{extrinsic} parameters, e.g. the embedding functions $X^{\mu}_\perp$\footnote{Note that, because of reparametrisation invariance along the worldvolume directions, we only have to consider the portion of the embedding functions that is normal to the branes, i.e. $\p^a X_{\mu} \, X^{\mu}_{\perp} = 0$.} hidden inside the induced metric. 

It is easy to see that if we fix the parameters of \eqref{bf27} to any constant values $(r_0)^*$,  $(u^a)^*$, $(X^{\mu}_{\perp})^*$, it will simply be a solution corresponding to black p-branes with such horizon radius, boost velocity, and embedding. However, if we turn these parameters into functions varying with respect to the worldvolume coordinates $\sigma$, \eqref{bf27} will no longer be a solution and we expect to have to add corrections correcting for the differences between constant and varying. Such corrections are of course derivatives of the parameter functions with respect to the worldvolume coordinates. The procedure where one turns the parameters of a family of gravity solutions into worldvolume functions and adds derivative corrections to it is called a derivative expansion. As we want our expansion to be well-behaved, we restrict our scope to long-wavelength physics so that, as things only change after going a large ``distance'', each order of derivative comes with a subduing factor. This is the reason why doing a derivative expansion can be alternatively understood as introducing long-wavelength perturbations to a seed solution. Note that the long-wavelength requirement of the derivative expansion is compatible with the MAE requirement that $\mathcal{R}, \mathcal{R}_E \gg r_b$. As such, we can use the profile of the branes in flat space as seed and figure out what long-wavelength corrections are needed to turn the seed into a profile that asymptotically match to the required background.

For concreteness, let us consider the example of bending black p-branes in flat space \cite{Camps:2012hw}. In such case, we take our ansatz to be the long-wavelength expansion of \eqref{bf27}:
\begin{multline}
\label{bf28}
ds^2_{ansatz} = \left( \gamma_{ab} (X^{\mu} (\sigma)) + \frac{r_0 ^n (\sigma)}{r^n} u_a(\sigma) u_b(\sigma) \right) d \sigma^a d \sigma^b\\
+ \frac{d r^2}{1 - \frac{r_0^n(\sigma)}{r^n}} + r^2 d \Omega_{n+1}^2 + \varepsilon f_{\mu \nu}  + ...
\end{multline}
where the induced metric $\gamma_{ab}$ is given by $\gamma_{ab} = \eta_{\mu \nu} \p_a X^\mu (\sigma) \p_b X^\nu (\sigma)$, $\varepsilon$ is a subduing factor of the same order as the first order worldvolume derivatives of the parameter functions, and the dots refer to higher derivative corrections. The subduing factor $\varepsilon$ can also be thought of as the ratio $r_b/\mathcal{R}$\footnote{From here and on wards, we shall not be distinguishing clearly the background scale $\mathcal{R}$ and the bending scale $\mathcal{R}_E$ but only use a representative scale $\mathcal{R}$. Note that, in our immediate example of bending black p-branes, the representative scale $\mathcal{R}$ should really be understood as the bending scale $\mathcal{R}_E$. }. In \eqref{bf28}, the intrinsic long-wavelength perturbations are expressed explicitly through the parameter functions $r_0 (\sigma)$ and $u^a(\sigma)$. The extrinsic long-wavelength perturbations are expressed implicitly through the induced metric $\gamma_{ab}$. Note that, as we bend our branes in flat space, the background metric $g_{\mu \nu}$ is still $\eta_{\mu \nu}$. However, this generalises straightforwardly to more general background metric.

By plugging \eqref{bf28} into the Einstein vacuum equations, $G_{\mu \nu} = 0$, and requiring that they are satisfied up to leading order in $\varepsilon$, we obtain a set of differential equations for the parameter functions $r_0, u_a, X^{\mu}$ and the corrections term $f_{\mu \nu}$. However, as our purpose is only to recover the blackfold equations from MAE, we focus solely on a subset of these differential equations for which only the parameter functions appear. These turn out to be the constraint equations, a subset of gravity equations obtained by projecting the Einstein equations along the worldvolume surface. 

As in these equations the intrinsic and extrinsic parameter functions decouple, we can consider them separately. In particular, considering the intrinsic parameter functions, keeping track of the subduing factors from the worldvolume derivatives, we obtain to first order in $\varepsilon$ \cite{Camps:2010br}:
\be
\label{bf30}
\nabla_a T^{ab} = 0
\ee
where $T^{ab}$ is given by
\be
\label{bf29}
T^{ab} = \frac{\Omega_{n+1}}{16 \pi G}r_0^n (n u^a u^b - \eta^{ab}) ~ .
\ee
Note that \eqref{bf29} is exactly the Brown-York tensor obtained from the flat black p-branes solution and, thus, \eqref{bf30} is nothing but the intrinsic blackfold equations. 

Considering the extrinsic (embedding) parameter functions, it is convenient to transform our background, here the Minkowski space, to an adapted coordinates system where it is readily compatible with our near zone ansatz \eqref{bf28}. In particular, we write our Minkowski background to leading order in $1/\mathcal{R}$ as \cite{Camps:2012hw}:
\be
\label{bf31}
ds^2_{flat} = \left( \eta_{a b} - 2 K_{ab}^{\ \ i } y_i \right) d \sigma^a d \sigma^b + d y_i dy^i + \mathcal{O} \left( y^2/\mathcal{R}^2 \right)
\ee
where the spacetime coordinates are $(\sigma^a, y^i)$. We can think of the metric \eqref{bf31} as being adapted to a $p+1$ dimensional worldvolume $\mathcal{W}_{p+1}$ with $y^i = 0$ denoting the position of $\mathcal{W}_{p+1}$. As such, $y^i$ acts as the orthogonal embedding parameter describing the long-wavelength extrinsic perturbations of the worldvolume $\mathcal{W}_{p+1}$ away from its trivial embedding, e.g. $\delta X^\mu_\perp (\sigma) = y^i$. 

Let us define our orthogonal coordinates $y^i$ such that the bending of our branes occurs along a specific orthogonal direction $y^{\hat{i}}$. By transforming the orthogonal coordinates $y^i$ into spherical coordinates with
\begin{align}
r &= \sqrt{y^i y_i} ~ ,\\
y_{\hat{i}} &= r \cos \theta   ~ ,
\end{align}
we can write the Minkowski background metric to leading order in $1/\mathcal{R}$ as
\be
ds^2_{flat} = \left( \eta_{a b} - 2 K_{ab}^{\ \ \hat{i}} \, r \cos \theta \right) d \sigma^a d \sigma^b + r^2 d \Omega_{(n+1)}^2  ~ .
\ee
Requiring that our ansatz \eqref{bf28} matches asymptotically in the far zone to the background to leading order in $1/\mathcal{R}$, we have
\begin{multline}
\label{bf33}
ds^2_{ansatz} = \left( \eta_{a b} - 2 K_{ab}^{\ \ \hat{i}} \, r \cos \theta + \frac{r_0^n }{r^n} u_a u_b \right) d \sigma^a d \sigma^b\\
+ \frac{d r^2}{1 - \frac{r_0^n(\sigma)}{r^n}} + r^2 d \Omega_{n+1}^2 + \varepsilon f_{\mu \nu}  
\end{multline}
where we have expressed the induced metric explicitly as
\be
\label{bf32}
\gamma_{ab} = \eta_{a b} - 2 K_{ab}^{\ \ \hat{i}} r \cos \theta ~ .
\ee
Before continuing, let us explain that the appearance of extrinsic curvature $ K_{ab}^{\ \ \hat{i}}$ in the induced metric is not as surprising as it might first appear. The leading order deviation away from $\eta_{ab}$ of the induced metric $\gamma_{ab}$ due to bending of the branes can be computed as 
\be
\delta \gamma_{ab} = - 2 K_{ab}^{\ \ i} \delta X_i = - 2 K_{ab}^{\ \ \hat{i}} \, y_{\hat{i}} = - 2 K_{ab}^{\ \ \hat{i}} \, r \cos \theta
\ee
where we have used \eqref{d1} in the first equality. Therefore, it is rather obvious that to leading order in $1/\mathcal{R}$ we have \eqref{bf32}.

Pushing our ansatz \eqref{bf33} to the overlap zone, to leading order in $r_0^n/r^n$ and $1/ \mathcal{R}$, we have
\begin{multline}
ds^2_{ansatz} = \left( \eta_{a b} - 2 K_{ab}^{\ \ \hat{i}} \, r \cos \theta + \frac{r_0 ^n }{r^n} u_a u_b \right) d \sigma^a d \sigma^b\\
+ \left(1 + \frac{r_0^n}{r^n} \right) d r^2 + r^2 d \Omega_{n+1}^2 + \varepsilon f_{\mu \nu}   ~ .
\end{multline} 
Plugging this into the Einstein equations, we note that we can extract an equation that doesn't involve $f_{\mu \nu}$, which can be written as \cite{Camps:2012hw}
\be
\label{extra4}
T^{ab} K_{ab}^{\ \ \hat{i}} = 0
\ee
where $T^{ab}$ is the Brown-York tensor \eqref{bf29}. This is simply the extrinsic blackfold equations. 

Before continuing, let us count the number of equations and degrees of freedom. For a p-brane in D dimensions, we have $p+1$ intrinsic blackfold equations \eqref{bf30} and $D- (p+1)$ extrinsic blackfold equations \eqref{extra4}\footnote{In equation \eqref{extra4}, we only consider the normal direction $y^{\hat{i}}$ as we only bend the brane in such direction, i.e. for all other directions, the extrinsic curvature $K_{ab}^{\ \ i}$ vanishes. However, in general, one would have $T^{ab}K_{ab}^{\ \ i} = 0$ where $i$ runs over all normal directions.}. Thus, in total, we have $D$ blackfold equations. Let us now count the number of degrees of freedom. The zeroth order long-wavelength deformations of the p-brane are given by the parameter functions $r_0, u_a, X^\mu$ \eqref{bf28}. As deformations in $X^\mu$ along the worldvolume directions can be undone by a worldvolume coordinates reparametrisation\footnote{Given that the branes have no edges.}, we are only interested in the deformations normal to the branes\footnote{This is the reason why it is often convenient to put $X^\mu$ in static gauge.}. Therefore, from $X^\mu$, we have $D-(p+1)$ degrees of freedom. As $u_a$ has to satisfy the unitary condition, from $u_a$, we have $p$ degrees of freedom. Lastly, we have 1 degree of freedom from $r_0$. Thus, altogether, we also have $D$ degrees of freedom. When we consider charged branes, there will be extra degrees of freedom associated to the ``electromagnetic'' currents. However, then, the current conservation equations \eqref{extra6}-\eqref{extra7} will allow us to define conserved charges. Once we fix these charges, we can write the extra degrees of freedom in terms of the old ones. 

We have shown explicitly through the example of bending black p-branes in flat space that the blackfold equations are part of the gravity equations for the leading order backreacted matched asymptotic description. As such, they can be viewed as the \textit{necessary conditions} for a matched asymptotic solution. This result can be straightforwardly extended to generic branes in generic backgrounds in supergravity theories \cite{Armas:2016mes}. 

On the other hand, the question whether the blackfold equations provide also the \textit{sufficient conditions} for a matched asymptotic description, i.e. solving only the subset of blackfold equations guarantee the satisfaction of all (super)gravity equations, is more tricky. In our immediate example of bending black p-branes in flat space, it was proven in \cite{Camps:2012hw} that, indeed, solving the blackfold equations guarantee a regular solution that matches the seed in the near zone and the background in the far zone to first order in $1/\mathcal{R}$. However, despite partial evidence in all cases that have been worked out in detail, the generalisation of this statement to generic branes and background in supergravity remains a conjecture, known as the \textit{blackfold conjecture} \cite{Niarchos:2015moa}. This conjecture is almost analogous to the statement in Fluid/Gravity that there is a one to one map between a solution of the fluid equations and a regular solution of the gravitational equations \cite{Bhattacharyya:2008jc}.

\section{Blackfold equations in supergravity theories}\label{BF5}
Going through the procedure explained in the previous sections, one can derive the blackfold equations of generic branes in generic background in a general (super)gravity theory \cite{Armas:2016mes}. For our later convenience, in this section, let us collect the blackfold equations for systems in type IIB supergravity and eleven-dimensional supergravity.

\paragraph{Type IIB supergravity}
The action for the bosonic sector of the type IIB supergravity theory in the Einstein frame can be written in the democratic formulation as \cite{Bergshoeff:2001pv}
\be
\label{bf40}
\mathcal{I}_{IIB} = \frac{1}{16 \pi G} \int_{\mathcal{M}_{10}} \left[ \star R - \frac{1}{2} d \phi \wedge \star d \phi - \frac{1}{2 } e^{- \phi} H_3 \wedge \star H_3 - \frac{1}{4} \sum_q e^{a_q \phi} \tilde{F}_{q+2} \wedge \star \tilde{F}_{q+2} \right]
\ee
where the dilaton coupling $a_q$ is given by $a_q = (3 - q)/2$ and the values of $q$ are $q = -1, 1, 3, 5, 7$. Of course, the equations of motion of the action \eqref{bf40} must be supplemented by the duality relations $\tilde{F}_{D - q - 2} = \star \tilde{F}_{q+2}$. 

Letting our degrees of freedom be represented by the lower form fields, from conservation considerations, we have the blackfold equations for probe branes with $H_3$ electric current $\bold{j}_2$; $H_3$ magnetic current $j_6$; $\tilde{F}_{1}$, $\tilde{F}_{3}$, $\tilde{F}_{5}$ electric currents $J_{0}$, $J_{2}$, $J_{4}$; $\tilde{F}_{1}$, $\tilde{F}_{3}$ magnetic currents $\mathcal{J}_{8}$, $\mathcal{J}_{6}$; and dilaton current $j_\phi$ are given by the energy-momentum tensor conservation equations\footnote{Our equations have some sign differences compared to \cite{Armas:2016mes}. These are simply due to differences in conventions.}
\be
\label{bf41}
\begin{split}
\nabla_\mu T^{\mu \nu} = &\frac{1}{2!} H_3^{\nu \mu_1 \mu_2} \bold{j}_{2 \mu_1 \mu_2} + \frac{e^{- \phi}}{6!} H_7^{\nu \mu_1 ... \mu_6} j_{6 \mu_1 ... \mu_6} + j_\phi \p^{\nu} \phi \\
&+ \sum_q \frac{1}{(q+1)!} \left( \tilde{F}_{q+2}^{\nu \mu_1 ... \mu_{q+1}} + \frac{q(q+1)}{2!} H_3^{\nu \mu_1 \mu_2} C_{q-1}^{\mu_3 ... \mu_{q+1}} \right) J_{q+1 \mu_1 ... \mu_{q+1}} \\
&- \sum_q \frac{e^{a_q \phi}}{(\tilde{q} +1)!} \left( \tilde{F}_{\tilde{q} +2}^{\nu \mu_1 ... \mu_{\tilde{q}+1}} + \frac{\tilde{q} (\tilde{q}-1) }{2!} H_3^{\nu \mu_1 \mu_2} C_{\tilde{q}-1}^{\mu_3 ... \mu_{\tilde{q}+1}} \right) \mathcal{J}_{\tilde{q} + 1 \mu_1 ... \mu_{\tilde{q}+1}} \\
&+ \frac{1}{4!} \left( \tilde{F}_5^{\nu \mu_1 ... \mu_4} + 3 H_3^{\nu \mu_1 \mu_2} C_2^{\mu_3 \mu_4} \right) J_{4 \mu_1 ... \mu_4} \\
&- \sum_q \frac{e^{a_q} \phi}{(q+2)!} \tilde{F}_{q+2}^{\mu_1 ... \mu_{q+2}} \left( \star j_6 \wedge C_{q-1} \right)^{\nu}_{\ \mu_1 ... \mu_{q+2}}  
\end{split}
\ee
and the current conservation equations
\begin{align}
&d \star J_{0} + \star J_{2}\wedge H_3 + e^{\phi} \tilde{F}_{7} \wedge \star j_6 = 0 ~ ,& &d \star J_{4} + e^{- \phi} \tilde{F}_3 \wedge \star j_6 = 0 ~ ,\\
&d \star J_{2} + \star J_{4} \wedge H_3 + \tilde{F}_{5} \wedge \star j_6 = 0~ , & &\\
&d \star \mathcal{J}_{8} = 0 ~ , & &d \star \mathcal{J}_{6} = H_3 \wedge \star \mathcal{J}_{8} ~ , \\
\label{bf42}
&d \star \bold{j}_2 = 0 ~ , & &d \star j_6 = 0
\end{align}
where we have defined $H_7 \equiv \star H_3$. The sum over $q$ takes value over $q = -1, 1, 3$ and we have introduced $\tilde{q} = D - q - 4$.

\paragraph{Eleven-dimensional supergravity}
The action of the bosonic sector of eleven-dimensional supergravity is given by \cite{becker_becker_schwarz_2006}
\be
\mathcal{I}_{M} = \frac{1}{16 \pi G} \int_{\mathcal{M}_{11}} \left[ \star R - \frac{1}{2} G_4 \wedge \star G_4 - \frac{1}{6}  A_3 \wedge G_4 \wedge G_4  \right] ~ .
\ee
The blackfold equations for probe branes with electric current $J_3$ and magnetic current $\mathcal{J}_6$ are given by 
\begin{align}
\label{bf50}
&\nabla_{\mu} T^{\mu \nu} = \frac{1}{3!} G_3^{\nu \mu_1 \mu_2 \mu_3} J_{3 \mu_1 \mu_2 \mu_3} - \frac{1}{6!} G_7^{\nu \mu_1 ... \mu_6} \mathcal{J}_{6 \mu_1 ... \mu_6} ~ ,\\
&d \star J_3 + \star \mathcal{J}_6 \wedge G_4 = 0 ~ ,\\
\label{bf51}
&d \star \mathcal{J}_6 = 0 
\end{align}
where we have defined the dual field strength $G_7 \equiv \star G_4$.

\chapter{On the existence of metastable antibranes}
\label{EX}
The purpose of this chapter is to demonstrate an application of the blackfold approach in the argument for the existence of metastable states of antibranes in warped throats. Our exemplar candidate is the Kachru-Pearson-Verlinde (KPV) \cite{Kachru:2002gs} metastable state of \textit{polarised} anti-D3 branes, or equivalently wrapped NS5 branes with dissolved anti-D3 brane charge (wrapped anti-D3-NS5 branes), at the tip of the Klebanov-Strassler (KS) throat \cite{Klebanov:2000hb}. As such, we are interested in the type IIB supergravity solution describing wrapped anti-D3-NS5 branes at the tip of the KS throat. Because an exact description of such configuration is technically unfeasible, we turn to approximate descriptions. In particular, a perturbative description of the configuration can be obtained through the technique of \textit{matched asymptotic expansion}, where the solution is approximated in the far zone by the background solution of interest (here the KS throat) and in the near zone by an uniform flat-space p-brane solution (here the D3-NS5 bound state). As the blackfold equations provide the \textit{necessary conditions} and perhaps also the \textit{sufficient conditions} for the leading order matched asymptotic solution, we make use of such equations to learn about the configuration of anti-D3-NS5 branes in KS throat.

We begin, in section \ref{extra2}, with an introduction to the KPV state and a brief review of the discussions regarding its existence. 
In section \ref{EX1}, we derive the blackfold equations for nonextremal anti-D3-NS5 branes at the tip of the KS throat, recovering the KPV DBI probe analysis at extremality. In section \ref{EX2}, we provide a discussion on the regime of validity of our analysis. Subsequently, in section \ref{EX3}, we describe effective potentials for nonextremal anti-D3-NS5 branes and study the effects of nonextremality on the conjectured KPV metastable state. In particular, we provide quantitative evidence for a thermal metastability-losing mechanism that is driven by horizon geometry. Lastly, in section \ref{EX4}, we discuss how these findings, together with previously known results, constitute to a strong argument for the existence of antibranes metastable state. 

\section{Introduction to the KPV state}
\label{extra2}
Let us begin our story with the Klebanov-Strassler (KS) throat \cite{Klebanov:2000hb}, a 10-dimensional type IIB supergravity solution. The KS throat involves a 6-dimensional deformed conifold, a 4-dimensional Minkowski space, and non-trial $F_3, F_5, H_3$ fluxes that in turn induce warping effects to the throat. The principal difference between our 6-dimensional deformed cone and a regular cone is that, instead of having a point-like tip, our deformed cone has an $S^3$ tip. One can intuitively think of the KS throat as the solution resulting from placing D3 and D5 brane charge at the tip of a Ricci-flat (deformed) conifold. The D3 and D5 brane charge will induce non-trivial $F_3, F_5, H_3$ fluxes as well as warping effects to the conifold, giving us a warped, fluxed throat geometry. For a detailed introduction to the KS throat, we refer readers to appendix \ref{KS}. In appendix \ref{KS}, we focus on aspects of the KS throat that are immediately relevant for its role as background geometry of metastable antibranes. For further discussion, we refer readers to the original paper \cite{Klebanov:2000hb} or the review \cite{Herzog:2001xk}. 

\begin{figure}[t]
\begin{subfigure}{0.45\textwidth}
\includegraphics[width=0.95\linewidth]{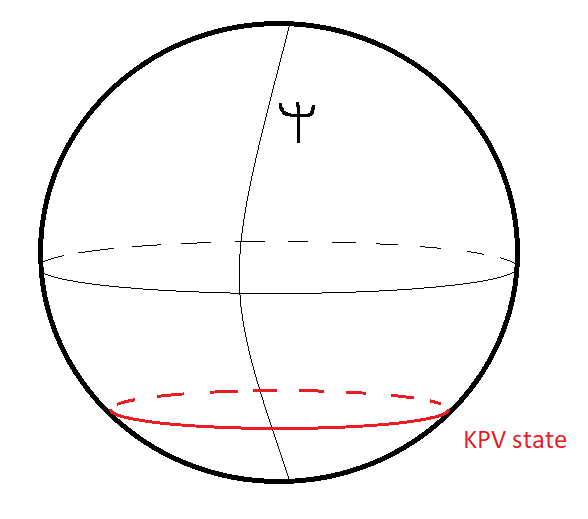} 
\end{subfigure}
\begin{subfigure}{0.55\textwidth}
\includegraphics[width=0.95\linewidth]{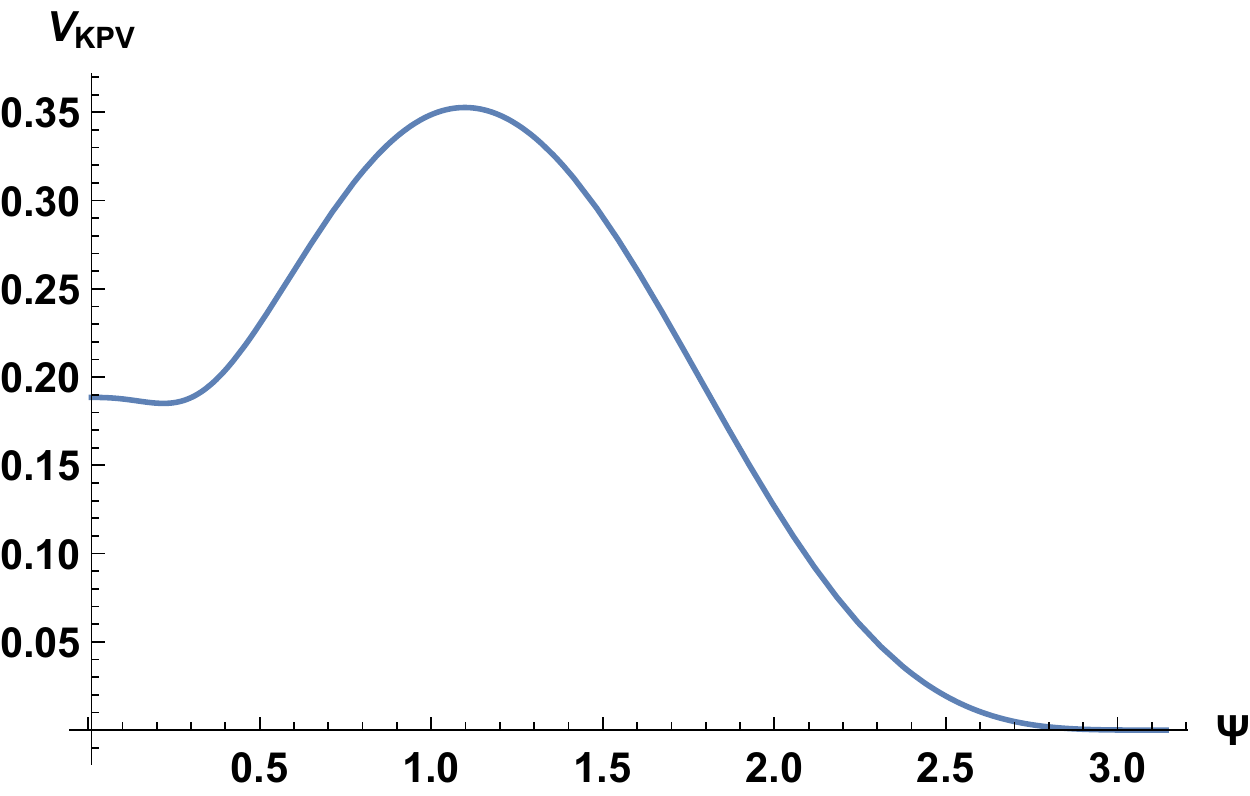}
\end{subfigure}
\caption{On the left, we have an illustrative picture of the tip of the KS throat, which is an $S^3$.  The coordinate $\psi \in (0, \pi)$ is the azimuth angle of this $S^3$. The red circle illustrates the KPV state, which is a spherical NS5 state (wrapping an $S^2$ of the $S^3$) that experiences a balance of force along the azimuth angle $\psi$. On the right, we have the effective potential of a spherical NS5 state at the tip of the KS throat for $p/M = 0.03$. As we can from the plot, the potential has a metastable minimum at $\psi \approx 0.3$. This is the KPV state.}
\label{extra1}
\end{figure}

Let us take some anti-D3 branes and place them near the tip of the KS throat. As one expects (or via explicit calculations), these anti-D3 branes will be attracted to the tip via both gravitational and ``electromagnetic'' forces. Through considering the non-abelian worldvolume theory of the stack of anti-D3 branes at the tip of the KS throat, \cite{Kachru:2002gs} argued that these anti-D3 branes will polarise via the Myers effects \cite{Myers:1999ps} into a spherical NS5 brane with dissolved anti-D3 brane charge, wrapping an $S^2$ of the $S^3$ at the tip. From the perspective of the NS5 brane, \cite{Kachru:2002gs} further argued that, for some regime of parameters, this NS5 brane can experience a balance of force that allow it to form a metastable state. In particular, they found that for $p/M$ between 0 and $p_{crit}$ with $p$ denotes the number of the anti-D3 branes, $M$ denotes the strength of the KS fluxes, and $p_{crit} \approx 0.080488$, the effective potential of the NS5 brane has a metastable minimum, see figure \ref{extra1}. This metastable state of spherical NS5 brane at the tip of the KS throat is what we call the KPV state. Let us note that the KPV configuration is classically stable but can tunnel quantum mechanically to the true minimum at $\psi = \pi$. However, the rate of decay is exponentially suppressed \cite{Kachru:2002gs}. 

Discussions on the existence of the KPV state were started from two main observations: First, in writing down the worldvolume action for the antibranes, \cite{Kachru:2002gs} utilised S-duality. The concern is that such step would take their analysis outside of its regime of validity, thus, invalidating the results \cite{Bena:2014jaa}. Second, in attempts to construct the supergravity description of the KPV configuration, people encountered unexpected difficulties which cast doubt on the state's existence. In \cite{Bena:2009xk}, unphysical singularities in the 3-form fluxes were observed in the supergravity description of anti-D3 branes at the tip of the KS throat where the anti-D3 branes are smeared all over the $S^3$ of the tip. By requiring a consistent UV/IR gluing of antibranes to the KS throat \cite{Cohen-Maldonado:2015ssa}, a set of no-go theorems were constructed. These no-go theorems can be summarised as below: 
\begin{itemize}
\item Smeared anti-D3 and extremal localised anti-D3 solutions are not permitted.
\item Non-extremal localised anti-D3 solutions are permitted but have no extremal limit. 
\item Spherical NS5 solutions are permitted.
\end{itemize}

An intuitive interpretation of these no-go theorems can be obtained via the Smarr relations \cite{Cohen-Maldonado:2015ssa, Cohen-Maldonado:2016cjh}. One starts by deriving the Smarr relations for antibranes at the tip of the throat geometry. This thermodynamic equation carries the gluing conditions by relating the UV (e.g. the ADM energy) and the IR (e.g. horizon area, surface gravity, chemical potential, charge) information of the solution. From the Smarr relations, we can quickly see that smeared anti-D3 and extremal localised anti-D3 solutions are not permitted. Furthermore, the Smarr relations offer an explanation for why the spherical NS5 escapes the no-go theorem while the localised anti-D3 does not. The key reason is that the spherical NS5 has a non-trivial horizon geometry which allows it to balance the non-zero ADM energy with dipole charge.    

In the rest of the chapter, we will study the KPV state using the blackfold approach. After obtaining some new results, we will come back and discuss the existence of KPV state more thoroughly in section \ref{EX4}.

\section{Blackfold equations for anti-D3-NS5 branes in KS throat}
\label{EX1}
\paragraph{The KS throat}
For the purpose of deriving the anti-D3-NS5 blackfold equations, we present here only the metric and the flux components of the KS throat that contribute in our derivation. For more discussions on the KS throat, we refer readers to appendix \ref{KS}. For our convenience, let us set the string scale $l_s = 1$. As the dilaton of the KS solution is a constant, we further set $g_s = 1$. As discussed in the appendix, in adapted coordinates, the KS metric near the apex is given by
\begin{multline}
\label{ex3}
g_{\mu \nu} d x^\mu d x^\nu = M b_0^2  \Big ( - dt^2 + (dx^1)^2 + (dx^2)^2 + (dx^3)^2  + dr^2 \\
+ d \psi ^2 + \sin^2 \psi \left( d \omega^2 + \sin^2 \omega d \varphi^2 \right)  + r^2 (d \theta_2^2 +  \sin^2 \theta_2 d \phi_2^2) \Big) + ...
\end{multline}
and the relevant fluxes are given by
\begin{align}
F_3 &= 2 M \sin^2 \psi \sin \omega \ d \psi \wedge d \omega \wedge d \varphi + ... ~ ,\\
\label{ex4}
H_7 &= - 2 M^3 b_0^4 \sin^2 \psi \sin \omega \ d t \wedge dx^1 \wedge dx^2 \wedge dx^3 \wedge d \psi \wedge d\omega \wedge d \varphi + ...
\end{align}
where $b_0^2 \approx 0.93266$ and the dots refer to components of the metric/fluxes that do not contribute in our derivation\footnote{Some terms in the dots are important to our stability analysis in chapter \ref{STA} and shall be discussed appropriately later on.}. The factor $M$, which counts the units of $F_3$ flux piecing the $S^3$ at the tip of the throat \eqref{ks1}, is a characteristic parameter of the KS throat. As we shall see shortly in the discussion on the regime of validity, our blackfold approach is applicable only for KS throat with large $M$.

\paragraph{Anti-D3-NS5 branes} 
As discussed in chapter \ref{BF}, the anti-D3-NS5 blackfold equations can be obtained as the conservation equations of equivalent sources induced by the branes onto the background. Equivalent sources are sources that mimic the far-zone effects of the seed solution on the background profile. For discussions of the anti-D3-NS5 branes and their equivalent sources, we refer readers to appendix \ref{BS1}. Here, we simply present the results. For nonextremal anti-D3-NS5 branes, the equivalent energy-stress tensor is
\begin{equation}
\label{ex5}
T^{ab} =  C \left[ r_0^2 \left( u^a u^b - \frac{1}{2} \gamma^{a b} \right ) - r_0^2 \sin^2 \theta \sinh^2 \alpha (\gamma^{a b} - v^a v^b - w^a w^b) - r_0^2 \cos^2 \theta \sinh^2 \alpha \gamma^{a b} \right]
\end{equation}
and the equivalent currents are
\begin{align}
J_2 &= C r_0^2 \sinh^2 \alpha \sin \theta \cos \theta \ v \wedge w ~ ,\\
J_4 &=  C r_0^2 \sinh \alpha \cosh \alpha \sin \theta \ * (  v \wedge w) ~ ,\\
\label{ex6}
j_6 &= - C r_0^2 \sinh \alpha \cosh \alpha \cos \theta \ * ( 1) 
\end{align}
where $C = \displaystyle \frac{\Omega_3}{8 \pi G}$ with $\Omega_3 = 2 \pi^2$ the volume of the unit radius $S^3$, and $*$ is the worldvolume Hodge dual operator. In our description, $r_0,\, \alpha,\, \tan \theta,\, v^a,\, w^a,\, u^a$ are the parameters of the anti-D3-NS5 branes, describing respectively the Schwarzschild horizon radius, the boost rapidity, the brane charge distribution, the orientation of the anti-D3 dissolved charge (with two orthogonal vectors $v$ and $w$)\footnote{To specify the orientation of the 4-dimensional anti-D3 branes in the 6-dimensional worldvolume of the NS5 brane, we can specify its 2 codimensions. These are described by the vectors $v$ and $w$.}, and the thermal flow (with one directional vector $u$).

\paragraph{Anti-D3-NS5 branes in KS throat}
In a blackfold set-up of anti-D3-NS5 branes in KS background, the variables of the system are
\be
\label{ex50}
r, \, \theta_2, \, \phi_2,\, \psi,\, r_0,\, \alpha,\, \tan \theta,\, v^a,\, w^a,  \, u^a ~ .
\ee
Variables $r,\, \theta_2,\, \phi_2,\, \psi$ are the embedding parameters of the branes to the background. Variables $r_0,\, \alpha,\, \tan \theta,\, v^a,\, w^a,\, u^a$ are the characteristic parameters of the seed solution.

In order to construct a metastable state, we place the anti-D3-NS5 branes at the tip of the KS throat in such way that 4 of the 6 dimensions of the D-brane bound state lie along the Minkowskian directions $t, x^1, x^2, x^3$, and the other 2 wrap around the 2-cycle $\omega, \varphi$. For simplicity, let us further restrict our attention to $t$ dependent configurations with dissolved anti-D3 brane charge lying along the Minkowski directions and thermal flow along the $t$ direction. This means we have set 
\be
r = 0 \, , 
\ee
restricted our scalar variables to
\be
\psi(t), r_0(t), \alpha(t), \tan \theta(t) \, , 
\ee
and specified our vectors as
\begin{align}
\label{ex1}
v^a \partial_a  &= \frac{\partial_\omega}{\sqrt{M} b_0 \sin \psi } ~ ,  \\
w^a \partial_a &= \frac{\partial_\varphi}{\sqrt{M} b_0 \sin \psi \sin \omega} ~ , \\
u^a \p_a &= \frac{\p_t}{ \sqrt{M} b_0 \sqrt{1 - \psi'^2}}
\label{ex2}
\end{align}
where $\psi' \equiv  \p_t \, \psi$. The factors in \eqref{ex1}-\eqref{ex2} are to make sure that $v$, $w$, $u$ satisfy the unitary condition, i.e. $v^a v_a = 1$. Note that, even when we ask $v$ and $w$ to be orthogonal to each other, i.e. $v^a w_a = 0$, there is still a rotational gauge symmetry in the specification of $v$ and $w$ due to the redundancies in the language we used to describe the orientation of the dissolved anti-D3 branes. Such gauge symmetry is helpful later in chapter \ref{STA} where we consider generic long-wavelength perturbations to the configuration.

\paragraph{The blackfold equations} Substituting the KS background profile \eqref{ex3}-\eqref{ex4} and the anti-D3-NS5 equivalent currents \eqref{ex5}-\eqref{ex6} into the blackfold equations in type IIB supergravity \eqref{bf41}-\eqref{bf42}, we obtain the anti-D3-NS5 blackfold equations. These blackfold equations can be written in term of energy-momentum conservation equations and current conservation equations as presented explicitly below.  
\begin{enumerate}
\item Energy-momentum conservation equations:
\begin{align}
\label{ex7}
\nabla_a T^{a b} &= \p^b X_\mu \, \mathcal{F}^\mu ~, \\
\label{ex8}
T^{ab} K_{ab}^{\,\,\, \,\,\, (i)}  &= \mathcal{F}^\mu \, n^{(i)}_\mu
\end{align}
where $n_\mu^{(i)}$ denotes the normal vectors of the anti-D3-NS5 blackfold, $K_{ab}^{\,\,\, \,\,\, (i)} = K_{ab}^{\ \ \, \rho} n_\rho^{(i)}$,  and the force term $\mathcal{F}^\mu$ is given by
\be
\label{ex9}
\mathcal{F}^\mu = - \frac{1}{6!}  H_{7}^{\mu a_1 ... a_6} j_{6 a_1 ... a_6} + \frac{1}{2!} \tilde{F}_3^{\mu a_1 a_2} J_{2 a_1 a_2}  ~ .
\ee

\item Current conservation equations 
\begin{align}
\label{ex10}
d * j_6 &= 0 ~ ,\\
d * J_4 + * j_6 \wedge F_3  &= 0 ~ ,\\
\label{ex11}
d * J_2 + H_3 \wedge * J_4 &= 0 
\end{align}
where $F_3, H_3$ are the projected background fluxes and $*$ is the 6-dimensional Hodge dual of the worldvolume directions.
\end{enumerate}

Considering the \textit{intrinsic equation} \eqref{ex7}, one can show that the equation is non-trivial only when $b = t$ and such equation can be simplified to 
\begin{multline}
\label{ex12}
 \partial_t \left( r_0^2 \left( \frac{3}{2} + \sinh^2 \alpha \right)  \right)  + 2 \cot \psi \psi' r_0^2 \Bigg( 1 + \sinh^2 \alpha  \sin^2 \theta  \Bigg) \\
= - \frac{2}{b_0^2} \psi' r_0^2 \sinh^2 \alpha \sin \theta \cos \theta ~ .
\end{multline}

Considering the \textit{extrinsic equation} \eqref{ex8}, as we are focusing on dynamics at the tip, the only normal vector is 
\be
n^{(1)} = \frac{\sqrt{ M }  b_0}{\sqrt{1 - \psi'^2}} \left( -  \psi' d t + d \psi  \right) ~ .
\ee
Plugging this into \eqref{ex8} yields
\begin{multline}
\label{ex13}
 \frac{\psi''}{ (1- \psi'^2)^2} \Bigg( \frac{3}{2} + \sinh^2 \alpha  \Bigg) +  \frac{\cot \psi}{ 1- \psi'^2} \Bigg( 1 + 2 \cos^2 \theta \sinh^2 \alpha  \Bigg)  \\
= \frac{2}{b_0^2} \frac{1}{\sqrt{1 - \psi'^2 }} \cos \theta \sinh \alpha \cosh \alpha +  \frac{2}{b_0^2} \frac{1}{1 - \psi'^2} \sinh^2 \alpha \sin \theta \cos \theta  ~ .
\end{multline}

From the \textit{current conservation equations} \eqref{ex10}-\eqref{ex11}, we can define conserved Page charges $\mathbb{Q}_3$ and $\mathbb{Q}_5$ that keep track of the number of anti-D3 branes and NS5 branes:
\begin{align}
\mathbb{Q}_5 &= * j_6 = C r_0^2 \sinh \alpha \cosh \alpha \cos \theta ~  , \\
\mathbb{Q}_3 &= \int_{S^2} * \left( J_4 + *(* j_6 \wedge C_2 ) \right) \\
&= -  4  \pi M \, C r_0^2 \sinh \alpha \cosh \alpha \left( \sin \theta \,  b^2_0 \sin^2 \psi +  \cos \theta \, \left( \psi - \frac{1}{2} \sin 2 \psi \right) \right)  
\end{align}
where we have used $C_2 = M (\psi - \frac{1}{2} \sin 2 \psi) \sin \omega d \omega \wedge d \varphi$. It follows immediately that we can write $\tan \theta$ as
\be
\label{ex14}
\tan \theta = \frac{1}{b_0^2 \sin^2 \psi} \left(  \frac{\pi p}{M} - \left(\psi - \frac{1}{2}\sin 2 \psi \right) \right)
\ee
where we have made the identification
\be
\label{ex15}
\frac{- \mathbb{Q}_3}{4 \pi \mathbb{Q}_5} = \pi p ~ .
\ee

Equations (\ref{ex12}),  (\ref{ex13}), and (\ref{ex14}) form our set of \textit{blackfold equations}.

\paragraph{Recovery of the KPV state at extremality} 
In the extremal limit, the intrinsic equation (\ref{ex12}) becomes obsolete as it is simply the derivative of the equation for $\tan \theta$ (\ref{ex14}). With some algebraic manipulations, the set of anti-D3-NS5 blackfold equations can be written as
\be
\label{ex19}
\cot \psi = \frac{1}{b_0^2} \sqrt{1 - \psi'^2} \sqrt{1 + \tan^2 \theta} + \frac{1}{b_0^2} \tan \theta - \frac{1}{2} (1 + \tan^2 \theta) \frac{\psi''}{1 - \psi'^2}
\ee 
where $\tan \theta$ is given by
\be
\tan \theta = \frac{1}{b_0^2 \sin^2 \psi} \left(  \frac{\pi p}{M} - \left(\psi - \frac{1}{2}\sin 2 \psi \right) \right) ~ .
\ee

On the other hand, from the DBI action in \cite{Kachru:2002gs}, we have the equation of motion
\be
\cot \psi =  \frac{1}{b_0^2} \sqrt{1 - \psi'^2} \sqrt{1 + \mathcal{P}^2} + \frac{1}{b_0^2} \mathcal{P} - \frac{1}{2} (1 + \mathcal{P}^2) \frac{\psi''}{1 - \psi'^2}
\ee
where $\mathcal{P}$ is given by
\be
\label{ex20}
\mathcal{P} = \frac{1}{b_0^2 \sin^2 \psi} \left(  \frac{\pi p}{M} - \left(\psi - \frac{1}{2}\sin 2 \psi \right) \right) ~ .
\ee
Thus, we have shown that the blackfold equations at extremality recover the results obtained from DBI analysis by KPV in a regime very different from theirs $p g_s \ll 1$. In particular, we observe directly from the supergravity regime $p g_s \gg 1$ that, for $0 < p/M < p_{crit}$ where $p_{crit} \approx 0.080488$, the anti-D3-NS5 branes can form a metastable state at the tip of the KS throat. For further discussions on the connections between the blackfold description of branes in supergravity and their DBI description in the probe regime, we refer readers to \cite{Niarchos:2015moa,Grignani:2016bpq,Armas:2016mes}.

\section{Regimes of validity} \label{EX2}
Validity of the blackfold analysis requires a large separation of scales $r_b\ll \mathcal R, L$ where $r_b$ is the characteristic near horizon scale of the seed branes, $R$ is the scale of the curvature radius of the bending in the branes, and $L$ is the characteristic length scale of the background. In the case at hand, the length scale $r_b$ is the largest scale among the energy density radius $r_\varepsilon \sim r_0\sinh\alpha $, the length scale associated to the conserved NS5 Page charge $r_h^{(NS5)} \sim \mathbb{Q}_5/ C $, and the length scale associated to the conserved anti-D3 Page charge $r_h^{(\overline{\rm D3})} \sim - \mathbb{Q}_3/ C$. The scale $\mathcal R \sim  \sqrt{g_s M} l_s \sin \psi $ is controlled by the size of the $S^2$ that the NS5 wraps, while the background scale $L \sim  \sqrt{g_s M} l_s $ is set by the size of the $S^3$. For our purposes, between the scale $\mathcal{R}$ and $L$,  it is sufficient to only consider the bending scale $\mathcal{R}$ as $\mathcal{R}$ is always smaller than $L$.

We have the condition $r_{\varepsilon} \ll \mathcal{R}$ leads to 
\be
\label{ex165}
r_0 \sinh \alpha \ll \sqrt{g_s M} l_s \sin \psi
\ee
which constrains the product $r_0 \sinh \alpha$. The condition $r_h^{(NS5)} \ll \mathcal{R}$ leads to 
\be
\sqrt{N_5} \ll g_s \sqrt{ M} \sin \psi 
\ee
where $N_5$ is the number of NS5 branes defined through $\mathbb{Q}_5 = \mathcal{Q}_5 = N_5 \, \mu_5$ with $\mu_5 = (2 \pi)^{-5} l_s^{-6} $ the charge of a single NS5 brane. Using \eqref{ex15}\footnote{With appropriate factors of $g_s$ and $l_s$ added.}, the condition $r_h^{(\overline{\rm D3})} \ll \mathcal{R}$ leads to
\be
\label{ex175}
\sqrt{\frac{p}{M}} \ll g_s \sqrt{M} \sin^2 \psi ~ .
\ee
From equations \eqref{ex165}-\eqref{ex175}, it is easy to see that our requirements fail at the North and South poles, $\sin \psi=0$. For sufficiently large $M$, however, our calculations will be valid everywhere except a small region around the poles. 

In addition, since the NS5 brane has a running dilaton one may worry whether regions of spacetime with large values of string coupling $e^\phi $ invalidate our analysis. We note that the running of the dilaton is capped off at the horizon for nonextremal solutions at the value $e^\phi (r_0)= g_s \protect \sqrt {\protect \qopname \relax o{sin}^2\theta + \protect \qopname \relax o{cosh}^2\alpha \protect \qopname \relax o{cos}^2\theta }$. Hence, by suitably tuning the asymptotic value of $g_s$ we can achieve wide areas in
  parameter space where our solutions are everywhere weakly coupled.
  Admittedly, this tuning is not possible for extremal solutions. However,
  since it is understood how to treat the strong coupling singularity of NS5
  branes in flat space, and since the constraint (blackfold) equations can be
  obtained in a far-zone analysis of the solution, where the string coupling is
  weak, we anticipate that a large dilaton in the bulk of the solution does not
  invalidate the conclusions of our analysis even at extremality.

\section{Nonextremal effective potentials}\label{EX3}
As noted in equations \eqref{ex19}-\eqref{ex20}, the equations of motions of the KPV DBI action are identical to the extremal blackfold equations. As such, we can take the KPV action as our extremal blackfold action:\footnote{This action is, up to some constant factors in front, simply the spherical NS5 action in \cite{Kachru:2002gs}.} 
\be
\label{extra3}
\mathcal{S}_{KPV} = \int_{\mathcal{M}_4} d x^4 \,  \sqrt{b_0^4 \sin^4 \psi + \left( \pi \frac{p}{M} - \psi + \frac{1}{2} \sin(2 \psi) \right)^2} \sqrt{1 - \psi'^2} - \left( \psi - \frac{1}{2} \sin 2 \psi \right)  
\ee
with $\mathcal{M}_4$ the 4-dimensional Minkowski space. For peace of mind, one can write down the equations of motion of this action, which is given by the usual Euler-Lagrange equations 
\be
\frac{\partial \mathcal{L}}{\partial \psi} - \frac{\partial }{\partial t} \left( \frac{\partial \mathcal{L}}{\partial \psi'} \right) = 0 ~ ,
\ee
and note that indeed we get back our extremal blackfold equations \eqref{ex19}. From the action \eqref{extra3}, we can easily read off the extremal NS5 effective potential:
\begin{equation}
\label{ex16}
V_{\text{KPV}} (\psi) = \sqrt{b_0^4 \sin^4 \psi + \left( \pi \frac{p}{M} - \psi + \frac{1}{2} \sin(2 \psi) \right)^2} - \left( \psi - \frac{1}{2} \sin 2 \psi \right)  ~ .
\end{equation}

The KPV effective potential gives rise to the extremal static blackfold equations. With the nonextremal static blackfold equations in hand, we ask what are the effective potentials that give rise to these equations. These potentials can, thus, be thought of as generalisations of the KPV effective potential for nonextremal antibranes. For constructing nonextremal effective potentials, we are interested in the static blackfold equation\footnote{In the time-independent case, the intrinsic equation \eqref{ex12} becomes trivial.}: 
\be
\label{ex18}
\cot \psi = \frac{1}{b_0^2} \left( \frac{\coth \alpha}{\cos \theta} + \tan \theta \right) \frac{2 \cos^2 \theta}{2 \cos^2 \theta + (\sinh \alpha)^{-2}} 
\ee
where $\tan \theta $ is given by \eqref{ex14}.

Since the local temperature of the NS5 brane does not vanish at extremality, it is appropriate to choose an effective potential that holds some other nonextremality parameters fixed, e.g. the constant boost rapidity $\alpha$ or the global entropy $S$. In this section, we take a look at both. 

\subsection{Effective potential at fixed boost rapidity}
Recall that the boost rapidity $\alpha$ is a measure of nonextremality for the seed anti-D3-NS5 bound state (see appendix \ref{BS1}). In particular when $\alpha \rightarrow \infty$, we have extremal anti-D3-NS5 branes and we can increase their nonextremality by pushing $\alpha$ closer and closer to 0. An obvious finite $\alpha$ effective potential is given by
\be
\label{ex17}
V_{\alpha} (\psi) = b_0^2 \sin^2 \psi \frac{1}{\cos \theta} - \coth \alpha \left( \psi - \frac{1}{2} \sin (2 \psi) \right) + \frac{1}{ \sinh^2 \alpha} H(\psi)
\ee
where $H(\psi)$ is 
\be
H(\psi) = \int^\psi_0 d \chi \cot \chi \, \sqrt{b_0^4 \sin^4 \chi + \left( \pi \frac{p}{M} - \chi + \frac{1}{2} \sin(2 \chi) \right)^2}  ~ .
\ee
When $\alpha \rightarrow \infty$, the constant $\alpha$ effective potential $V_{\alpha}$ \eqref{ex17} reduces to the KPV effective potential $V_{KPV}$ \eqref{ex16}\footnote{In $V_{\alpha} (\psi)$, we use $\theta$ but it is trivial to substitute in $\tan \theta$ \eqref{ex14} to see that they indeed agree.}. Away from extremality, as one can check explicitly, this potential gives rise to the static blackfold equation \eqref{ex18}. Let us note also that the $V_\alpha$ potential near $\pi$ has an unfortunate feature of being “infinitely” sensitive to nonextremality as $H(\pi) = \infty$ due to a $\cot \chi$ in its definition

\begin{figure}
\begin{subfigure}{0.5\textwidth}
\includegraphics[width=0.95\linewidth]{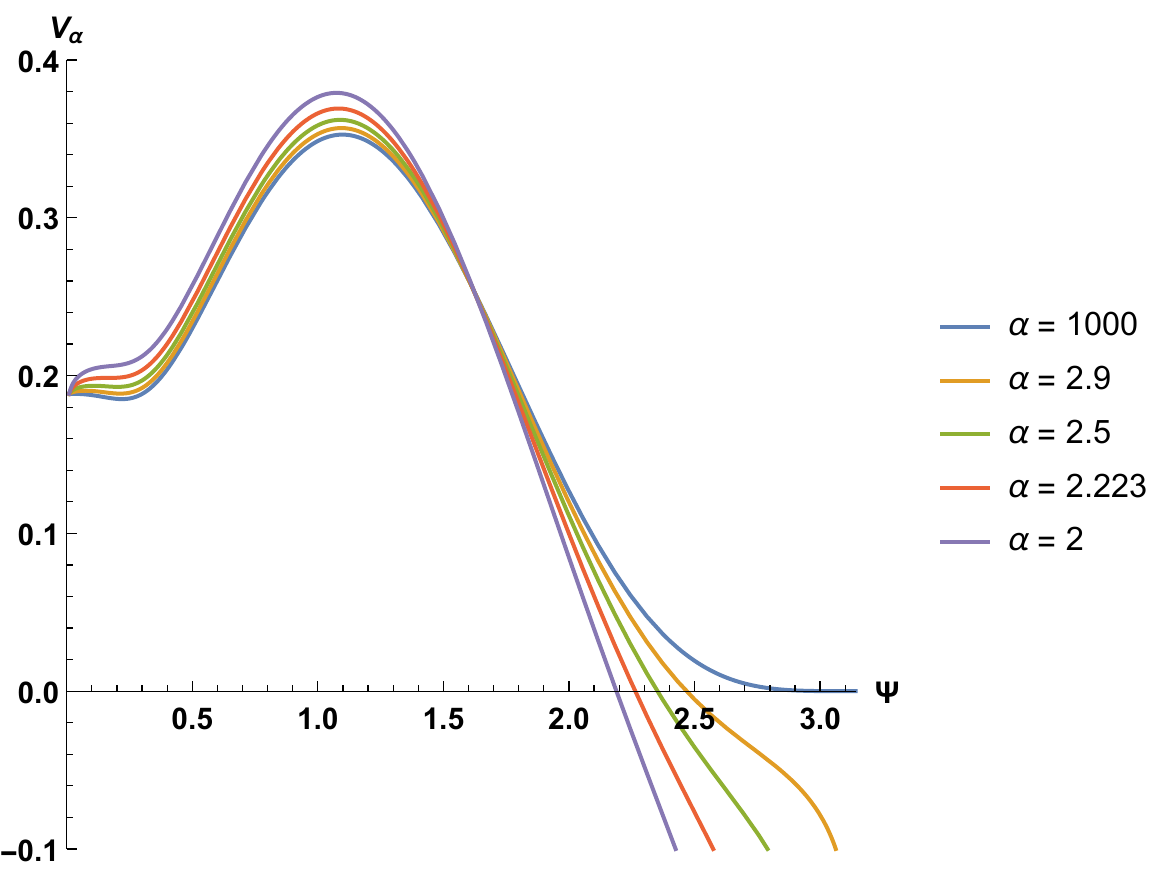} 
\end{subfigure}
\begin{subfigure}{0.5\textwidth}
\includegraphics[width=0.95\linewidth]{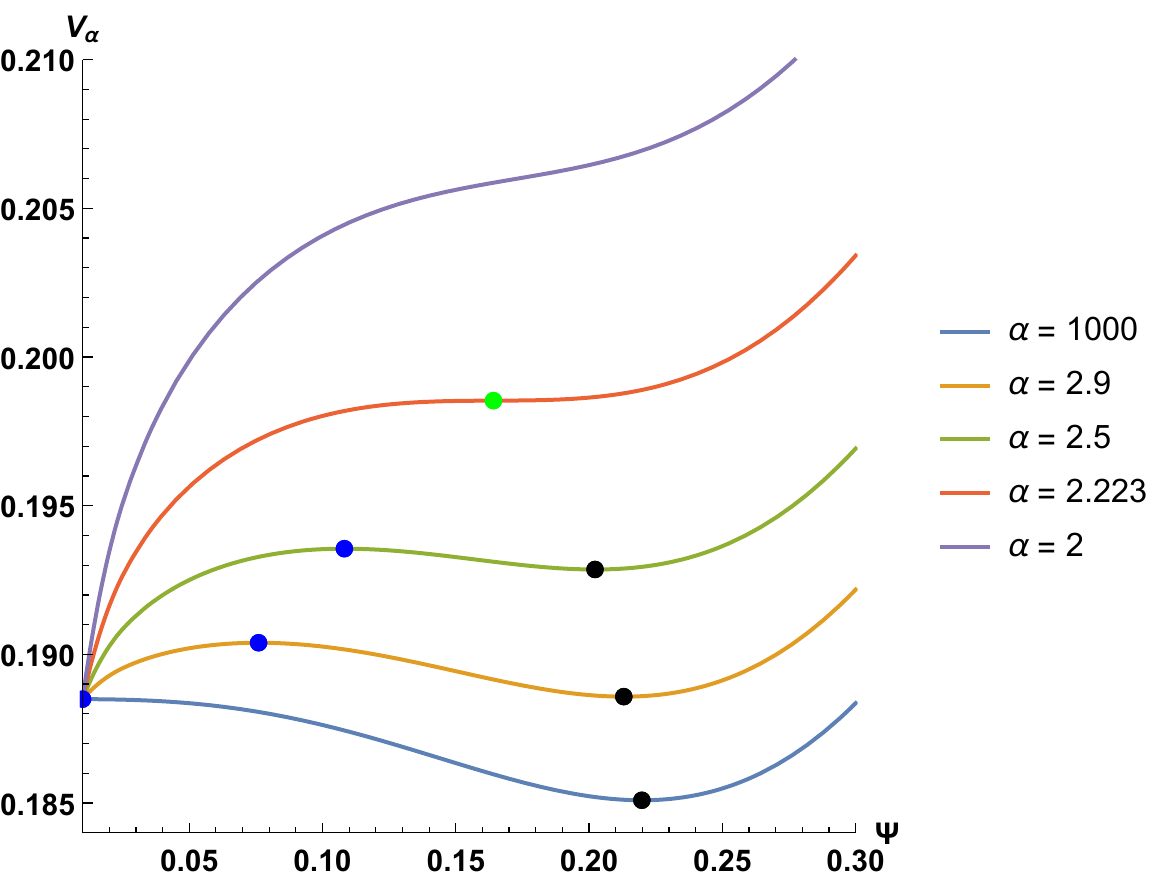}
\end{subfigure}
\caption{Plots of the effective potential at fixed boost rapidity $\alpha$, as a function of the angle $\psi$ on $S^3$. Both figures represent plots at $p/M = 0.03$. The right plot zooms into the region near the North pole of the $S^3$. As we increase nonextremality, we encounter a critical value $\alpha^* \approx 2.223$, where the metastable vacuum of KPV (black dots on the right) merges with a new unstable vacuum (blue dots on the left) to create a saddle point (green dot)}
\label{exfig1}
\end{figure}

Consider first the regime $p/M < p^*/M \simeq 0.08$ where the extremal solutions have a metastable vacuum. In figure \ref{exfig1}, we show how the effective  potential $V_\alpha$ changes as we vary the boost rapidity $\alpha$ for a fixed value of $p/M$. We observe two interesting new features. Firstly, as soon as $\alpha$ is turned on, a new unstable vacuum emerges (black dots on the right plot of figure \ref{exfig1}) near the North pole, $\psi \simeq 0$. For sufficiently low level of extremality, there are three extrema: two unstable and one metastable. Secondly, as we increase the nonextremality further, i.e. lowering $\alpha$, the new unstable extremum comes closer to the metastable vacuum and the two merge at a critical value of the rapidity $\alpha^*$, which is a function of $p/M$. Above this value the metastable vacuum is lost. 

The new unstable state represents a fat black NS5 with a highly pinched $\mathbb{R}^3\times S^5$ horizon geometry that resembles a black anti-D3. Instead, the metastable state starts life near extremality as a thin black NS5 with $\mathbb{R}^3\times S^2 \times S^3$ horizon topology. At the merger, the metastable black NS5 turns effectively into a black anti-D3. The picture of a merger driven by horizon geometry is reinforced by the following observation. 

\begin{figure}
\begin{subfigure}{0.5\textwidth}
\includegraphics[width=0.95\linewidth]{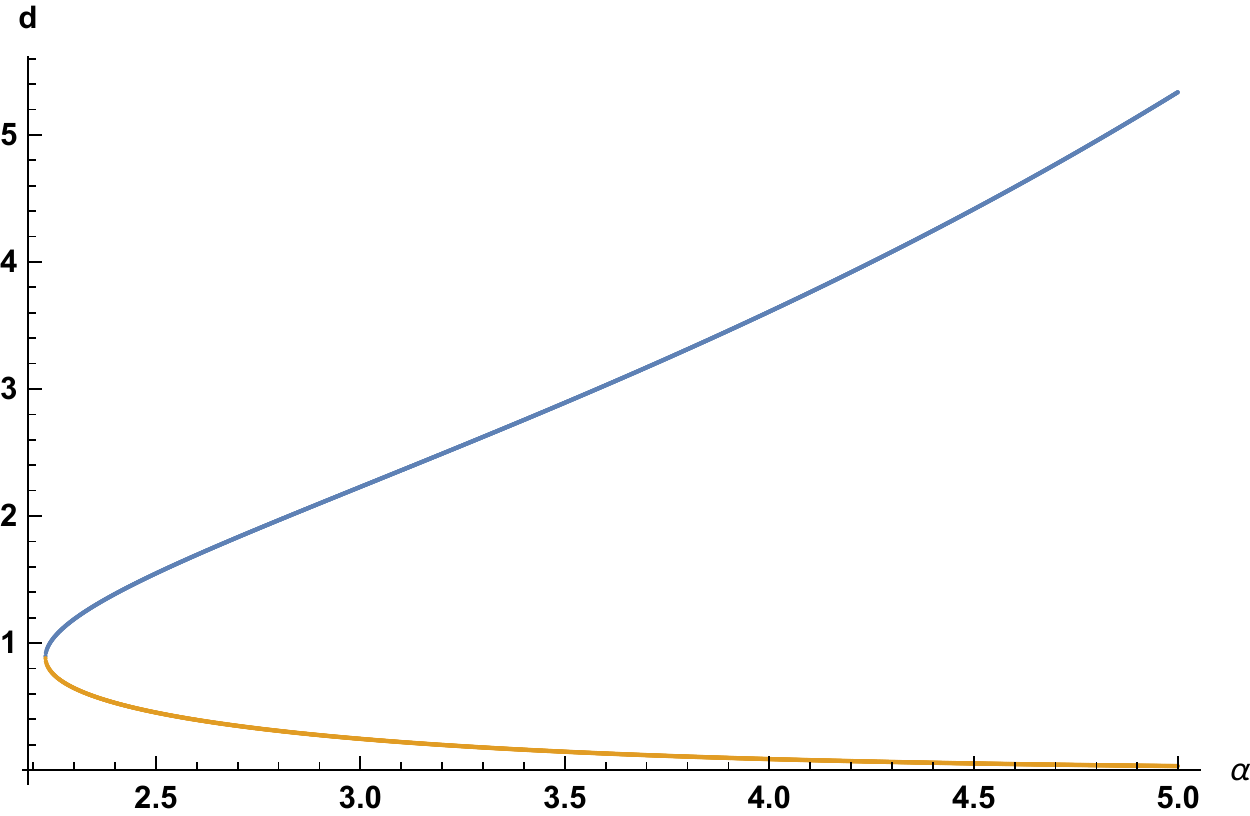} 
\end{subfigure}
\begin{subfigure}{0.5\textwidth}
\includegraphics[width=0.95\linewidth]{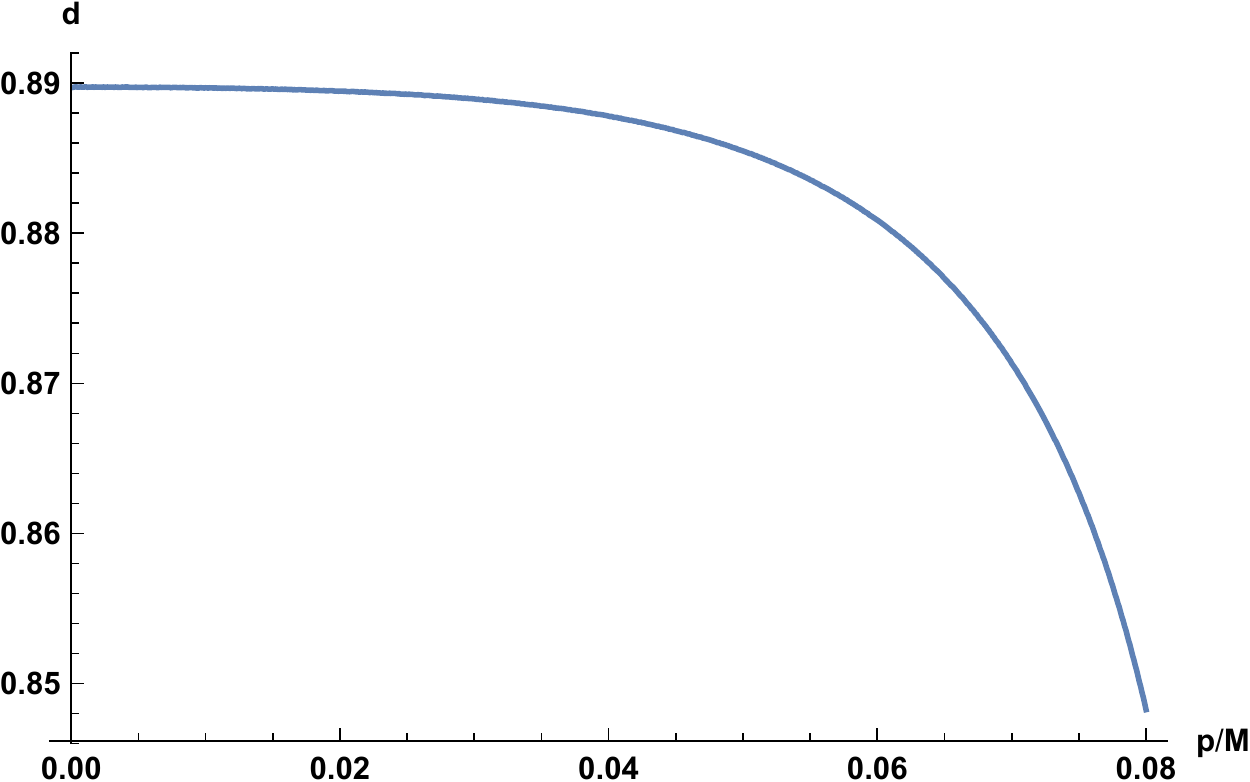}
\end{subfigure}
\caption{Plots of the ratio $d$ that expresses how `fat' a nonextremal $\overline{\rm D3}$-NS5 bound state is. On the left plot we depict the dependence of $d$ on the nonextremality parameter $\alpha$ for the unstable (blue) and metastable (orange) branches for $p/M=0.03$. On the right plot, we depict $d$ at the critical merger point as a function of $p/M$.}
\label{exfig2}
\end{figure}

A quantitative measure of the `fatness' of a black NS5 wrapping an $S^2$ is provided by the ratio
\be
d = \frac{2\sqrt{p} \, \hat r_0}{\sqrt{M}\sin \psi} ~ ,
\ee
where $\hat r_0 \equiv \sqrt{\mathcal C} r_0/\sqrt{Q_5}$ is dimensionless. The ratio $d$, which is a natural function of $p/M$ and the equilibrium $\psi$, compares the scale $2\sqrt{g_s p} \hat r_0 \ell_s= 2(g_s p N_5^{-1})^{1/2} r_0$ associated to the Schwarzschild radius and the scale of the $S^2$ wrapped by the NS5 worldvolume $\sqrt{g_s M}\ell_s \sin \psi$. As an illustration, on the left plot in figure \ref{exfig2} we see how $d$ behaves in the unstable branch (blue colour) and the metastable branch (orange colour). The unstable branch has visibly higher values of $d$, expressing the dominance of the Schwarzschild radius. The metastable branch captures a thin black NS5 with small values of $d$. The merger occurs at a value of $d$ notably close to 1. 

On the right plot of figure \ref{exfig2}, we show how $d$ at the merger point behaves as a function of $p/M$. Remarkably, the ratio remains effectively constant, near the value $0.89$ over a significant range. It deviates slightly from this value in the vicinity of the upper bound of $p/M$, where effects from the second unstable state (already visible in KPV \cite{Kachru:2002gs}) become important. The characteristically weak dependence of $d$ on $p/M$ is a clear signal that the properties of the merger point are closely tied to the properties of the horizon geometry. 
Finally, by increasing $p/M$ further, above the critical value $p^*/M\simeq 0.08$, we observe the complete loss of the metastable vacuum exactly as in the extremal KPV analysis \cite{Kachru:2002gs}. The unstable vacuum in the vicinity of the North pole, however, remains even above $p^*/M$ and constitutes the single vacuum of the nonextremal static blackfold equations.

\subsection{Effective potential at fixed entropy}
\begin{figure}[t]
\begin{subfigure}{0.5\textwidth}
\includegraphics[width=0.95\linewidth]{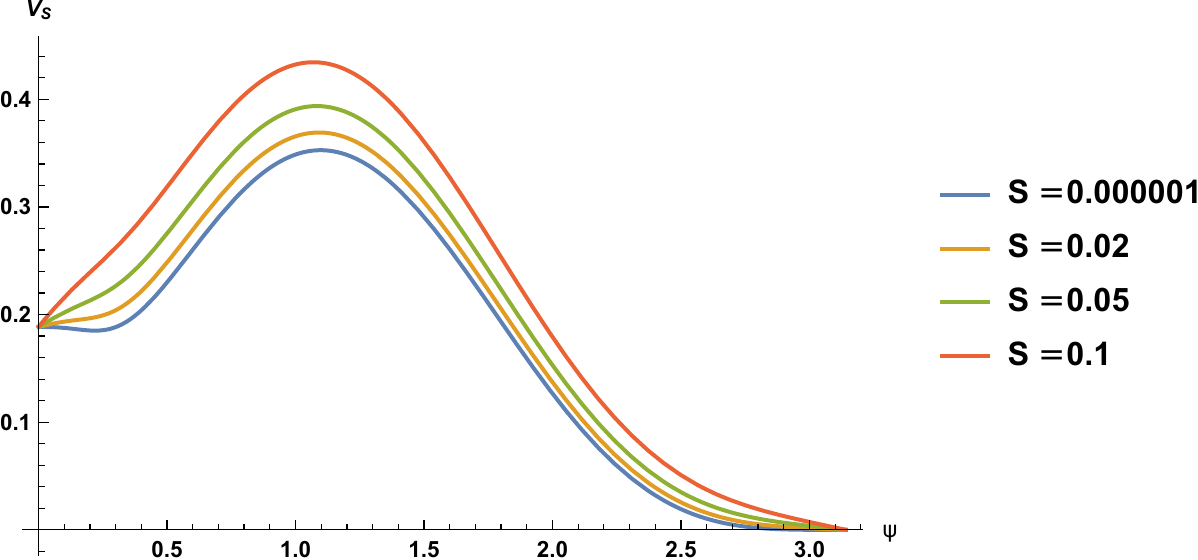}
\end{subfigure}
\begin{subfigure}{0.5\textwidth}
\includegraphics[width=0.95\linewidth]{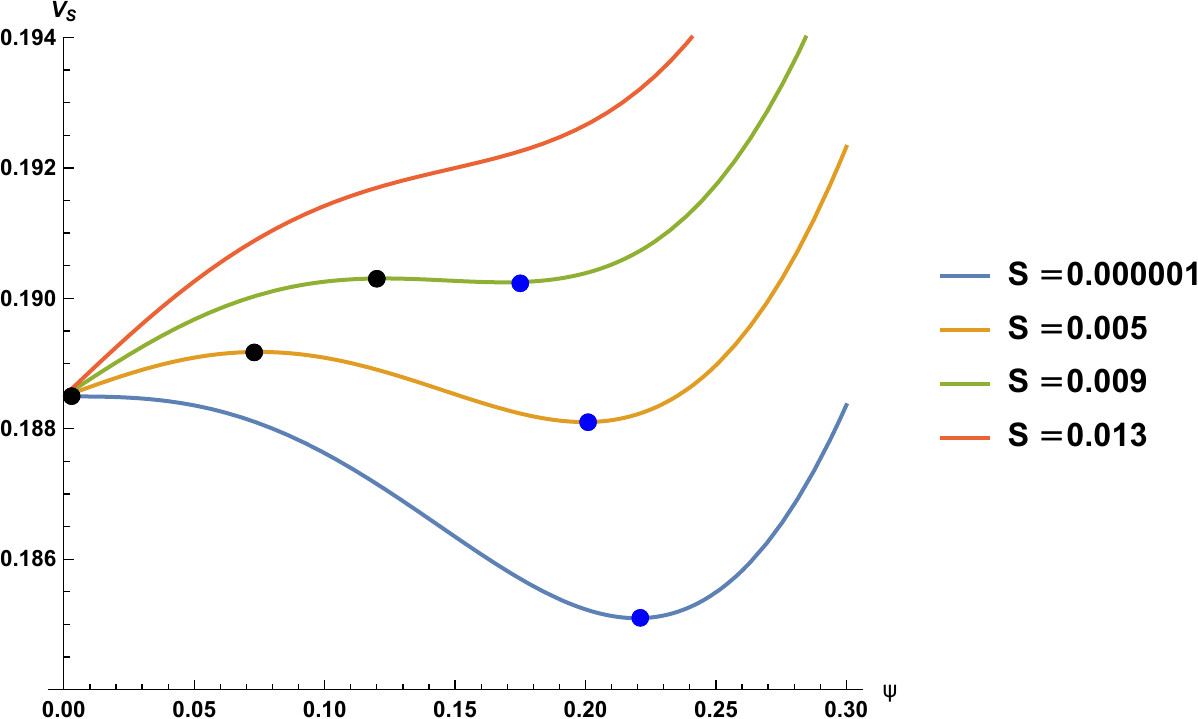}
\end{subfigure}
\caption{Plots of the effective potential at fixed entropy, $V_S$, as a function of the angle $\psi$ on $S^3$. Both figures represent plots at $p/M=0.03$. The right plot zooms into the region near the North pole of the $S^3$. 
As we increase the entropy we encounter a critical value $S^*$, where the metastable vacuum of KPV (blue dots on the right) merges with a new unstable vacuum (black dots on the left). 
}
\label{exfig3}
\end{figure}

As the non-thermodynamic nonextremal parameter boost rapidity $\alpha$ might not be everybody's cup of tea, let us consider effective potentials that hold the global entropy $S$ fixed and demonstrate that they give the same picture. 

We define the global constant entropy as $S \equiv \int_{\mathcal{B}_5} d V_5 \, s \sim r_0^3 \cosh \alpha \sin^2 \psi $ where $\mathcal{B}_5$ is the spatial part of the NS5 worldvolume, and $s$ is the local entropy \eqref{bs2}. Note that, by dividing appropriate factors of constant charge $\mathbb{Q}_5$, we can eliminate $r_0$ to get
\be
\label{ex21}
S \sim \hat{S} = \frac{\sin^2 \psi}{ \sqrt{ \cos^{3} \theta \sinh^{3} \alpha \cosh \alpha}} ~ .
\ee
For convenience, let us from now on use $\hat{S}$ instead of the actual value of $S$. The constant $S$ effective potential is given by
\be
V_S (\psi) = \frac{b_0^2 \sin^2 \psi}{ \cos \theta } \left( \frac{3 + 2 \sinh^2 \alpha}{2 \sinh \alpha \cosh \alpha } \right) - \left( \psi - \frac{1}{2} \sin (2\psi) \right) 
\ee
where $\alpha (\psi, \hat{S})$ should be thought of as implicit function of $\hat{S}$. By obtaining an explicit expression for $ \alpha (\psi, \hat{S})$ from \eqref{ex21}, one can check directly that $V_S (\psi)$ gives rise to our nonextremal blackfold equation \eqref{ex18}. As demonstrated in the plot of the constant entropy potentials in figure \ref{exfig3}, all features described in the previous section using the constant boost rapidity $\alpha$ stay true, with the added perk of having the potentials well-behaved near the South pole $\psi = \pi$.

\section{Discussions on the existence of anitbranes metastable states}\label{EX4}

The initial discovery of the anti-D3-NS5 metastable state was performed from probe computations in \cite{Kachru:2002gs} in two complementary ways: $(i)$ using the worldvolume theory of the anti-D3 branes and $(ii)$ using a worldvolume theory for the NS5 branes. In the D3 perspective $(i)$, the non-abelian DBI action is best understood in the super-Yang-Mills limit, which, effectively, restricts the description close to the north pole of the $S^3$ $(\psi \approx 0)$. The NS5 brane perspective $(ii)$ does not have this restriction but the formulation of an effective worldvolume theory for NS5 branes is more challenging. KPV employed an abelian DBI action that arises by S-duality from the DBI action of the D5 brane. This step is dubious (see e.g.\ \cite{Bena:2014jaa}), because it is in conflict with the regime of validity of the probe approximation that requires $g_s p\ll 1$.

The fate of the metastable state beyond the probe approximation involves higher levels of complexity. Considerable effort has been devoted to understand the properties of backreaction in the supergravity regime where one needs to construct backreacted anti-D3 brane solutions with KS asymptotics. Many works, starting with \cite{Bena:2009xk}, revealed solutions of the supergravity equations that involved unphysical singularities in the 3-form fluxes.\footnote{Related earlier work includes \cite{DeWolfe:2008zy,McGuirk:2009xx}. Subsequent developments after \cite{Bena:2009xk} include \cite{Bena:2010ze,Dymarsky:2011pm,Bena:2011hz,Gautason:2013zw,Blaback:2012nf}.} The presence of these singularities was viewed by some authors as evidence that backreaction can change dramatically the conclusions of the probe approximation, casting doubt on the very existence of the metastable state originally discovered by KPV (and its subsequent applications to string phenomenology, e.g.\ \cite{Kachru:2003sx}). This conclusion was challenged, however, by the authors of \cite{Michel:2014lva} who argued that the inclusion of backreaction effects in the effective field theory of a {\it single} anti-D3 brane are mild and under control, as one would naively expect.

The nonextremal properties of anti-D3 branes can provide further information about the physics of the system. The thermal properties of anti-D3 black branes in the Klebanov-Strassler background were discussed in a series of papers \cite{Bena:2012ek,Bena:2013hr,Blaback:2014tfa,Hartnett:2015oda}.

In the overwhelming majority of the supergravity constructions the discussion centred around the physics of the backreaction of point-like anti-D3 branes. There are several reasons why the NS5-brane point of view is more appropriate: 
\begin{itemize}
\item[1)] The metastable state in the probe computation of \cite{Kachru:2002gs} is a spherical NS5 state.
\item[2)] A natural candidate for the resolution of the observed supergravity singularities involves the formation of a spherical NS5-brane state \'a la Polchinski-Strassler \cite{Polchinski:2000uf}.
\item[3)] The exact supergravity arguments of \cite{Cohen-Maldonado:2015ssa,Cohen-Maldonado:2016cjh}, which are based on Smarr relations, provided a natural explanation for the observed singularities and indicated that the no-go theorems for point-like anti-D3 branes can be evaded for spherical NS5 branes in agreement with the probe computation. In particular, there are no regular extremal solutions of point-like anti-D3 branes. Here and in what follows, the terminology `point-like anti-D3' refers to solutions with vanishing NS5 brane dipole charge and spherical horizon topology (more precisely, $\IR^3 \times S^5$ horizon topology). Regular solutions of NS5 branes wrapping a 2-cycle, with horizon topology $\IR^3 \times S^2 \times S^3$, can evade the no-go theorem. At finite temperature, point-like anti-D3 black brane solutions can in principle exist but they cannot have a regular extremal limit. Regular black NS5 brane solutions wrapping a 2-cycle are in principle allowed.
\end{itemize}
For all these reasons, a proper understanding of anti-D3 backreaction in the KS background requires information about spherical NS5 states. Without this information, previous indications, either from the probe or the supergravity computations with point-like or smeared anti-D3 brane solutions, remain inconclusive.

By utilising the blackfold equations, the necessary conditions (and perhaps sufficient conditions) for the first order matched asymptotic solution, we have made the following key observations of the backreacted spherical NS5 brane states in KS throat.
\begin{itemize}

\item[$1)$] As discussed in section \ref{EX1}, extremal spherical NS5 branes should obey at leading order in the derivative expansion the same equations that were employed by KPV, namely the equations that arise from the S-dual of the DBI action for the D5 brane. Since our blackfold analysis are consistent in the supergravity regime of large $g_s p$, there is no clash between different regimes of validity and, thus, that removes one of the criticisms against the KPV metastable state.

\item[$2)$] In the appropriate regime of parameters, $p/M\lesssim 0.08$, the extremal spherical NS5 branes exhibit two vacua away from the north and south poles: one metastable and one unstable. The study of nonextremal effective potentials in section \ref{EX3} reveals that, as soon as the branes become nonextremal, an additional unstable vacuum appears. This is a novel, `fat' unstable NS5-brane state. Increasing the entropy of the solutions leads to a merger of the fat unstable state with the thin metastable state, annihilating both vacua. As a function of the entropy, the nonextremal system exhibits a transition with the features of saddle-node bifurcation. The plot of the ratio $$d \sim (\text{NS5 Schwarzschild radius}/ \text{Radius of the wrapped } S^2)$$ (figure \ref{exfig2}) at the point of transition shows that $d$ is constantly close to 1 for the majority of the regime, signally that the origin of this transition is closely related to the geometric properties of the corresponding black hole horizon.

\item[$3)$] The emerging picture from our blackfold analysis is suggestively consistent with the exact analysis of \cite{Cohen-Maldonado:2015ssa}. In all cases where \cite{Cohen-Maldonado:2015ssa} lifted the no-go theorem, our blackfold analysis produce a go with concrete quantitative predictions. In particular, we observe that the anti-D3-NS5 branes can form a metastable state at the tip of the KS throat. However, such metastable solution would disappear as soon as we heat the anti-D3-NS5 state sufficiently that its horizon geometry resembles that of a point-like anti-D3 state.
\end{itemize}
We believe that these findings constitute strong evidence for the existence of the metastable states, since they have now been argued for in complementary regimes.

\chapter{On the stability of metastable antibranes}
\label{STA}
The claims regarding metastability of antibranes in warped throats from DBI analysis in \cite{Kachru:2002gs,Klebanov:2010qs} and from the blackfold approach in \cite{Armas:2018rsy,M2M5brane} are all with respect to \textit{specific} modes of deformations. For their numerous applications, particularly cosmological string de Sitter construction \cite{Kachru:2003aw}, it is important to determine whether these branes are truly metastable under generic perturbations. In this chapter, through an application of the blackfold approach, we study the classical stability of the exemplar Kachru-Pearson-Verlinde (KPV) state \cite{Kachru:2002gs} of polarised anti-D3 branes at the tip of the Klebanov-Strassler (KS) \cite{Klebanov:2000hb} throat with regards to generic long-wavelength deformations. As demonstrated later on, we observe that the blackfold equations (constraints on long-wavelength deformations) prohibit the existence of tachyonic modes.

We begin with a review of the blackfold description of the KPV state in section \ref{STA1}. Subsequently,  in section \ref{STA2}, we present a long-wavelength stability analysis of the state. In this section, we also provide a discussion on how the different notions of charge give rise to an interesting dynamical property of the anti-D3-NS5 branes.

\section{Blackfold description of the KPV state}\label{STA1}
The purpose of this section is to review the blackfold description of the KPV state. Various aspects of the metastable anti-D3-NS5 blackfold, including recovery of the KPV state, have already been discussed in chapter \ref{EX}. Nevertheless, we find it useful to revisit the starting point of our stability analysis.

\paragraph{Blackfold equations}
In the blackfold description of extremal anti-D3-NS5 branes in KS background, the variables of the system are
\be
r, \theta_2, \phi_2, \psi, r_h,\tan \theta, v^a, w^a  ~ .
\ee    
Variables $r, \theta_2, \phi_2, \psi$ are embedding parameters of the anti-D3-NS5 branes to the background. Variables $r_h,\tan \theta, v^a, w^a$ are characteristic parameters describing respectively extremal horizon radius, charge distribution, and the orientation of the dissolved charge.

For describing the KPV state, as we are only interested in static and spatially homogeneous configurations, we can already fix variables $r, \theta_2, \phi_2, v^a, w^a$ (see equations (\ref{80})-(\ref{81}) for detailed expressions) and set the remaining variables $r_h, \psi, \tan \theta$ to constants with respect to the worldvolume coordinates. In our conventions, the extremal anti-D3-NS5 blackfold equations are given by
\begin{enumerate}
\item The energy-momentum conservation equations:
\begin{align}
\label{3000}
\nabla_a T^{a b} &= \p^b X_\mu \, \mathcal{F}^\mu ~ , \\
\label{3001}
T^{ab} K_{ab}^{\,\,\, \,\,\, (i)}  &= \mathcal{F}^\mu \, n^{(i)}_\mu
\end{align}
where $n_\mu^{(i)}$ denotes the normal vectors of the anti-D3-NS5 blackfold, $K_{ab}^{\,\,\, \,\,\, (i)} = K_{ab}^{\ \ \, \rho} n_\rho^{(i)}$,  and the force term $\mathcal{F}^\mu$ is given by
\begin{multline}
\label{90}
\mathcal{F}^\mu = - \frac{1}{6!}  H_{7}^{\mu a_1 ... a_6} j_{6 a_1 ... a_6} + \frac{1}{2!} \tilde{F}_3^{\mu a_1 a_2} J_{2 a_1 a_2} + \frac{3}{4!} H_3^{\mu a_1 a_2} C_2^{a_3 a_4}J_{4 a_1 ... a_4} \\
+ \frac{1}{4!} \tilde{F}_5^{\mu a_1 ... a_4} J_{4 a_1 ... a_4}  ~ .
\end{multline}
For the purpose of describing the KPV state, the terms with $H_3$ and $\tilde{F}_5$ in \eqref{90} are not relevant because they vanish at the tip of the throat. Nevertheless, as they will play a role when we consider perturbations away from the tip, we present them explicitly here.

\item The current conservation equations:
\begin{align}
\label{34}
d * j_6 &= 0 ~ , \\
\label{345}
d * J_4 + * j_6 \wedge F_3  &= 0 ~ ,\\
\label{35}
d * J_2 + H_3 \wedge * J_4 &= 0 
\end{align}
where $F_3, H_3$ are the projected background fluxes and $*$ is the 6-dimensional Hodge dual of the worldvolume directions.
\end{enumerate}

From the current conservation equations, we can define the conserved Page charges $\mathbb{Q}_3$ and $\mathbb{Q}_5$ that keep track of the number of anti-D3 and NS5 branes:
\begin{align}
\label{sta1}
\mathbb{Q}_5 &= * j_6 = C r_h^2 \cos \theta ~ ,\\
\label{71}
\mathbb{Q}_3 &= \int_{S^2} * \left( J_4 + *(* j_6 \wedge C_2 ) \right) \\
&= -  4  \pi \Big( C r_h^2 \sin \theta M b^2_0 \sin^2 \psi + C r_h^2 \cos \theta M (\psi - \frac{1}{2} \sin 2 \psi  ) \Big)  
\end{align}
where we have used $C_2 = M (\psi - \frac{1}{2} \sin 2 \psi) \sin \omega d \omega \wedge d \varphi$. It follows immediately that we can write $\tan \theta$ as
\be
\label{1}
\tan \theta = \frac{1}{b_0^2 \sin^2 \psi} \left(  \frac{\pi p}{M} - \left(\psi - \frac{1}{2}\sin 2 \psi \right) \right)
\ee
where we have made the identification
\be
\label{1010}
\frac{- \mathbb{Q}_3}{4 \pi \mathbb{Q}_5} = \pi p ~ .
\ee
From the energy-momentum tensor conservation equations, after some algebraic acrobatics, we can write all variables in term of $\psi$ and obtain the equation 
\be
\label{2}
\cot \psi - \frac{1}{b_0^2} \sqrt{1 + \tan^2 \theta} - \frac{1}{b_0^2} \tan \theta = 0 ~ .
\ee

\paragraph{The KPV state} 
We can numerically determine that equation (\ref{2}) has a metastable solution for $0 < p/M < p_{crit}$ where $p_{crit} \approx 0.080488$. These metastable solutions are the KPV states. For our convenience, let us note down some explicit information of the configuration. With respect to our variables, the KPV states are specified by
\begin{align}
\label{80}
r &= 0, & \psi &= \psi_0,& \tan \theta &= \frac{1}{b_0^2 \sin^2 \psi_0} \left( \frac{\pi p}{M} - \psi_0 + \frac{1}{2}\sin (2 \psi_0)\right) ~ ,
\end{align}

\begin{align}
\label{81}
r_h &= \sqrt{\frac{\mathbb{Q}_5}{C \cos \theta }}, &v^a \p_a &=  \frac{1}{\sqrt{ M} b_0 \sin \psi_0 } \partial_\omega ,& w^a \p_a &= \frac{1}{\sqrt{M} b_0 \sin \psi_0 \sin \omega} \partial_\varphi 
\end{align}
where $\psi_0$ is the metastable solution of
\be
\cot \psi - \frac{1}{b_0^2} \sqrt{1 + \tan^2 \theta} - \frac{1}{b_0^2} \tan \theta = 0  ~ .
\ee
We note also the induced metric on the worldvolume of the anti-D3-NS5 branes
\begin{equation}
\gamma_{a b} d\sigma^a d\sigma^b = M b_0^2 \left( - dt^2 + (dx^1)^2 + (dx^2)^2 + (dx^3)^2  + \sin^2 \psi_0 \left( d \omega^2 + \sin^2 \omega d \varphi \right)  \right ) \, ,
\end{equation}
the non-zero components of the worldvolume Christoffel symbol $\Theta^a_{bc}$
\begin{align}
\label{01}
\Theta^{\varphi}_{\omega \varphi} &= \Theta^{\varphi}_{\varphi \omega} = \cot \omega ~ , & \Theta^{\omega}_{\varphi \varphi} &= - \cos \omega \sin \omega  \, ,
\end{align}
the relevant components of the  background Christoffel symbol $\Gamma^{\mu}_{\alpha \beta}$
\begin{align}
&\Gamma^{\psi}_{\omega \omega} = - \cos \psi_0 \sin \psi_0 ~ , & &\Gamma^{\psi}_{\varphi \varphi} = - \cos \psi_0 \sin \psi_0 \sin^2 \omega ~ ,
\end{align}
and the non-zero component of the extrinsic curvature $K_{ab}^{\ \ \, \rho}$
\begin{align}
\label{03}
K_{\omega \omega}^{\ \ \ \psi} &= - \cos \psi_0 \sin \psi_0~ , & K_{\varphi \varphi}^{\ \ \ \psi} &= - \cos \psi_0 \sin \psi_0 \sin^2 \omega  ~ .
\end{align}

\section{Stability of the KPV state}\label{STA2}
The goal of this section is to analyse generic deformations of the KPV state. Starting with the blackfold description of the configuration, we introduce generic perturbations by varying slightly all its variables. Subsequently, we derive the blackfold perturbation equations, and use these equations to identify allowed deformations. 
\subsection{Perturbation parameters}
\label{PerPara}
To introduce perturbations to our system, we vary slightly the variables of the configuration around their KPV values. Explicitly, we have
\begin{align}
r &= 0 + \delta r ~,& \psi &= \psi_0 + \delta \psi ~, & r_h &= \sqrt{\frac{\mathbb{Q}_5}{C \cos \theta(\psi_0)}} + \delta r_h ~,
\end{align}
\begin{align}
\tan \theta = \frac{1}{b_0^2 \sin^2 \psi_0} \left( \frac{\pi p}{M} - \psi_0 + \frac{1}{2}\sin (2 \psi_0)\right) + \delta \tan \theta  ~ ,
\end{align}
\begin{align}
v^a \p_a &=  \frac{1}{\sqrt{ M} b_0 \sin \psi_0 } \partial_\omega + \delta v^a \p_a ,\\
w^a \p_a &= \frac{1}{\sqrt{ M} b_0 \sin \psi_0 \sin \omega} \partial_\varphi + \delta w^a \p_a
\end{align}
where all variations are functions of the worldvolume coordinates, e.g. $\delta r_h (\sigma)$. To simplify our syntax, from here on we shall \textbf{denote the variable values at the KPV configuration by the variables themselves}, e.g. $\psi_0$ will be denoted as $\psi$, the value of $\tan \theta$ at KPV is denoted as $\tan \theta$, etc. 

Let us make use of symmetries and constraints to minimise the number of parameters we work with while still preserving all the relevant information for the stability analysis. Firstly, because of Lorentz symmetry of the blackfold equations and the original KPV configuration, without loss of generality, we can consider variations involving the worldvolume coordinate $t$ only instead of the full Minkowskian coordinates $t,x^1,x^2,x^3$. Secondly, using the unitary constraints on $v$ and $w$, i.e. $v^a v_a = w^a w_a = 1$, we can show that
\begin{align}
\label{5}
\delta v^\omega &= - \frac{\cos \psi }{\sqrt{M} b_0  \sin^2 \psi} \delta \psi ~ , \\
\label{6}
\delta w^\varphi &= - \frac{\cos \psi }{\sqrt{M} b_0  \sin^2 \psi \sin \omega} \delta \psi ~ .
\end{align}
Thirdly, as we use $v$ and $w$ together as normal vectors to specify the anti-D3 charge orientation inside the NS5 branes, it is obvious that we have a rotational gauge symmetry here. Making use of this gauge symmetry and the orthogonality constraint, i.e. $v^a w_a = 0$, we can set 
\be
\delta v^\varphi = \delta w^\omega = 0 ~ .
\ee
With the simplifications noted above, our relevant variation parameters are
\begin{gather}
\delta r (t, \omega, \varphi), \delta \psi (t, \omega, \varphi), \delta r_h (t, \omega, \varphi), \delta \tan \theta (t, \omega, \varphi),\\ 
\delta v^t (t, \omega, \varphi), \delta v^\omega (t, \omega, \varphi), \delta w^t (t, \omega, \varphi), \delta w^\varphi (t, \omega, \varphi)
\end{gather}
where $\delta v^\omega, \delta w^\varphi$ can be written in term of $\delta \psi$ as expressed in (\ref{5})-(\ref{6}).

\paragraph{Regime of validity}

As we utilise the blackfold equations, it is obvious that our stability analysis only makes sense when the blackfold equations are valid. As discussed thoroughly in section \ref{EX2}, this happens when the size of the wrapped $S^2$ is very large, i.e. when $\psi \neq 0, \pi$ and $M$ very large. Let us note also that, with our current approach, we can only study  long-wavelength perturbations despite the potentially deceiving form of the (\ref{5})-(\ref{6}).

When studying ripples on the spherical NS5, because we are looking at compactified space, one might be worried whether these modes can be slowly varying. However, as we require this wrapped $S^2$ to be very large, up to some order of spherical harmonics, there is no problem.   

\subsection{Blackfold perturbation equations}
In this subsection, we present the blackfold equations for perturbations around the KPV state. We relegate the exciting details on the derivation of these equations to appendix \ref{secB}. 
\subsubsection{Conservative Currents \& Charges}
As shown in (\ref{A}), the $j_6$ conservation equation implies 
\be
\p_a \, \delta \mathbb{Q}_5 = 0
\ee
where $\mathbb{Q}_5 = C r_h^2 \cos \theta$. This means $\delta \mathbb{Q}_5$ is a constant of motion. Recall that $\mathbb{Q}_5$ keeps track of the number of NS5 branes. As we are interested in the dynamical stability of the KPV configuration, we impose the condition that $ \delta \mathbb{Q}_5$ vanishes. Note that the imposition $ \delta \mathbb{Q}_5 = 0$ automatically fixes $\delta r_h$ in term of $\delta \tan \theta$
\be
\label{002}
\delta r_h = \frac{1}{2} r_h \cos \theta \sin \theta \delta \tan \theta \, .
\ee
As shown in (\ref{B}), the $J_4$ conservation equation implies 
\begin{multline}
\label{007}
- \mathbb{Q}_5 M b_0^2 \sin^2 \psi \sin \omega
\Bigg(  \p_t \delta \tan \theta + 2 \tan \theta \cot \psi \p_t \delta \psi + \frac{2}{b_0^2}  \p_t \delta \psi \Bigg) \\
=  \mathbb{Q}_5  M^{3/2} b^3_0 \tan \theta  \sin \psi \Big( \p_\varphi \delta w^t + \p_\omega  \left( \sin \omega \delta v^t \right) \Big) ~ .
\end{multline}
Integrating over $\omega$ and $\varphi$ and enforcing the periodicity conditions
\be
\delta w^t|_{\varphi =0} = \delta w^t|_{\varphi = 2 \pi} ~ ,
\ee
we obtain\footnote{Let us note that the equation keeps constant the $\mathbb{Q}_3$ Page charge while put no restrictions on the $\mathcal{Q}_3$ brane charge, which is free to vary.}
\be
\p_a \delta \mathbb{Q}_3 = 0
\ee
where
\be
\label{100}
\delta \mathbb{Q}_3 = \int_{S^2} \delta \left( * \tilde{J}_4 \right) = -  \mathbb{Q}_5 M b_0^2 \sin^2 \psi \int d \omega d \varphi \sin \omega \Bigg(  \delta \tan \theta + 2 \left( \tan \theta \cot \psi + \frac{1}{b_0^2} \right) \delta \psi \Bigg) ~ .
\ee
This means $\delta \mathbb{Q}_3$ is a constant of motion. In a similar fashion to how the $ \mathbb{Q}_5$ charge keeps track of the number of NS5 branes, the $ \mathbb{Q}_3$ charge keeps track of the number of anti-D3 branes. As we are interested in the dynamical stability of the KPV configuration, we shall impose that $\delta  \mathbb{Q}_3 = 0$. However, note that unlike the $ \mathbb{Q}_5$, the imposition $\delta  \mathbb{Q}_3 = 0$ doesn't automatically guarantee the satisfaction of the current perturbation equation. \\
Finally, as shown in (\ref{C}), the $J_2$ conservation equation implies 
\begin{align}
\label{003}
\cot \theta \cos^2 \theta \p_\omega \delta \tan \theta + \sqrt{M} b_0 \sin \psi \p_t \delta v^t  = 0 ~ ,\\
\cot \theta \cos^2 \theta  \p_\varphi \delta \tan \theta +  \sqrt{M} b_0 \sin \psi \sin \omega \p_t \delta w^t = 0 ~ ,\\
\label{004}
\p_\varphi \delta v^t - \p_\omega (\sin \omega  \delta w^t )= 0 ~ .
\end{align}

\subsubsection{Energy-momentum conservation equations}
Recall from (\ref{3000})-(\ref{3001}), the intrinsic and extrinsic blackfold equations 
\begin{align}
\nabla_a T^{a b} &= \p^b X_\mu \, \mathcal{F}^\mu  ~ ,\\
T^{ab} K_{ab}^{\,\,\, \,\,\, (i)}  &= \mathcal{F}^\mu \, n^{(i)}_\mu ~ .
\end{align}
Focusing on perturbations around the KPV state, as shown in (\ref{D}), the intrinsic equation implies for $b = t,\omega, \varphi$ respectively
\begin{enumerate}
\item The $t$ intrinsic perturbation equation: 
\begin{multline}
\label{005}
\p_t \delta \tan \theta +  \frac{\sqrt{M} b_0}{\sin \psi} \tan \theta \left( \p_\omega \delta v^t + \frac{1}{\sin \omega} \p_\varphi \delta w^t + \cot \omega \delta v^t \right) 
 \\
 + 2 \left( \cot \psi \tan \theta + \frac{1}{b_0^2} \right) \p_t \delta \psi = 0 ~ ,
\end{multline} 

\item The $\omega$ intrinsic perturbation equation:
\be
\label{9}
\sqrt{M} b_0 \sin \psi  \tan^2 \theta  \p_t \delta v^t +   \sin \theta \cos \theta  \p_\omega \delta \tan \theta    = 0 ~ ,
\ee

\item The $\varphi$ intrinsic perturbation equation: 
\be
\label{10}
\sqrt{M} b_0 \sin \psi \sin \omega \tan^2 \theta \p_t \delta w^t  + \sin \theta \cos \theta \p_\varphi \delta \tan \theta  = 0 ~ .
\ee
\end{enumerate}
Similarly, as shown in (\ref{E}), the extrinsic blackfold equation implies
\begin{enumerate}
\item The $\psi$ extrinsic perturbation equation:
\be
\label{008}
(\p_t)^2 \delta \psi -  \frac{ \cos^2 \theta}{ \sin^2 \psi}   \nabla^2 \delta \psi  = \frac{2  \cos^2 \theta}{ \sin^2 \psi}  \delta \psi + \frac{2}{b_0^2}  \cos^2 \theta \left( 1 + \sin \theta \right) \delta \tan \theta ~ ,
\ee

\item The $r$ extrinsic perturbation equation:
\begin{multline}
\label{001}
 (\p_t)^2 \delta r - \frac{\cos^2 \theta}{ \sin^2 \psi}  \nabla^2 \delta r   = \frac{8 a_2}{a_0} \sin \theta \delta r +  \frac{8 a_2 }{ a_0}  \delta r   -  \frac{  16 a_0 + 20 a_2  }{5  a_0} \cos^2 \theta  \delta r \\
 +  \frac{ 4  }{5} \cos^2 \theta   \sin^2 \omega  \delta r 
\end{multline}
where $a_0 \approx 0.71805$, $a_2 = - (3 \times 6^{1/3})^{-1} $ are the warping constants of the KS throat (\ref{A44}) and $\nabla^2$ is the normalised Laplacian, i.e. $\nabla^2 = (\p_\omega)^2 + 1/\sin^2 \omega (\p_\varphi)^2 + \cot \omega \p_\omega$. 

\end{enumerate}

Before continuing, let us note an interesting fact about the $r$ extrinsic equation. If one follows the details in paragraph \ref{extrinsic}, it can be easily seen that the term 
\be
\frac{8 a_2 }{ a_0}  \sin \theta \delta r
\ee
is the $\tilde{F}_5$ electromagnetic force term while the terms
\be
\label{asdf}
\frac{8 a_2 }{ a_0}  \delta r  -  \frac{  16 a_0 + 20 a_2  }{5  a_0} \cos^2 \theta  \delta r  +  \frac{ 4  }{5} \cos^2 \theta   \sin^2 \omega  \delta r
\ee
are the gravitational force terms coming from the warping of the throat. The direction of the electromagnetic force term depends on the sign of the D3 brane charge carried by the KPV state $\mathcal{Q}_3 = C r_h^2 \sin \theta$. As KPV is a polarised state of anti-D3 branes, one might naively expect that this force is always attractive. However, this is not the case. The reason is because, in a fluxed setting, the D3 Page charge (\ref{71}) and the D3 brane charge are not necessarily the same. In particular, for a range of $p/M$ near $p_{crit}$, the $\mathcal{Q}_3$ brane charge flips sign and, consequently, the electromagnetic force becomes repulsive. This effect can also be seen with the Klebanov-Pufu (KP)  configuration \cite{Klebanov:2010qs} of anti-M2 branes at the tip of the Cvetic-Gibbons-Lu-Pope (CGLP) throat \cite{Cvetic:2000db}.

\begin{figure}
\begin{centering}
\begin{subfigure}{0.32\textwidth}
\includegraphics[width=\linewidth]{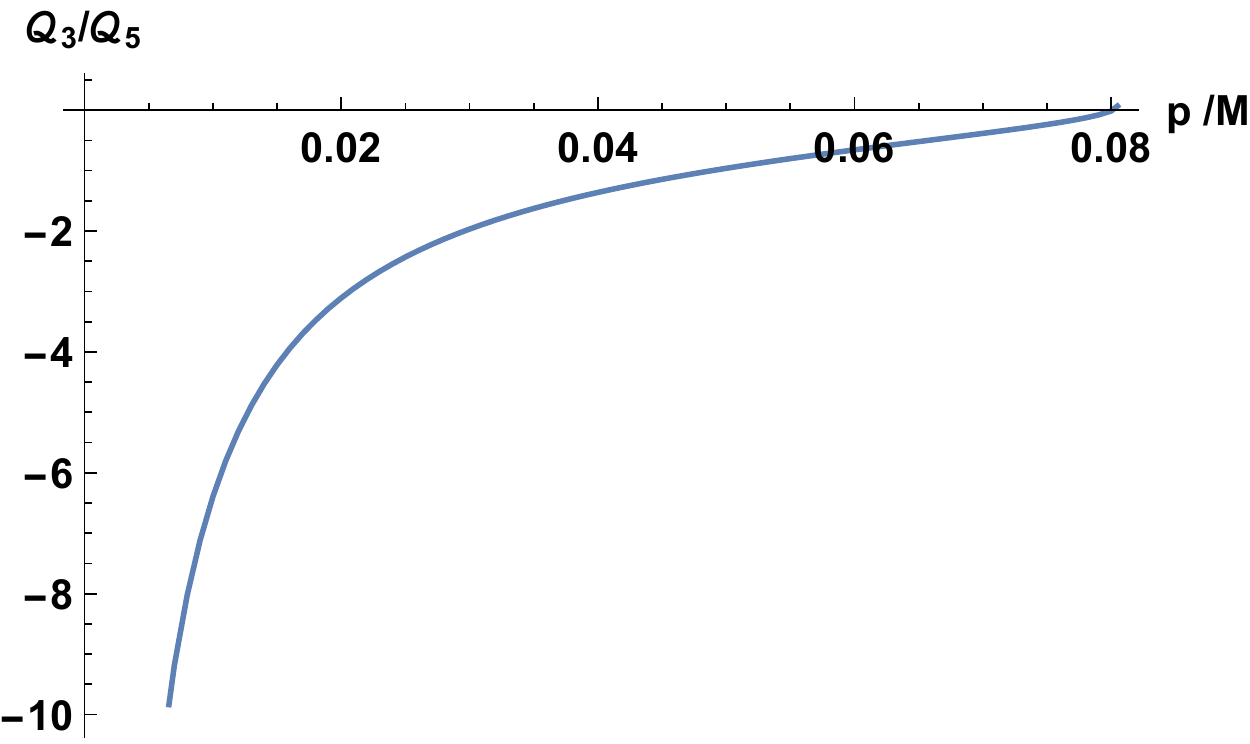} 
\end{subfigure}
\begin{subfigure}{0.32\textwidth}
\includegraphics[width=\linewidth]{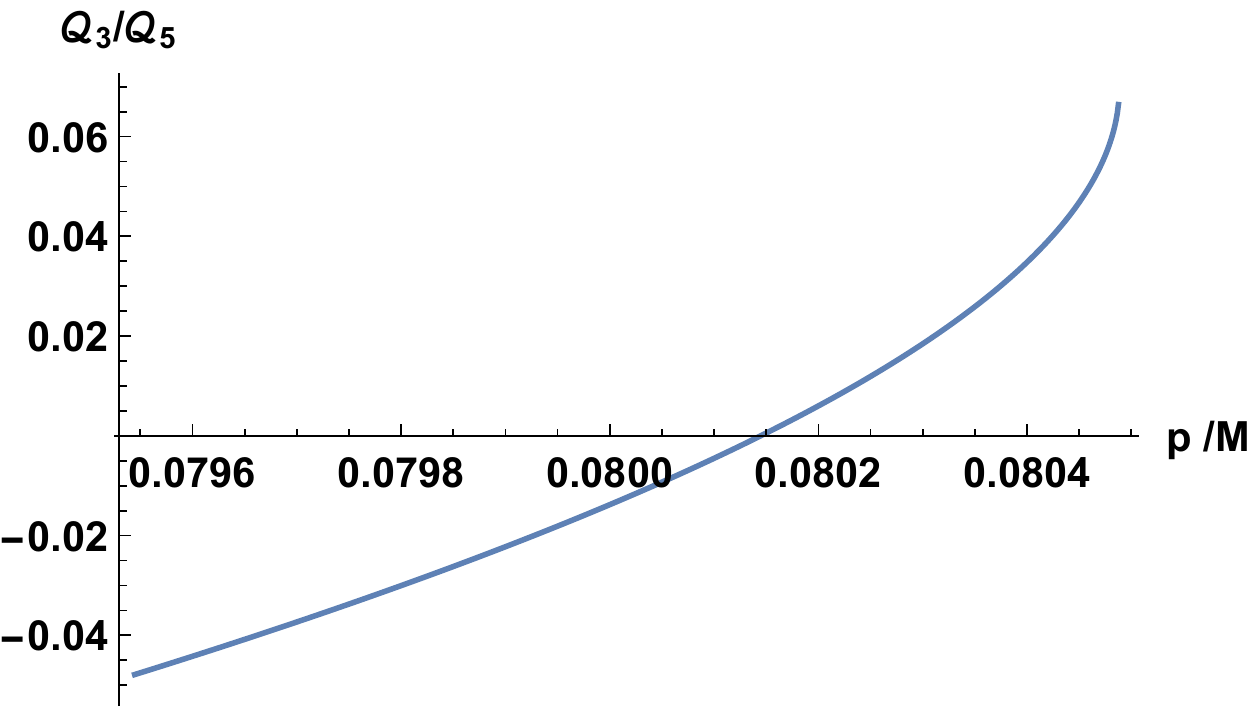}
\end{subfigure}
\begin{subfigure}{0.32\textwidth}
\includegraphics[width=\linewidth]{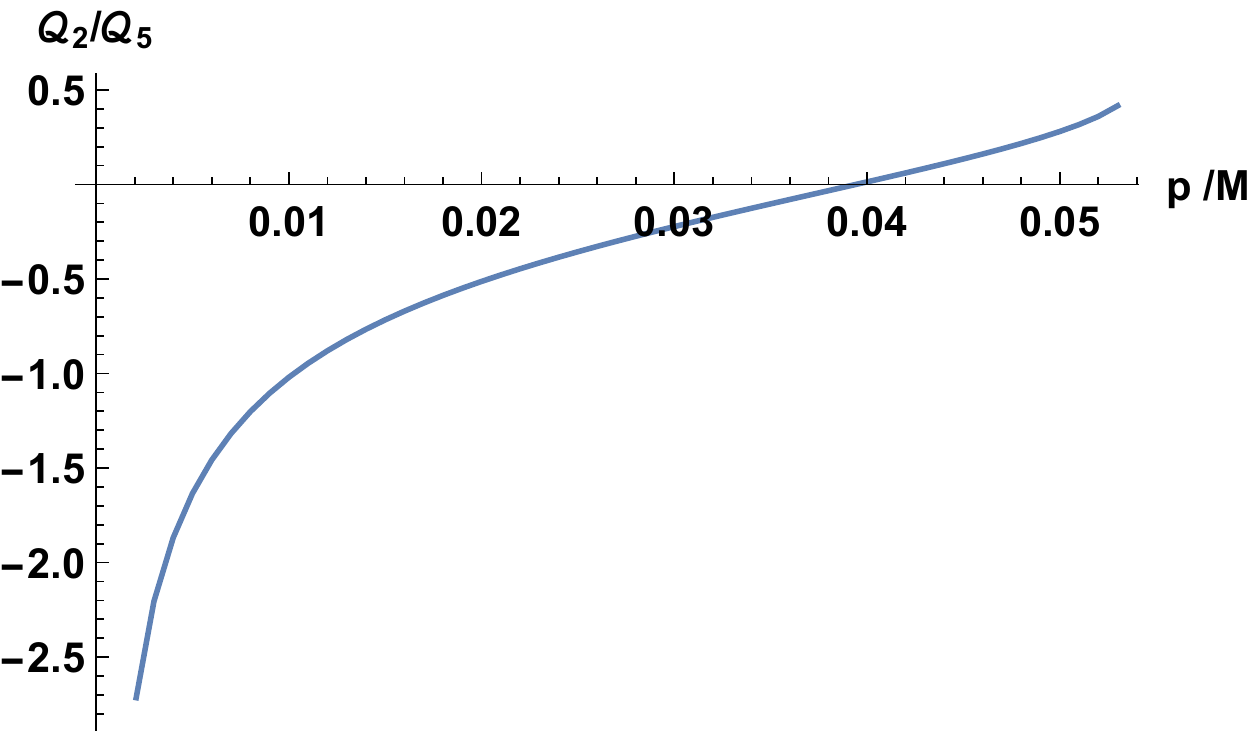}
\end{subfigure}
\caption{Plots of the ratio of brane charge density for the KPV and KP metastable state, as a function of $p/M$ from 0 to $p_{crit}$ where $p_{crit}$ is the point after which metastability is lost. The left plot is the ratio $\mathcal{Q}_3/\mathcal{Q}_5$ for the KPV state. The middle plot zooms into the regime near $p_{crit}$ of the left plot. The right plot is the ratio $\mathcal{Q}_2/\mathcal{Q}_5$ for the KP state, a metastable state of polarised anti-M2 branes in CGLP background.}
\label{stafig1}
\end{centering}
\end{figure}

In figure \ref{stafig1}, we plot the ratio of brane charge density for the KPV and KP metastable state. For the KPV state, this is the ratio $\mathcal{Q}_3 / \mathcal{Q}_5$. As the NS5 brane charge density $\mathcal{Q}_5 = C r_h^2 \cos \theta$ is the same as the conserved NS5 Page charge $\mathbb{Q}_5$ (\ref{sta1}) which we take to be positive, the sign flip in $\mathcal{Q}_3 / \mathcal{Q}_5$ indicates a sign flip of the D3 brane charge. This flip happens near $p_{crit}$ as one can see from the middle plot of figure \ref{stafig1}. For the KP metastable state of anti-M2 branes, the same flip happens as demonstrated in the right plot. Let us note also that the blowing up of the ratio $\mathcal{Q}_3 / \mathcal{Q}_5$ near $p/M = 0$ is nothing to be afraid of. Near $p/M = 0$, the metastable equilibrium value of $\psi$ is very small, and so the size of the spherical NS5 is also very small. As the conserved Page charge $\mathbb{Q}_3$ \eqref{71} is not a charge density but a total charge over the NS5 sphere, the blowing up of the charge density $\mathcal{Q}_3$ is simply to compensate for the small size. Of course, the same can be said for the $\mathcal{Q}_2/\mathcal{Q}_5$ plot of the KP state.  

\subsection{Stability analysis}
Immediately from the blackfold perturbation equations above, we see that the $\delta r$ variation decouples from other variations and is controlled only by equation (\ref{001}). This allows us to study separately stability of the non-radial perturbations and stability of the radial perturbations. For our convenience, before continuing, let us expand all our perturbations into momentum and spherical harmonic modes. We have
\begin{align}
\delta v^t &=  \int d \lambda \, e^{-i \lambda t} \sum^{\infty}_{l = 0} \sum^l_{m = -l} (S_{v^t})_l^m (\lambda) Y^m_l (\omega, \varphi) ~ ,\\
\delta w^t &=  \int d \lambda \, e^{-i \lambda t} \sum^{\infty}_{l = 0} \sum^l_{m = -l} (S_{w^t})_l^m (\lambda) Y^m_l (\omega, \varphi) ~ ,\\
\delta \tan \theta &=  \int d \lambda \, e^{-i \lambda t} \sum^{\infty}_{l = 0} \sum^l_{m = -l} (S_{\tan \theta})_l^m (\lambda) Y^m_l (\omega, \varphi) ~ ,\\
\delta \psi &=  \int d \lambda \, e^{-i \lambda t} \sum^{\infty}_{l = 0} \sum^l_{m = -l} (S_\psi)_l^m (\lambda) Y^m_l (\omega, \varphi) ~ ,\\
\delta r &=  \int d \lambda \, e^{-i \lambda t} \sum^{\infty}_{l = 0} \sum^l_{m = -l} (S_r)_l^m (\lambda) Y^m_l (\omega, \varphi)
\end{align}
where $Y^m_l(\omega, \varphi)$ are the standard spherical harmonics. Note that we do not write down the expansion for $\delta v^\omega$, $\delta w^\varphi$, and $\delta r_h$ because they can be expressed in term of other perturbations as shown in (\ref{5}), (\ref{6}), and (\ref{002}). 

\paragraph{Stability of non-radial perturbations}
Assuming $\lambda \neq 0$, expanding our perturbations in momentum and spherical harmonic modes, the $\omega$ intrinsic perturbation equation (\ref{9}) yields
\be
\sum^{\infty}_{l = 0} \sum^l_{m = -l} (S_{v^t})_l^m Y^m_l = - \frac{i \cot \theta \cos^2 \theta}{ \lambda \sqrt{M} b_0 \sin \psi}  \sum^{\infty}_{l = 0} \sum^l_{m = -l} (S_{\tan \theta})_l^m  \p_\omega  Y^m_l 
\ee
where $\lambda, \omega, \varphi$ dependence of $S^m_l (\lambda)$ and $Y^m_l (\omega, \varphi)$ have been subdued for syntactical simplicity. Similarly, from the $\varphi$ intrinsic perturbation (\ref{10}), we have
\be
\sum^{\infty}_{l = 0} \sum^l_{m = -l} (S_{w^t})_l^m Y^m_l = - \frac{i \cot \theta \cos^2 \theta}{ \lambda \sqrt{M} b_0 \sin \psi \sin \omega}   \sum^{\infty}_{l = 0} \sum^l_{m = -l} (S_{\tan \theta})_l^m  \p_\varphi  Y^m_l ~ .
\ee
Let us note that satisfying the $\omega$ and $\varphi$ intrinsic perturbation equation automatically guarantee the satisfaction of the $J_2$ conservation equations (\ref{003})-(\ref{004}). Turning our attention to the $t$ intrinsic perturbation equation (\ref{005}), making use of the expressions above along with the identity $\nabla^2 Y^m_l = - l ( l + 1) Y^m_l$, we can show that 
\be
(S_{\tan \theta})^m_l = - \frac{2  \lambda^2  \sin^2 \psi \left( \cot \psi \tan \theta + 1/ b_0^2 \right) }{\lambda^2 \sin^2 \psi  -  l ( l + 1) \cos^2 \theta  } (S_\psi)^m_l ~ .
\ee
Again, let us note that satisfying the $t$ intrinsic perturbation equation automatically guarantee the satisfaction of the $J_4$ conservation equation (\ref{007}) and the conservation of $\mathbb{Q}_3$ charge (\ref{100}). Plugging in the expression of $(S_{\tan \theta})^m_l$ in term of $(S_\psi)^m_l$ into the $\psi$ extrinsic perturbation equation (\ref{008}) yields a quadratic equation for $\lambda^2$
\be
\lambda^4 + b \lambda^2 + c = 0
\ee
where the constants $b$ and $c$ are given respectively by
\begin{align}
b &= - \frac{4}{b_0^2} \cos^2 \theta (\sin \theta +1) \left( \cot \psi \tan \theta + \frac{1}{b_0^2} \right)-2 \left(l^2+l-1\right)   \frac{\cos^2\theta}{\sin^2 \psi} ~ , \\
c &= (l-1) l (l+1) (l+2) \frac{\cos^4\theta}{\sin^4 \psi} ~ .
\end{align}
Then, it trivially follows that 
\be
\lambda^2 = \frac{- b \pm \sqrt{b^2 - 4 c}}{2} ~ .
\ee
It is important to remember that, as declared in the ``Perturbation parameters'' paragraph \ref{PerPara}, $\psi$ and $\theta$ denote the values of the variables evaluated at the KPV configuration. This means, for any KPV configuration, we can write down explicitly the values of $b$ and $c$, thus, the value of $\lambda^2$.

\begin{figure}\centering
\includegraphics[width= 0.8 \textwidth]{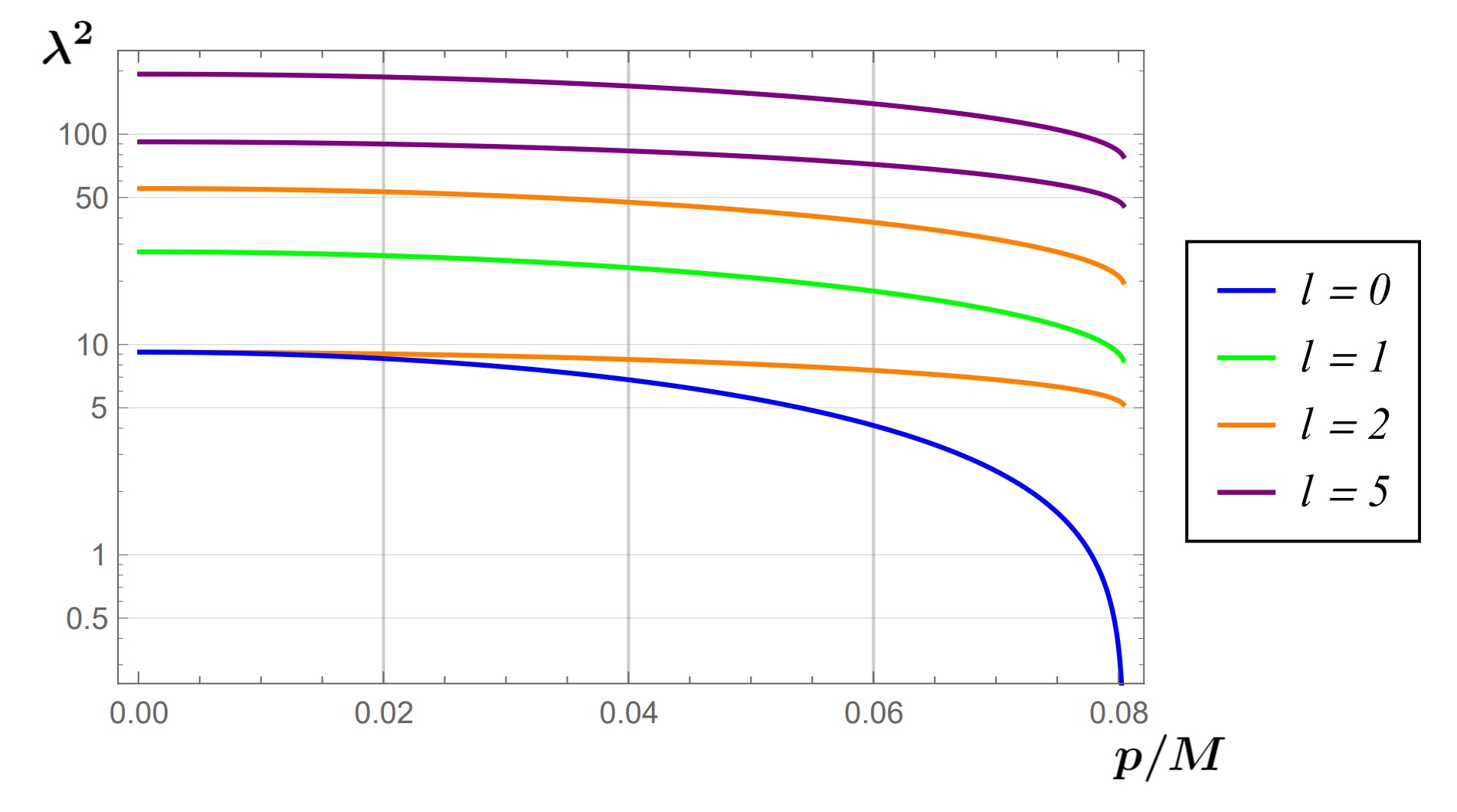}
\caption{\label{Fig1} Plot of $\lambda^2$ of non-radial perturbations against $p/M$. }
\end{figure}

It can be shown that $\lambda^2$ is positive for all KPV configurations. The case when $l = 0$ corresponds to having spherically homogeneous deformations around the KPV configuration and, as one would expect, it recreates the picture previously found. Including non-spherically homogeneous deformations does not change the statement regarding (meta)stability. In Figure \ref{Fig1}, we present the values of $\lambda^2$ for KPV configurations with $p/M \in (0, p_{crit})$ for $l$ equals $0$, $1$, $2$, and $5$.

Before continuing, let us ask the question: what happens if $\lambda = 0$? If $\lambda = 0$, the conservation of $\mathbb{Q}_3$ charge (\ref{100}) and the $\psi$ extrinsic perturbation equation (\ref{008}) both provide constraints on the $Y^0_0$ spherical harmonics mode of $\delta \tan \theta$ and $\delta \psi$. These conditions can only be simultaneously satisfied when
\be
\label{009}
\frac{1}{ \sin^2 \psi}   - \frac{2}{b_0^2}   \left( 1 + \sin \theta \right) \left( \tan \theta \cot \psi + \frac{1}{b_0^2} \right) = 0 ~ .
\ee
Recall that the KPV states exist when the parameter $p/M$ is in the range $p/M \in (0,p_{crit})$  where $p_{crit} \approx 0.080488$. As one can easily checked, equation (\ref{009}) cannot be satisfied with any KPV states strictly in the regime $p/M \in (0,p_{crit})$. It is only satisfied when $p/M = p_{crit}$ as one would expect.

\paragraph{Stability of radial perturbations}
Turning our attention to radial perturbations, it is instructive to consider the following intuition. Ignoring the gravitational effects of the NS5 branes (terms with $\cos \theta$), the radial force term is given by
\be
\frac{8 a_2 }{ a_0}  \sin \theta \delta r + \frac{8 a_2 }{ a_0}  \delta r ~ .
\ee
It is easy to see that the electromagnetic repulsion from the D3 brane charge is never stronger than its gravitational pull. With the expectation that the ignored NS5 gravitational effects are attractive, we expect the sign flip of the D3 brane charge to not pose a threat to the radial stability. Even though not explicitly discussed, following a similar argument, one can demonstrate the same phenomenon in the KP state.

Armed with this intuition, let us expand $\delta r$ in equation (\ref{001}) into momentum and spherical harmonic modes:
\begin{multline}
\label{avv}
- \lambda^2 \sum^{\infty}_{l = 0} \sum^l_{m = -l} (S_r)_l^m  Y^m_l  + \frac{\cos^2 \theta}{\sin^2 \psi}   \sum^{\infty}_{l = 0} \sum^l_{m = -l} (S_r)_l^m  l (l + 1)Y^m_l    \\
= \Bigg( \frac{8 a_2}{a_0} \sin \theta  +  \frac{8 a_2 }{ a_0} -  \frac{16 a_0 + 20 a_2}{5  a_0} \cos^2 \theta + \frac{8}{15} \cos^2 \theta \Bigg) \sum^{\infty}_{l = 0} \sum^l_{m = -l} (S_r)_l^m  Y^m_l \\
- \frac{ 16}{15} \sqrt{\frac{\pi}{5}} \cos^2 \theta     \sum^{\infty}_{l = 0} \sum^l_{m = -l} (S_r)_l^m  Y^0_2 \,  Y^m_l 
\end{multline}
where we have used 
\be
\sin^2 \omega = \frac{2}{3} - \frac{4}{3} \sqrt{\frac{\pi}{5}} Y^0_2  ~ .
\ee
Considering spherical harmonic modes $Y^m_l$, we note that even though equation (\ref{avv}) doesn't mix $m$ modes, because of the $Y^0_2 \, Y^m_l$ contraction in the last term, $l$ modes are coupled and have to be studied together. Recall that the contraction of spherical harmonics with the $Y^0_2$ mode can be expressed as a sum of harmonics
\be
Y^0_2 \, Y^m_l = \sqrt{\frac{5 (2l +1)}{4 \pi}} \sum_{l_3} (-1)^{m} \sqrt{2 l_3 + 1} \left( \begin{matrix}
2  &  \ l & l_3 \\
0 & \ m & - m
\end{matrix}\right) \left( \begin{matrix}
2  & \ l & \ l_3 \\
0 & \ 0 & \ 0
\end{matrix}\right) Y_{l_3}^{m} 
\ee
where $\left( \begin{matrix}
2  & \ l & l_3 \\
0 & \ m & - m
\end{matrix}\right)$
and
 $\left( \begin{matrix}
2  & \ l & \ l_3 \\
0 & \ 0 & \ 0
\end{matrix}\right)$
are the Wigner 3j-symbols, which vanish unless $|l - 2| \leq l_3 \leq l + 2$. By writing down the condition for each individual $l$ mode, equation (\ref{avv}) can be expressed as a set of linear equations of $(S_r)^m_l$.  

As $m$ modes decoupled, let us discuss in details the spherical harmonic modes with $m = 0$. The associated matrix of the linear system of $(S_r)^0_l$ is given by
\be
\begin{pmatrix} \mathbb{A} \ \vline \ 0
\end{pmatrix}
=
\begin{pmatrix}
\lambda^2 + d & 0 &  \frac{ - 8 \cos^2 \theta}{15 \sqrt{5}} & \dots &\vline & \ 0\\ 
0 & \lambda^2 + d - \frac{2 \cos^2 \theta}{\sin^2 \psi} - \frac{16}{75} \cos^2 \theta & 0 & \dots &\vline & \ 0\\ 
\frac{ - 8 \cos^2 \theta}{15 \sqrt{5}} & 0 & \lambda^2 + d - \frac{6 \cos^2 \theta}{\sin^2 \psi} - \frac{16}{105} \cos^2 \theta & \dots &\vline & \ 0\\

\vdots & \vdots  & \vdots  & \ddots &\vline & \ \vdots
\end{pmatrix} 
\ee
where, for convenience, we have defined a constant $d$ as
\be
d = \frac{8 a_2}{a_0} \sin \theta  +  \frac{8 a_2 }{ a_0} -  \frac{16 a_0 + 20 a_2}{5  a_0} \cos^2 \theta + \frac{8}{15} \cos^2 \theta  ~ .
\ee
The system of linear equations is only satisfied when the determinant of the associated matrix vanishes, i.e. $\det \mathbb{A} = 0$. Even though
$\mathbb{A}$ is not diagonal, as the contribution of the off-diagonal terms to the determinant of $\mathbb{A}$ is numerically much smaller than that of the diagonals, the determinant of $\mathbb{A}$ can be well-approximated by the product of the diagonal terms. With this approximation, it is trivial that $\lambda^2$ is always positive. Let us mention also that cases of $m \neq 0$ can be treated the same way and yield a similar conclusion. 

\begin{figure}\centering
\includegraphics[width= 0.8 \textwidth]{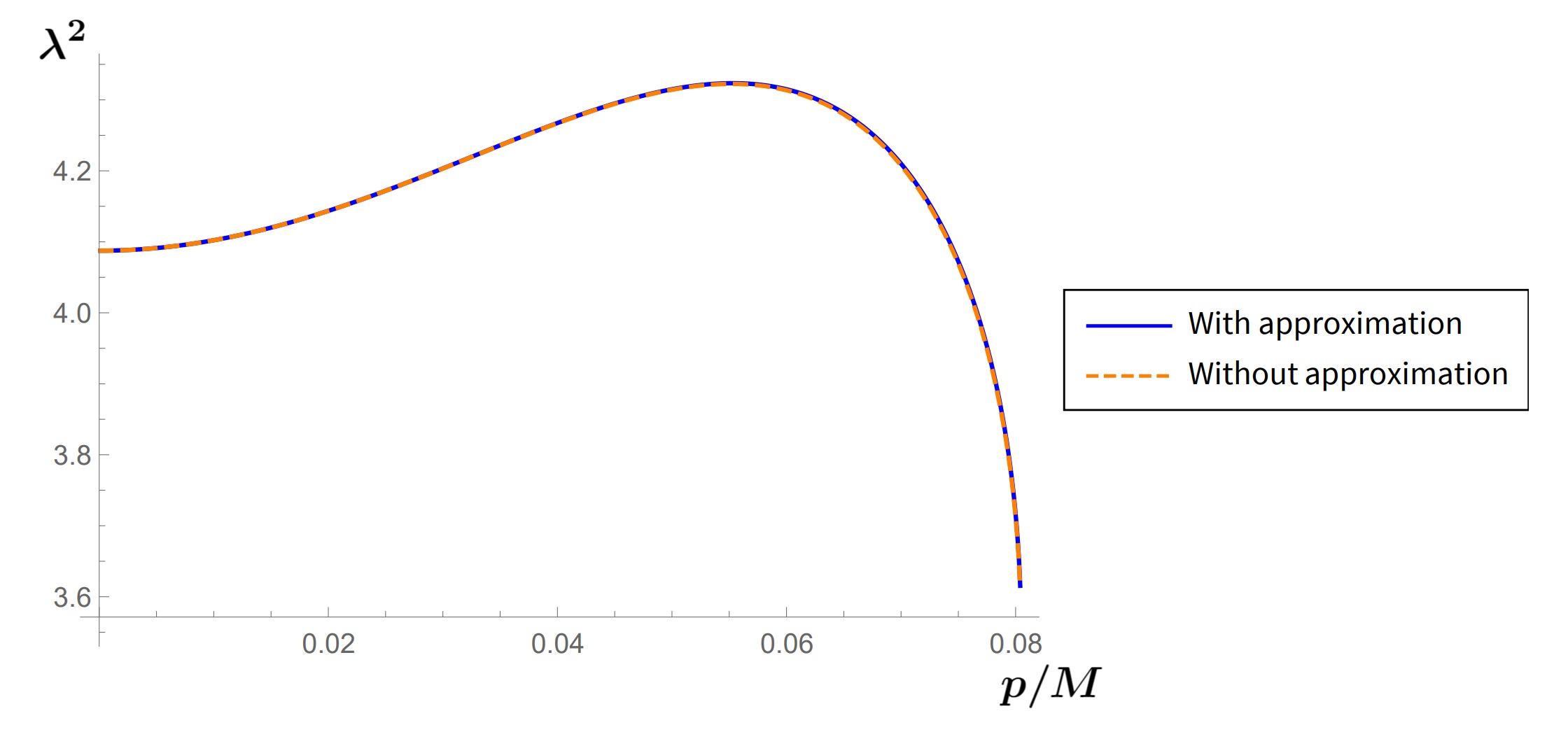}
\caption{\label{Fig2} Plot of $\lambda^2$ of radial perturbations against $p/M$. }
\end{figure}

In Figure \ref{Fig2}, we plotted the smallest $\lambda^2$ root computed both with the diagonal approximation\footnote{Practically, this is a plot of $\lambda^2 = -d$ as $-d$ is obviously the smallest root in such case.} and without the diagonal approximation, truncating $\mathbb{A}$ to be of order $21 \times 21$. From the plot, it can easily be seen that the off-diagonal corrections are indeed very minimal and don't affect the underlying physics of the system. Lastly, let us note that the dip in $\lambda^2$ near $p_{crit}$ is because of the effect mentioned in the discussion below equation (\ref{asdf}) where the $\mathcal{Q}_3$ charge flips sign and the electromagnetic force becomes repulsive. Nevertheless, as demonstrated here, this electromagnetic repulsion is outweighed by gravitational attraction. Even though not explicitly stated, from the blackfold treatment of the KP state in chapter \ref{THE}, one can easily show a similar picture for the polarised anti-M2 state.

\chapter{On the thermal transition of metastable antibranes}
\label{THE}
The purpose of this chapter is to study the effects of non-zero temperature on metastable states of antibranes in warped throat. Our exemplar candidate is the Klebanov-Pufu (KP) \cite{Klebanov:2010qs} metastable state of polarised anti-M2 branes, or equivalently M5 branes with dissolved anti-M2 brane charge (wrapped anti-M2-M5 branes), at the tip of the Cvetic-Gibbons-Lu-Pope (CGLP) throat \cite{ Cvetic:2000db}. As presented later on, we discovered that the KP state exhibit an exotic, previously inaccessible pattern of thermal transition different from that of the KPV.

We begin, in section \ref{THE1}, with a derivation of the blackfold equations for anti-M2-M5 branes at the tip of the CGLP throat, recovering KP DBI analysis at extremality. Subsequently, in section \ref{THE2}, we construct nonextremal effective potentials for these antibranes and, in section \ref{THE3}, study the effects of nonextremality on their metastable state. As the derivation of the anti-M2 blackfold equations is similar to that of the anti-D3 discussed in chapter \ref{EX}, such discussion is brief. On the other hand, the discussion on thermal potentials and metastability-losing mechanism is more comprehensive.  

\section{Blackfold equations for anti-M2-M5 branes in CGLP throat}
\label{THE1}

\paragraph{Anti-M2-M5 in CGLP throat}
The description of our background geometry, i.e. the CGLP throat, is provided in appendix \ref{CON2} while the description of the seed solution, i.e. the M2-M5 bound state, is provide in appendix \ref{C2}. In a blackfold set-up of anti-M2-M5 branes at the tip of the CGLP throat, the variables of the system are\footnote{For the variables of the anti-D3 blackfold in KS throat \eqref{ex50}, we allow also the parameters of the transverse $S^2$ away from the tip. However, for the anti-M2 blackfold, we shall not include such parameters because for our purposes, we don't care about dynamics off the tip.}
\be
r, \, \psi,\, r_0,\, \alpha,\, \tan \theta,\, v^a,\, w^a, \, k^a , \, u^a ~ .
\ee
The variables $r, \, \psi$ are the embedding degrees of freedom of the branes to the background which, from the worldvolume perspective, are also referred to as transverse scalars. The variables $r_0,\, \alpha,\, \tan \theta,\, v^a,\, w^a, \, k^a , \, u^a$ are the characteristic degrees of freedom of the seed solution, describing respectively the Schwarzschild horizon radius, the boost rapidity, the brane charge distribution, the orientation of the anti-M2 dissolved charge (with three orthogonal vectors $v$, $w$, and $k$), and the thermal flow (with one directional vector $u$).

In order to construct a metastable state, we place the anti-M2-M5 branes at the tip of the CGLP throat in such way that 3 of the 6 dimensions of the M-brane bound state lie along the Minkowskian directions $t, x^1, x^2$, and the other 3 wrap around the 3-cycle $\vartheta, \omega, \varphi$. For simplicity, let us further restrict our attention to $t$ dependent configurations with dissolved anti-M2 brane charge lying along the Minkowski directions and thermal flow along the $t$ direction. This means we have set 
\be
r = 0 \, , 
\ee
restricted our scalar variables to
\be
\psi(t), r_0(t), \alpha(t), \tan \theta(t) \, , 
\ee
and specified our vectors as
\begin{align}
\label{the10}
u^a \p_a &= \frac{1}{m^{1/3} b_0 \sqrt{1 - \psi'^2}} \p_t ~ ,\\
v^a \partial_a  &= \frac{1}{m^{1/3} b_0 \sin \psi } \partial_\vartheta ~ , \\
w^a \partial_a &= \frac{1}{m^{1/3} b_0\sin \psi \sin \vartheta} \partial_\omega ~ ,\\
\label{the11}
k^a \partial_a  &= \frac{1}{m^{1/3} b_0 \sin \psi \sin \vartheta \sin \omega} \partial_\varphi 
\end{align}
where $\psi' \equiv  \p_t \, \psi$. The factors in \eqref{the10}-\eqref{the11} are to make sure that $v$, $w$, $k$, $u$ satisfy the unitary condition, e.g. $v^a v_a = 1$. 

\paragraph{The blackfold equations} Applying the eleven-dimensional blackfold equations \eqref{bf50}-\eqref{bf51} to our configuration of anti-M2-M5 branes in CGLP background, we obtain the blackfold equations for our system. Analogous to the anti-D3-NS5 case, these blackfold equations can be written in term of energy-momentum and current conservation equations. While the energy-momentum conservation equations can be simplified to\footnote{At this point, there should also be an equation coming from the $t$ component of the intrinsic blackfold equation. However, such equation becomes trivial for the recovery of the KP state at extremality and the construction of nonextremal effective potentials so we shall not write it explicitly here.}
\begin{multline}
\label{the1}
\frac{\psi''}{1 - \psi'^2}\left( \frac{4}{9} + \frac{1}{3} \sinh^2 \alpha \right) + \cot \psi \left( \frac{1}{3} + \cos^2 \theta \sinh^2 \alpha \right) \\
=  - \frac{9}{4 b_0^3} \sin \theta \cos \theta \sinh^2 \alpha +  \sqrt{1 - \psi'^2} \frac{9}{4 b_0^3} \cos \theta \sinh \alpha \cosh \alpha ~ ,
\end{multline}
the current conservation equations give rise to the conserved Page charges
\be
\label{theqsaa}
\mathbb{Q}_5 = C r_0^3 \cos \theta \sinh \alpha \cosh \alpha ~ ,
\ee
\begin{multline}
\mathbb{Q}_2 = \frac{27 \pi^2 m}{2} C r_0^3  \sinh \alpha \cosh \alpha \cos \theta  \left(\frac{1}{3} \cos^3 \psi - \cos \psi + \frac{2}{3} \right) \\
- 2 \pi^2 C m b_0^3 \, r_0^3  \sinh \alpha \cosh \alpha \sin \theta \sin^3 \psi ~ .
\end{multline}
From these, we immediately have
\be
\label{the2}
\tan \theta  =  \frac{1}{b_0^3  \sin^3 \psi} \left( - \frac{9 p}{\tilde{M}} + \frac{27 }{4} \left(\frac{1}{3} \cos^3 \psi - \cos \psi + \frac{2}{3} \right)  \right)
\ee
where we have made the identification
\be
\frac{p}{\tilde{M}} \equiv \frac{\mathbb{Q}_2}{18 \pi^2 m \mathbb{Q}_5}
\ee
with $\tilde{M}$ defined in \eqref{setupab}.

\paragraph{Regime of validity} The validity of the blackfold analysis requires a large separation of scales $r_b\ll \mathcal R, L$ where $r_b$ is the characteristic near horizon scale of the seed branes, $\mathcal{R}$ is the scale of the curvature radius of the bending in the branes, and $L$ is characteristic length scale of the background. Following an analogous argument to that of section \ref{EX2}, we obtain the requirements
\begin{align}
\label{the4}
&r_0 \sinh \alpha \ll b_0 m^{1/3} \sin \psi, & &(N_5)^{1/3} \ll  \tilde{M}^{1/3} \sin \psi,   & &\left( \frac{p}{\tilde{M}} \right)^{1/3} \ll \tilde{M}^{1/3} \sin^2 \psi ~ .
\end{align}
It is clear that our requirements fail at the North and South pole, $\sin \psi=0$. For sufficiently large $\tilde{M}$ (or equivalently $m$), however, our calculations are valid everywhere except for a small region around the poles.

\paragraph{Recovery of the KP state at extremality}
At extremality, the set of blackfold equations  can be written as
\be
 \cot \psi =  - \frac{9}{4 b_0^3} \tan \theta + \frac{9}{4 b_0^3} \sqrt{1 - \psi'^2}  \sqrt{1 + \tan^2 \theta} - \frac{\psi''}{3 (1 - \psi'^2)} \left(1 + \tan^2 \theta \right)
\ee
where $\tan \theta$ is given by
\be
\tan \theta  =  \frac{1}{b_0^3  \sin^3 \psi} \left( - \frac{9 p}{\tilde{M}} + \frac{27 }{4} \left(\frac{1}{3} \cos^3 \psi - \cos \psi + \frac{2}{3} \right)  \right) ~ .
\ee

On the other hand, from the DBI action in \cite{Klebanov:2010qs}, we have the equation of motion
\be
\cot \psi =- \frac{9}{4 b_0^3} \mathcal{P} + \frac{9}{4 b_0^3} \sqrt{1 - \psi'^2} \sqrt{1 + \mathcal{P}^2}  - \frac{\psi''}{3 (1 - \psi'^2)} \left( 1 + \mathcal{P}^2 \right) 
\ee
where $\mathcal{P}$ is given by
\be
\mathcal{P} =  \frac{1}{ b_0^{3}  \sin^3 \psi }\left( - \frac{9 p}{\tilde{M}} + \frac{27}{4} \left(\frac{1}{3} \cos^3 \psi - \cos \psi + \frac{2}{3} \right) \right) ~ .
\ee
Thus, we have shown that the blackfold equations \eqref{the1}-\eqref{the2} at extremality recover in the supergravity regime the results obtained from DBI analysis in \cite{Klebanov:2010qs}.

\section{Heated antibranes}
\label{THE2}

From the point of view of holography, temperature can be incorporated in the system at hand in different ways. Let us list here three rough possibilities:
\begin{itemize}
\item One option is to begin by adding temperature to the supersymmetric vacuum of the dual QFT. In the bulk, this involves a CGLP black hole with positive M2-brane charge.\footnote{For a construction of smeared black M2-brane solutions that preserve an $SO(5)$ symmetry see \cite{Dias:2017opt}.} Then, one can analyse the existence and properties of a metastable state in this thermal environment. In the bulk, an analysis based on the probe approximation would entail at leading order the use of a DBI-type action for a wrapped M5 brane in the background of the CGLP black hole. 

\item A second option that focuses more directly on the thermal effects on the metastable state itself goes along the following lines. In the bulk description, we can either consider solutions in the probe approximation using a thermalised DBI-like effective action in CGLP, or in the supergravity regime we can attempt to construct a wrapped M5 black hole with negative M2 charge that asymptotes to the supersymmetric background. In this section, we will focus on the second approach using blackfold techniques. The fundamental difference between this bullet point and the previous one is that as one turns off the antibrane charge, in the first case one recovers a thermal state of the dual QFT, whereas in the second one recovers a supersymmetric ground state of the dual QFT.

\item A third more general option is to thermalise all the sectors of the system at the same time.%
\footnote{This approach was taken for example in \cite{Grignani:2010xm} which considers the thermalized version of the
BIon solution by analyzing a D3-F1 blackfold in hot flat space  and in \cite{Grignani:2012iw}
which considers  an F1 blackfold in the AdS black hole background to study finite temperature Wilson loops.}
This would entail in the bulk the construction of a black hole solution that describes the backreaction of a thermally excited wrapped M5 brane in the background of the CGLP black hole. One could also try to capture aspects of this case with blackfold techniques but we will not explore this possibility here. 
\end{itemize}

Adding temperature to the effective actions of weakly coupled open strings is a notoriously difficult problem that involves open string loop computations (we refer readers to \cite{Grignani:2013ewa} for relevant discussions). In that sense, implementing the option of the second bullet point with a DBI-like probe analysis is not a straightforward exercise. In the supergravity regime, however, the blackfold equations allow the incorporation of thermal effects rather easily. In this section, from the anti-M2-M5 blackfold equations \eqref{the1}-\eqref{the2}, we construct thermal effective potentials for the anti-M2-M5 branes.

In what follows, we focus on static configurations where the anti-M2-M5 blackfold equations simplify to
\be
\label{the3}
\cot \psi \left( \frac{1}{3} + \cos^2 \theta \sinh^2 \alpha \right) 
=  \frac{9}{4 b_0^3} \cos\theta \sinh\alpha \left( - \sin \theta \sinh \alpha +   \cosh \alpha \right)
\ee
with 
\be
\tan \theta  =  \frac{1}{b_0^3  \sin^3 \psi} \left( - \frac{9 p}{\tilde{M}} + \frac{27 }{4} \left(\frac{1}{3} \cos^3 \psi - \cos \psi + \frac{2}{3} \right)  \right) ~.
\ee

The solutions of equation \eqref{the3} at fixed $p/\tilde M$ are parametrised by a free constant. This could be a nonextremality parameter like $r_0$ or $\alpha$, or a more physically motivated thermodynamic parameter like the total entropy $S$ or the global temperature $T$.

The solutions of \eqref{the3} and their properties will be discussed in detail in section \ref{THE3}. In the rest of this section, we explain how to obtain these solutions as extrema of suitable effective potentials. A different potential is formulated for each parameter that we choose to keep fixed. We will discuss three kinds of potentials: $V_T$ where the global temperature $T$ is kept fixed, $V_{S}$ where the total entropy $S$ is kept fixed, and $V_{\alpha}$ where the parameter $\alpha$ is kept fixed. 

\subsection{Comments on effective thermodynamic potentials}
Before we present specific effective potentials for the anti-M2-M5 configurations of interest, it is useful to first comment on a slightly more general problem. The general blackfold equations describe an effective fluid on a dynamical hypersurface. Let us assume that we are interested in stationary solutions of these equations. For such solutions, there is a worldvolume Killing vector field ${\boldsymbol k}^a$, which is assumed to be the pullback of a background Killing vector field ${\boldsymbol k}^\mu$. It has been argued in \cite{Emparan:2009at,Emparan:2011hg,Armas:2018ibg} that by using standard thermodynamic quantities, it is possible to formulate effective actions of the transverse scalars whose extrema reproduce the profiles of the stationary solutions. In these actions, the intrinsic degrees of freedom of the effective fluid are integrated out and the variational problem is restricted to stationary configurations. These actions are guaranteed to produce correct stationary solutions if they recover the currents of the fluid under general variations of the background fields.

For concreteness, let us focus on the case of interest: anti-M2-M5 blackfolds in the CGLP background. We can obtain an effective action by varying over stationary configurations at a fixed global temperature $T$ in the following manner. By definition, the global temperature is related to the local temperature $\mathcal{T}$ \eqref{bs20} of the effective fluid as $T= |{\boldsymbol k}| \mathcal{T}$. For the Killing vector ${\boldsymbol k}^a$ we have ${\boldsymbol k}^a \p_a = \partial_t$, $|{\boldsymbol k}|=m^{1/3}b_0$, and $u^a = {\boldsymbol k}^a/|{\boldsymbol k}| = \p_t / (m^{1/3} b_0)$. Variations of the background lead to variations of the induced metric $\delta \gamma_{ab}$, the pulled-back three-form gauge potential $\delta A_{3abc}$ and its dual $\delta A_{6abcdef}$, defined such that $G_7=dA_6+A_3\wedge G_4/2$. These variations are performed keeping $\mathbb{Q}_2$, $\mathbb{Q}_5$ and $T$ fixed. Let us denote $\gamma^\perp_{ab} \equiv v_a v_b + w_a w_b + z_a z_b$ the projector onto worldvolume directions perpendicular to the dissolved M2 directions. Keeping $\mathbb{Q}_2$, $\mathbb{Q}_5$ and $T$ fixed under variations imply the variational properties
\be
\begin{split}
\label{thermoaa}
&\delta \tan\theta = (v \wedge w \wedge z)^{abc} \delta A_{3abc} - \frac{1}{2} \tan\theta \, \gamma^{\perp ab} \delta \gamma_{ab}~~, \\
&\delta r_0 = -\frac{1}{2} r_0 u^a u^b \delta \gamma_{ab} - r_0 \sinh\alpha \cosh\alpha \, \delta(\tanh\alpha)~~, \\
&\delta (\tanh\alpha) = \frac{3}{2} \frac{\tanh \alpha}{1-\sinh^2\alpha} u^a u^b \delta \gamma_{ab}
+ \frac{\tanh \alpha}{1-\sinh^2\alpha} \sin\theta \cos\theta\, \delta (\tan \theta)
\end{split}
\ee
respectively. Under such variations one can easily show that the thermodynamic effective action 
\be
\label{thermoad}
\begin{split}
{\mathcal S}_T =& - \int_{\mathcal{M}_6} d^6 \sigma \sqrt{-\gamma} \, \mathcal{F} + \mathbb{Q}_5 \int_{\mathcal{M}_6} \mathbb{P}[A_6]+\mathbb Q_2\int_{\mathcal{M}_3} \mathbb{P}^{||}[A_3]~~,\\
=& - \int_{\mathcal{M}_6} d^6 \sigma \sqrt{-\gamma} \, \mathcal{F} + \mathbb{Q}_5 \int_{\mathcal{M}_6}\left(\mathbb{P}[A_6]+\frac{1}{2}\mathbb{P}[A_3\wedge A_3]+\frac{\Phi_2}{\Phi_5}dV_\perp\wedge\mathbb{P}^{||}[A_3]\right)
\end{split}
\ee
reproduces the correct currents
\be
\label{thermoae}
\delta {\mathcal S}_T = \int_{\mathcal{M}_6} d^6 \sigma \sqrt{-\gamma} \left( \frac{1}{2} T^{ab}\delta \gamma_{ab} + J_3^{abc}\delta A_{3abc} + \mathcal{J}_6^{a_1\cdots a_6} \delta A_{6a_1\cdots a_6} \right)
~.
\ee
In \eqref{thermoad} $\mathcal{F} = \varepsilon- \mathcal{T} s$ is the free energy \eqref{forceaed}, $\mathbb{P}[A_6]$ is the pullback of the background six-form $A_6$, $\mathcal{M}_6$ is the six-dimensional worldvolume of the effective theory, $\mathbb{P}^{||}[A_3]$ is the pullback of the background $A_3$ onto $\mathcal{M}_3$ and $dV_\perp=\sqrt{\gamma_\perp}dv\wedge d w\wedge dz$ is the volume form on $\mathcal{M}_3^\perp$. 

Invariance of \eqref{thermoad} under gauge transformations $\delta A_3=d\Lambda_2$ and $\delta A_6=d\Lambda_5$ for gauge parameters $\Lambda_2,\Lambda_5$ leads to the conservation equations for $ J_3$ and $\mathcal{J}_6$ respectively. The last term in \eqref{thermoad} vanishes for the specific configurations that we are interested in, since $\mathbb{P}^{||}[A_3]=0$. However, in order to extract the correct currents via a variational principle, it is required.

In \eqref{thermoad}, it is implicitly assumed that we have implemented all constraints from the constant $\mathbb Q_2$, $\mathbb{Q}_5$ and $T$ together with a stationary ansatz for the vectors $u^a, v^a, w^a, z^a$ and that we have expressed $r_0$, $\alpha$, $\theta$ in terms of the transverse scalars. The resulting action is an action of the transverse scalars alone.\footnote{It is also possible to write an action for M2-M5 branes that does not make assumptions about the background or how the M2 branes are embedded into the M5. In this case, additional dynamical fields must be introduced as in \cite{Armas:2018atq, Armas:2018zbe}.} By varying it with respect to the transverse scalars, we are guaranteed to obtain equations that lead to the correct stationary solutions of the blackfold equations. Explicit formulae for wrapped M5 branes will appear in the next subsection. 

The effective action \eqref{thermoad} has a well defined extremal limit $T\to0$, reducing to the PST action \cite{Pasti:1996vs, Pasti:1997gx} (multiplied by the number of M5-branes  $N_5$) when all worldvolume gauge fields have been integrated out. However, the existence of a maximum temperature \eqref{potTad}, or in cases of bound states with a Hagedorn temperature such as that of the anti-D3-NS5 configuration in chapter \ref{EX} for which $T=0$ does not describe all extremal solutions, the potential at fixed $T$ is unsuitable for describing the entire phase space of off-shell configurations. Instead defining $\mathcal{B}_5$ as the spatial part of the worldvolume $\mathcal{M}_6$, a closely related effective action where we keep the total entropy 
\be
\label{thermoaf}
S =  \int_{\mathcal{B}_5} \sqrt{-\gamma} \, s \, u^t
\ee
fixed is more appropriate and can be obtained by Legendre transforming \eqref{thermoad} yielding
\be
\label{thermoag}
{\mathcal S}_S = -\int_{\mathcal{M}_6} d^6 \sigma \sqrt{-\gamma} \, \varepsilon + \mathbb{Q}_5 \int_{\mathcal{M}_6} \mathbb{P}[A_6]+\mathbb Q_2\int_{\mathcal{M}_3} \mathbb{P}^{||}[A_3]
~.
\ee

\subsection{Potential at fixed temperature}
\label{fixedT}
In this subsection, we present the precise form of the potential $V_T$ for anti-M2-M5 branes wrapping an $S^3$ at the tip of the CGLP throat. The black M5s of interest are characterised by the local temperature $\mathcal{T} = \frac{3}{4\pi r_0 \cosh\alpha}$. Eliminating $r_0$ with the use of equation\ \eqref{theqsaa} we can write
\be
\label{potTaa}
\mathcal{T}^3 = \frac{27 \mathcal{C}}{64\pi^3 \mathbb{Q}_5} \T^3~, ~~~
\T^3 \equiv \frac{\cos\theta \sinh\alpha}{\cosh^2\alpha}
~.
\ee
We will use $\T$ to express all the relevant formulae. Dividing ${\mathcal S}_T$ by the infinite volume of the $\IR^{3,1}$ part of the M5 worldvolume and an overall constant factor of $36 \pi^2 m^2 b_0^3 \mathcal{Q}_5$ we obtain the potential
\be
\label{potTab}
V_T(\psi) = \frac{b_0^3 \sin^3\psi (1 + 3 \sinh^2 \alpha )}{54 \cos\theta \sinh\alpha \cosh\alpha } - \frac{3}{8} f(\psi)
\ee
with 
\be
f(\psi) = \frac{1}{3} \cos^3 \psi - \cos \psi + \frac{2}{3} ~ .
\ee
In this formula, $\alpha$ and $\theta$ are implicit functions of $\psi$ where $\theta(\psi)$ can be obtained explicitly from equation\ \eqref{the2} and $\alpha(\psi)$ from combining \eqref{potTaa} and \eqref{the2}. The potential $V_T$ depends parametrically on $p/\tilde M$ and $\T$. One can show by direct evaluation that the equation $\frac{dV_T}{d\psi}=0$ is equivalent to the blackfold equations \eqref{the1}-\eqref{the2}. In particular, when $\T=0$ we recover the extremal vacua of Klebanov and Pufu.

There are two intricacies of the fixed-$T$ potential that are worth highlighting. The first one is that equation\ \eqref{potTaa} has in general two solutions of $\alpha$ for a given angle $\psi$ at a fixed value of $\T$. 
The two solutions are
\be
\label{potTac}
\tanh\alpha_\pm (\psi) = \sqrt{\frac{1}{2} \pm \sqrt{\frac{1}{4} - \frac{\T^6}{\cos^2\theta(\psi)}}}
~.
\ee
The branch of $\alpha_+$ is valid for $\alpha \geq \alpha_*$ and the branch of $\alpha_-$ for $\alpha\leq \alpha_*$. The critical value $\alpha_*$ defines a point where $\alpha_+ = \alpha_-$, i.e.\ a point where $\cos^2\theta(\psi) = 4 \T^6$. Numerically, $\alpha_* \simeq 0.881374$. Notice that these solutions are real only when 
\be
\label{potTad}
|\cos\theta(\psi)| \geq 2 \T^3
~.
\ee
This inequality imposes a constraint on the domain of $\psi$ where the potential $V_T(\psi)$ in \eqref{potTab} can be defined sensibly.

\begin{figure}[t!]
\begin{center}
\includegraphics[width=10cm]{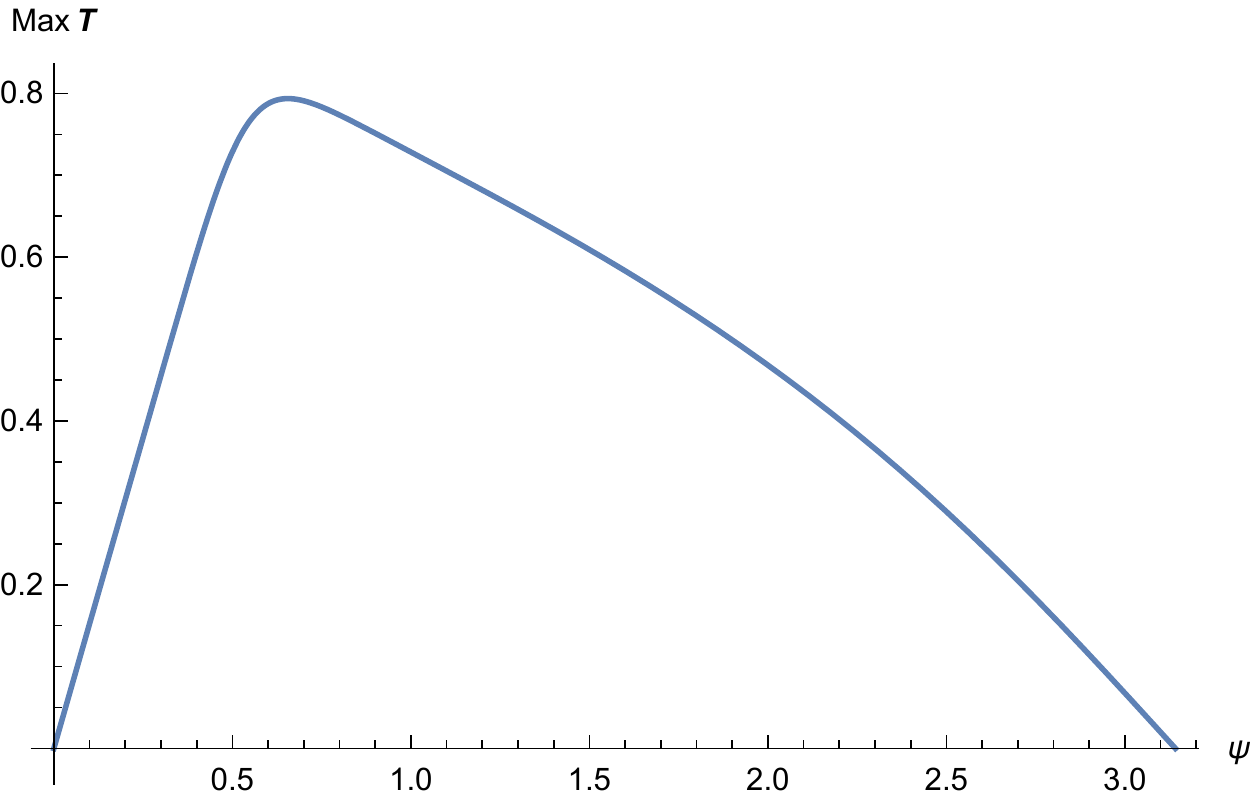}
\end{center}
\caption{A plot of the maximum possible value of the temperature $\T$ of the M2-M5 branes at each angle $\psi$ for $p/\tilde{M} = 0.03$.}
\label{Fig_maxtemp}
\end{figure}

The second related feature is that the temperature of the wrapped M5s at a given angle $\psi$ has a maximum possible value. This follows immediately from \eqref{potTaa} and the fact that the function $\sinh\alpha/\cosh^2\alpha$ has a maximum value of $1/2$. A plot of the maximum temperature at a given angle $\psi$ appears in Figure\ \ref{Fig_maxtemp}.

\subsection{Potential at fixed entropy}

The action \eqref{thermoag} allows us to formulate a potential whose extrema determine the equilibria of the wrapped M5 branes at a fixed total entropy. In the case at hand the total entropy \eqref{thermoaf} is given by the expression
\be
\label{potSaa}
S = \frac{8\pi^3 m^{\frac{5}{3}} b_0^5 \mathcal{Q}_5^{\frac{4}{3}}}{3\, \CC^{\frac{1}{3}}} \bS
~,~~
\bS^3 \equiv \frac{\sin^9\psi}{\cos^4 \theta \sinh^4\alpha \cosh\alpha}
~.
\ee
We express all relevant quantities using the properly normalised entropy $\bS$. Dividing ${\mathcal S}_S$ in \eqref{thermoag} by the infinite volume of the $\IR^{3,1}$ part of the M5 worldvolume and the overall factor $36 \pi^2 m^2 b_0^3 \mathcal{Q}_5$ we obtain the potential
\be
\label{potSab}
V_S (\psi) = \frac{b_0^3 \sin^3\psi (4 + 3 \sinh^2 \alpha  )}{54 \cos\theta \sinh\alpha \cosh\alpha } - \frac{3}{8} f(\psi) 
~.
\ee
with $\alpha$ and $\theta$ being implicit functions of $S$ and $\psi$. Again, one can verify by direct computation that the extrema of this potential reproduce the correct static solutions of the blackfold equations at fixed total entropy $S$. The potential $V_S$ depends parametrically on $p/\tilde M$ and the entropy $\bS$. At $\bS=0$ the potential \eqref{potSab} reduces to the potential that follows from the DBI action.

In this case the potential is well defined in the whole range of angles $\psi$. We will present numerical plots of the potential in different regimes of parameters in the next section. The same type of fixed-entropy potential was computed for the wrapped NS5 branes in the Klebanov-Strassler background of type IIB string theory in chapter \ref{EX}.

\subsection{Potential at fixed boost rapidity}

The above discussions demonstrate that one can consider effective potentials in different ensembles. All of them reproduce the same static configurations as the original blackfold equations but the off-shell shape of the potential in each case is different. It is natural to ask whether it is possible to define a potential that keeps some other quantity constant, possibly one that does not have a straightforward thermodynamic interpretation. When we solve the combination of equations\ \eqref{the1}-\eqref{the2}, technically the most convenient choice would be to solve them at a fixed value of $\alpha$. $\alpha=\infty$ would be the extremal case and $\alpha=0$ the exact opposite.

One can show by direct computation that the following fixed-$\alpha$ potential does the job:
\be
\label{potAaa}
V_\alpha (\psi) = \frac{1}{18} b_0^3 \sin^3 \psi \frac{1}{\cos \theta(\psi)} - \frac{3}{8} \coth \alpha\, f(\psi) + \frac{1}{\sinh^2 \alpha} H (\psi)
\ee
with
\be
\label{potAab}
H(\psi) = \int_{\psi_0}^\psi d \chi \left( \cot \chi \sqrt{ \frac{\hat{H}_0}{96} \sin^6 \chi + \left(\frac{3}{8} f(\chi) - \frac{p}{2 \tilde{M}} \right)^2 } \right)
~.
\ee
where the constant $\hat{H}_0$ is given in \eqref{setupac}. The constant $\psi_0$ in the lower limit of the integration in \eqref{potAab} is arbitrary. Its value determines an arbitrary additive constant to the potential. As before, we obtain $\theta(\psi)$ by solving the equation\ \eqref{the2}. As a trivial check, notice that $V_\alpha$ reduces to the effective potential of \cite{Klebanov:2010qs} when $\alpha \to \infty$ (in that case the last term in $V_\alpha$ vanishes).

By varying the values of $\alpha$, we obtain the full range of static wrapped M5-brane configurations that we would obtain directly from the blackfold equations. The same overall set of static configurations can be obtained by extremising either of the potentials $V_T$ or $V_S$ for different values of $T$ and $S$. In that sense, all the potentials that we described above are equivalent. 

The off-shell shape of each potential is different. One should exercise some caution when employing the full shape of the potential to make statements about, say, the stability of the different vacua. Since the entropy current is conserved in our leading order ideal hydrodynamic effective theories, time-dependent solutions will naturally evolve conserving the total entropy. This suggests that the off-shell shape of the fixed-$S$ potential contains correct information about the stability of the vacua we find (at least within the homogeneous ansatz of wrapped M5s that we used). In the next section, we plot the fixed-$\alpha$ potential and show that it shares the same qualitative features as the fixed-$S$ potential.

\section{Antibranes metastability at finite temperature}
\label{THE3}

We are now in position to determine in detail what happens to the KP vacua once we turn on the temperature. We will discuss the nonextremal physics from the perspective of all the potentials presented in the previous section.

\begin{figure}[t!]
\begin{center}
\includegraphics[width=12cm]{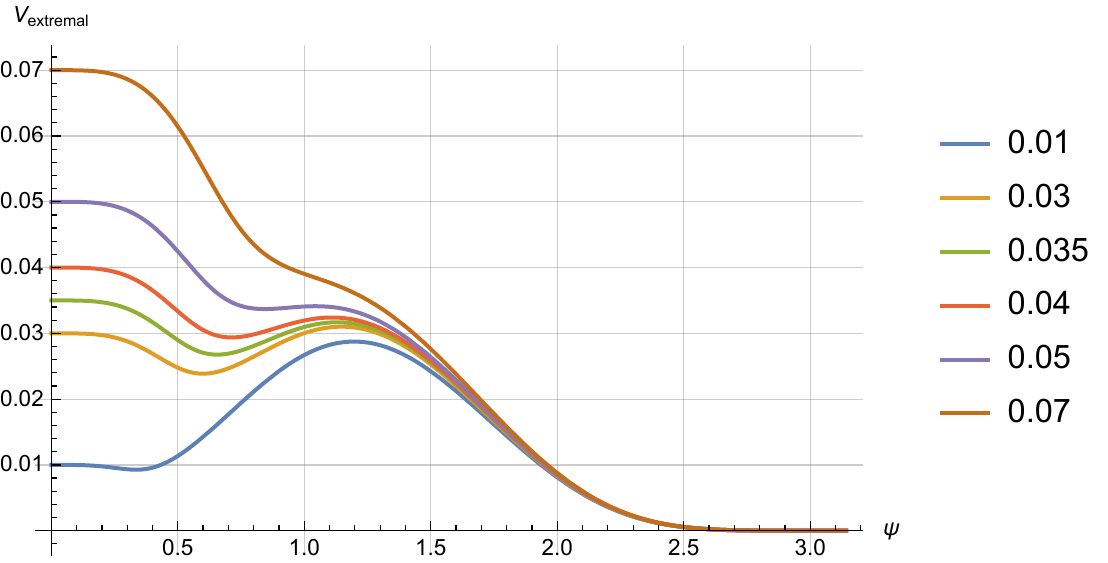}
\end{center}
\caption{A plot of the extremal potential $V_{\rm extremal}$ as a function of $\psi$ as discovered in \cite{Klebanov:2010qs}. Different colours depict the plot for different values of the $p/\tilde M$ (the values of $p/\tilde M$ for each colour are quoted in the legend on the right).}
\label{extremal}
\end{figure}

\subsection{Vacua and transitions}
\label{metavacua}

For reference, Figure\ \ref{extremal} depicts the extremal potential, first obtained in \cite{Klebanov:2010qs}. There is a clearly visible metastable vacuum for $p/\tilde M \leq \mathfrak p_* \simeq 0.0538$. In this regime, there are also two unstable extrema: one at $\psi=0$ and another in the vicinity of $\psi\simeq 1.2$. The point $\psi=0$ is outside the regime of validity of our long-wavelength approximations.

\begin{figure}
\begin{center}
\includegraphics[width=10cm]{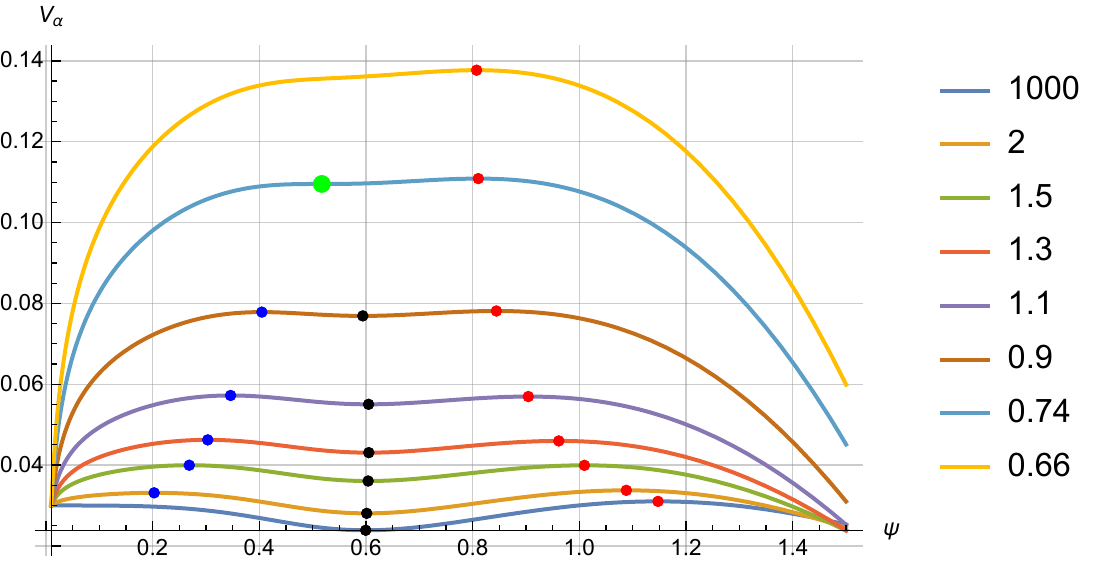}
\end{center}
\caption{Plots of the nonextremal potential $V_\alpha$ as a function of $\psi$ for $p/\tilde M =0.03$. The range of the plot is restricted in the region $\psi\in (0,1.5)$ where the most interesting physics occurs. Different colours depict the potential at different values of the nonextremality parameter $\alpha$ (the specifics of these values are listed in the legend). The blue dots indicate the unstable fat M5 vacua near the north pole $(\psi=0)$. The black dots indicate the metastable vacuum. The red dots indicate a second unstable vacuum (thin M5 branch). The green dot at $\alpha \simeq 0.7424$ is a merger point of the blue and black vacua.}
\label{Valpha_3}
\end{figure}

\begin{figure}
\begin{subfigure}{0.5\textwidth}
\includegraphics[width=0.95\linewidth]{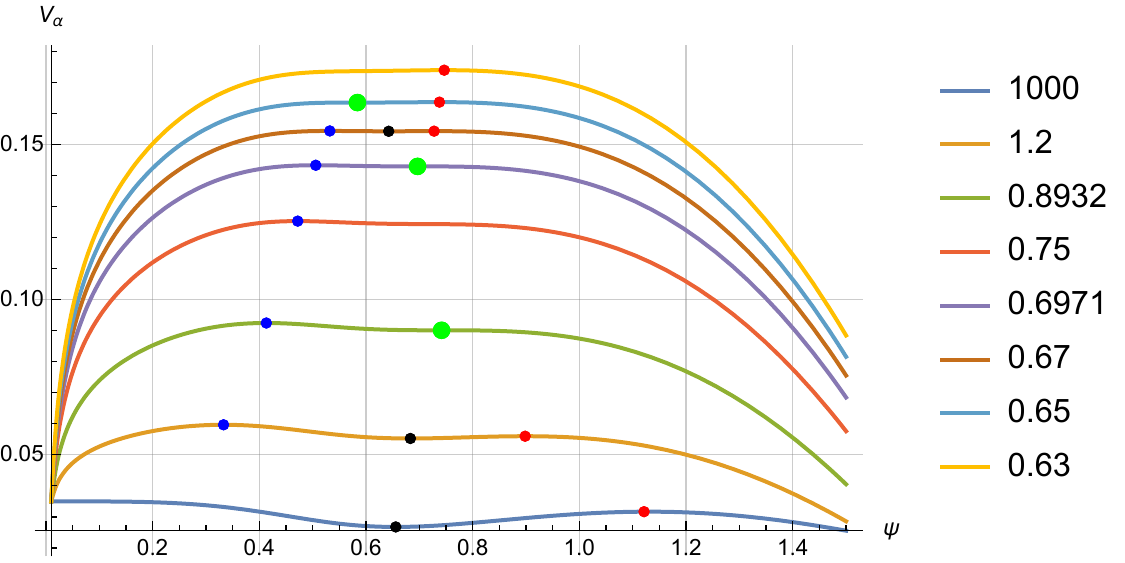} 
\end{subfigure}
\begin{subfigure}{0.5\textwidth}
\includegraphics[width=0.95\linewidth]{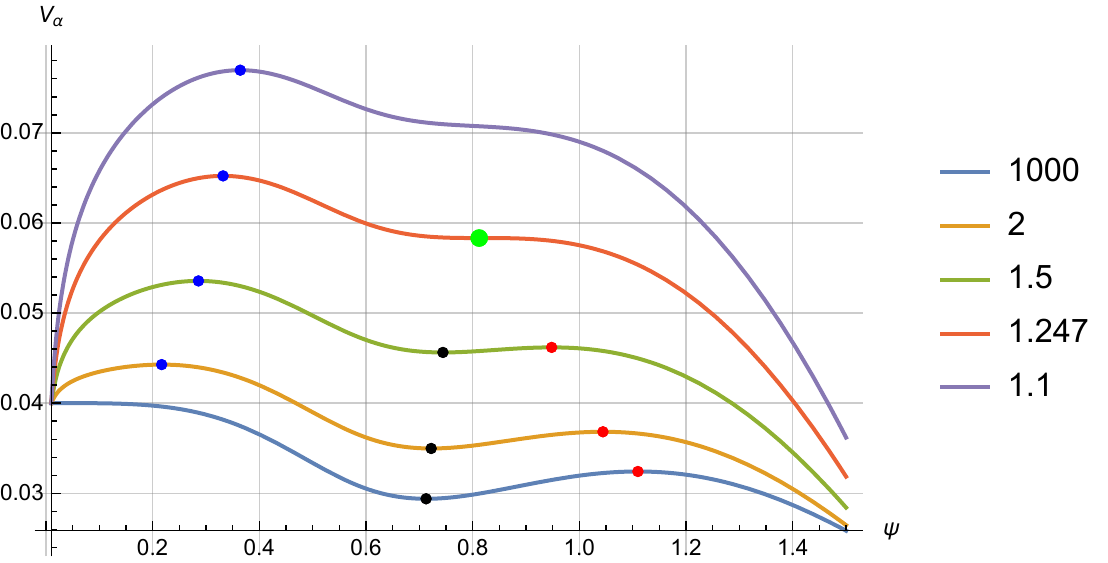}
\end{subfigure}
\caption{On the left, we plot the nonextremal potential $V_\alpha$ as a function of $\psi$ for $p/\tilde M =0.035$. Different colours depict the potential at different values of the nonextremality parameter $\alpha$. Once again, the blue dots indicate an unstable vacuum near the north pole $(\psi=0)$ (fat M5 branch) and the black dots the metastable vacuum. The red dots indicate a second unstable vacuum (thin M5 branch). In this case there are three green dots. At $\alpha \simeq 0.8932$ and $\alpha \simeq 0.6971$ they represent merger points of a metastable state with a red unstable thin M5 state. At $\alpha \simeq 0.65$ the green dot represents a merger with a blue unstable fat M5 state. On the right, we plot the nonextremal potential $V_\alpha$ as a function of $\psi$ for $p/\tilde M =0.04$. The plotted values of $\alpha$ are listed in the legend. In this regime there is a single green dot at $\alpha \simeq 1.247$, which is a merger point of the black metastable state with the red unstable thin M5 state.}
\label{Valpha_35_4}
\end{figure}

In what follows we focus on the `metastable regime' $p/\tilde M\in (0,\mathfrak p_*)$ and examine how thermal effects modify the stable and unstable vacua. It is technically convenient to start with the analysis of the blackfold equations at fixed $\alpha$, where the plots of the potential $V_\alpha$ (in \eqref{potAaa}, \eqref{potAab}) exhibit the extrema most clearly. In Figs.\ \ref{Valpha_3} and \ref{Valpha_35_4},  we present plots of $V_\alpha$ at three different values of $p/\tilde M$: 0.03, 0.035 and 0.04. Curves with different colours represent the form of the potential at the same $p/\tilde M$ for different values of $\alpha$. The dots indicate the locations of the extrema and the colour of the dots the nature of the solution at those extrema. We use the following conventions:
\begin{itemize}
\item A blue dot represents an unstable solution in the vicinity of the north pole at $\psi=0$. This is a black M5 brane solution wrapping an $S^3$ with a small radius compared to the Schwarzschild radius (further details on this aspect will appear in the next subsection). We call this type of solutions {\it fat M5 branes}.  
\item A black dot represents a metastable solution. These solutions are thermalised versions of the KP metastable state. 
\item A red dot represents an unstable wrapped M5 black brane whose $S^3$ radius is large compared to the Schwarzschild radius. We call this type of solutions {\it thin M5 branes}.
\item A green dot represents the merger of an unstable state with a metastable state.
\end{itemize}

Depending on the regime of $p/\tilde M$ the system exhibits three different types of bifurcations. 

\vspace{0.2cm}
\noindent
{\bf Regime I: small $p/\tilde M$.}
The first type occurs for $p/\tilde M \in (0,\mathfrak p_1)$. Numerically, we have determined $\mathfrak p_1 \simeq 0.0345$. The characteristic behaviour of this regime appears in Figure\ \ref{Valpha_3}. The bottom blue curve is a near-extremal curve at $\alpha=1000$. As we decrease $\alpha$ (and therefore increase the nonextremal effects) we observe the gradual convergence of the fat unstable branch towards the metastable branch. They merge at a small value of $\alpha$ ($\alpha\simeq 0.7424$ in the case of Figure\ \ref{Valpha_3}) at $\psi \simeq 0.5$ which corresponds to the renormalised temperature $\T \simeq 0.73873$. At even smaller values of $\alpha$ only the unstable thin M5 brane branch (red dot) remains. In this regime we observe the same saddle-node type bifurcation that was observed in the case of polarised anti-D3 branes in the Klebanov-Strassler background in chapter \ref{EX}. In the next subsection we will present quantitative evidence that suggests that this type of merger is driven by properties of the horizon geometry.

\vspace{0.2cm}
\noindent
{\bf Regime II: intermediate $p/\tilde M$.}
Interestingly, unlike the polarised anti-D3s in Klebanov-Strassler, in the M-theory case at hand there are two additional types of transitions that point towards qualitatively different properties of the dual three-dimensional QFT. A more involved transition pattern occurs for $p/\tilde M \in (\mathfrak p_1, \mathfrak p_2)$. Numerically, we obtain $\mathfrak p_2 \simeq 0.0372$. This is a small window where, as we decrease $\alpha$, {\it three} consecutive saddle-node-type bifurcations occur. First, the metastable state merges with the red thin black M5 state on the right of the plot (in Figure\ \ref{Valpha_35_4} on the left this occurs at $\alpha \simeq 0.8932$, $\psi\simeq 0.74173$ and $\T \simeq 0.78917$). This type of transition is qualitatively different compared to the transition in Figure\ \ref{Valpha_3}. It bears a strong resemblance to the zero temperature transition in Figure\ \ref{extremal} when $p/\tilde M$ crosses the threshold for the existence of a metastable vacuum. For a range of lower values of $\alpha$ only the fat unstable M5 brane state (blue dots) exists. Then, at another saddle-node-type bifurcation a metastable state and a thin unstable state re-appear out of nothing (in Figure\ \ref{Valpha_35_4} on the left this occurs at $\alpha\simeq 0.6971$, $\psi \simeq 0.69662$ and $\T\simeq 0.78318$). Subsequently, at even lower values of $\alpha$ the new metastable state starts moving closer to the fat unstable state. A third and final merger between the fat and metastable states occurs, which is qualitatively of the same character as in regime I. In Figure\ \ref{Valpha_35_4} on the left this merger occurs at $\alpha\simeq 0.65$, $\psi\simeq 0.58389$ and $\T\simeq 0.75475$.

\vspace{0.2cm}
\noindent
{\bf Regime III: high $p/\tilde M$.}
There is a third regime, where $p/\tilde M\in (\mathfrak p_2, \mathfrak p_*)$. In this case, there is only one merger, which is a merger between the metastable state and the red thin unstable state. In Figure\ \ref{Valpha_35_4} on the right this merger occurs at $\alpha\simeq 1.247$ at $\psi\simeq 0.81220$, $\T\simeq 0.75445$. As we noted above, this type of thermal transition does not occur in the case of polarised anti-D3 branes in the Klebanov-Strassler background. In the next subsection we present evidence suggesting that properties of the horizon geometry play a less important role in this type of merger.

\begin{figure}
\begin{subfigure}{0.5\textwidth}
\includegraphics[width=0.95\linewidth]{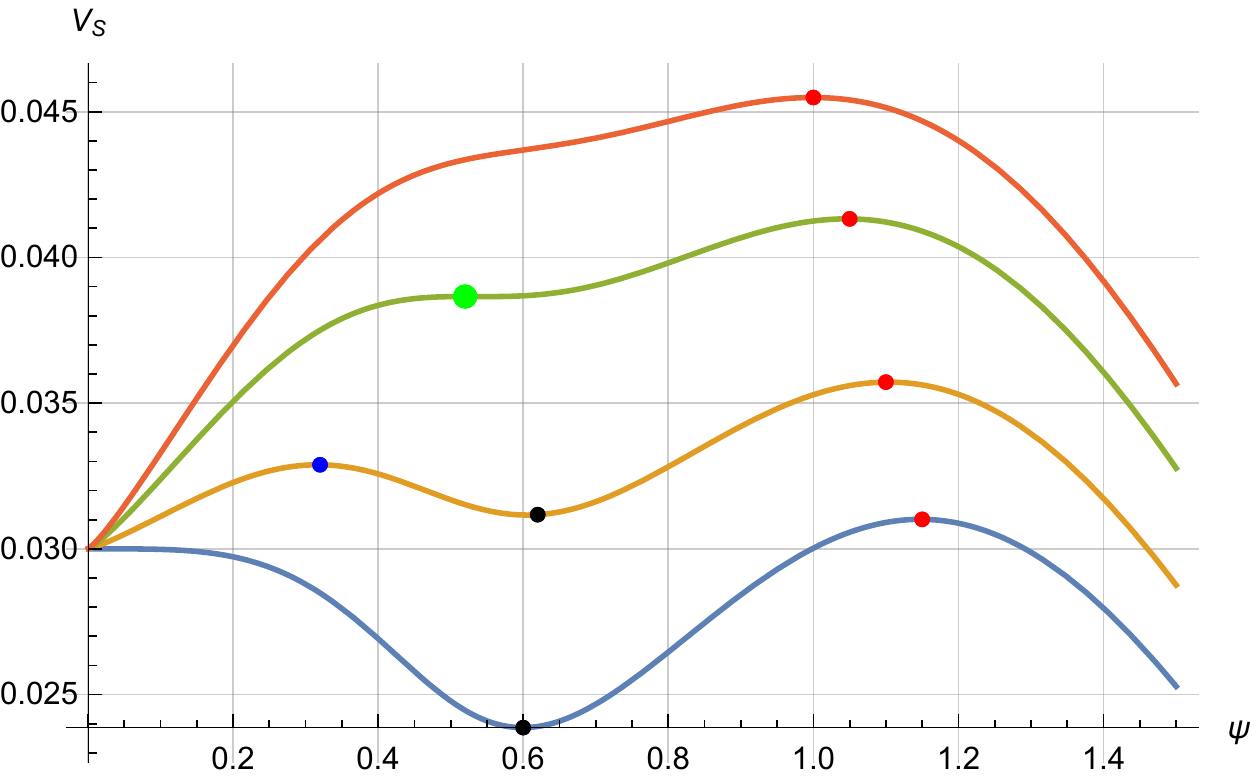}
\end{subfigure}
\begin{subfigure}{0.5\textwidth}
\includegraphics[width=0.95\linewidth]{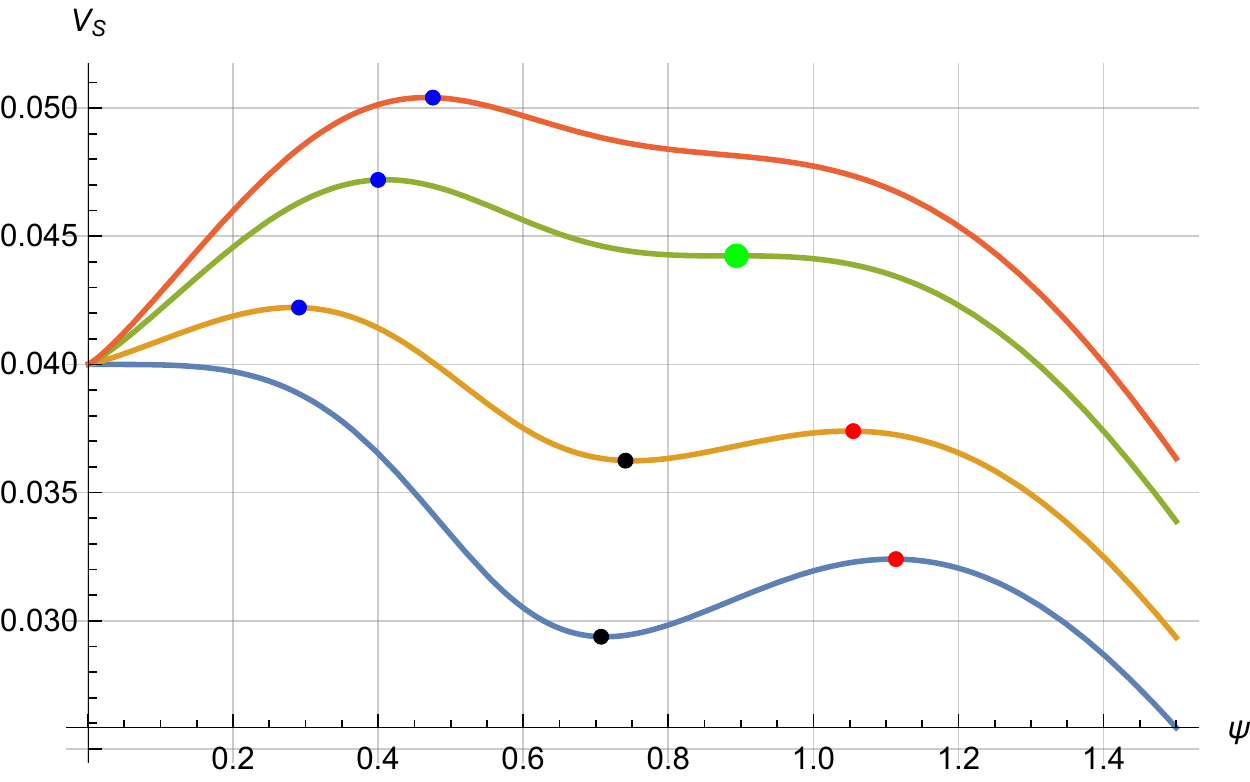}
\end{subfigure}
\caption{Plots of the fixed-$S$ counterparts of the plots in Figs.\ \ref{Valpha_3} and \ref{Valpha_35_4} on the right. The left plot depicts $V_S$ at $p/\tilde M =0.03$ and the right plot $V_S$ at $p/\tilde M=0.04$.}
\label{Ventro}
\end{figure}

The corresponding analysis of the system at fixed total entropy $S$ with the use of the potential $V_S$ reveals exactly the same qualitative and quantitative features. In Figure\ \ref{Ventro} we present the fixed-$S$ counterparts of the plots in Figs.\ \ref{Valpha_3} and \ref{Valpha_35_4} on the right. It is visually harder to observe the three transitions in the plot of $V_S$ at $p/\tilde M = 0.035$, so we did not include this value in Figure \ref{Ventro}. As we noted previously, the $V_S$ potential is better motivated physically compared to the $V_\alpha$ potential.

\begin{figure}[t!]
\begin{subfigure}{0.5\textwidth}
\includegraphics[width=0.95\linewidth]{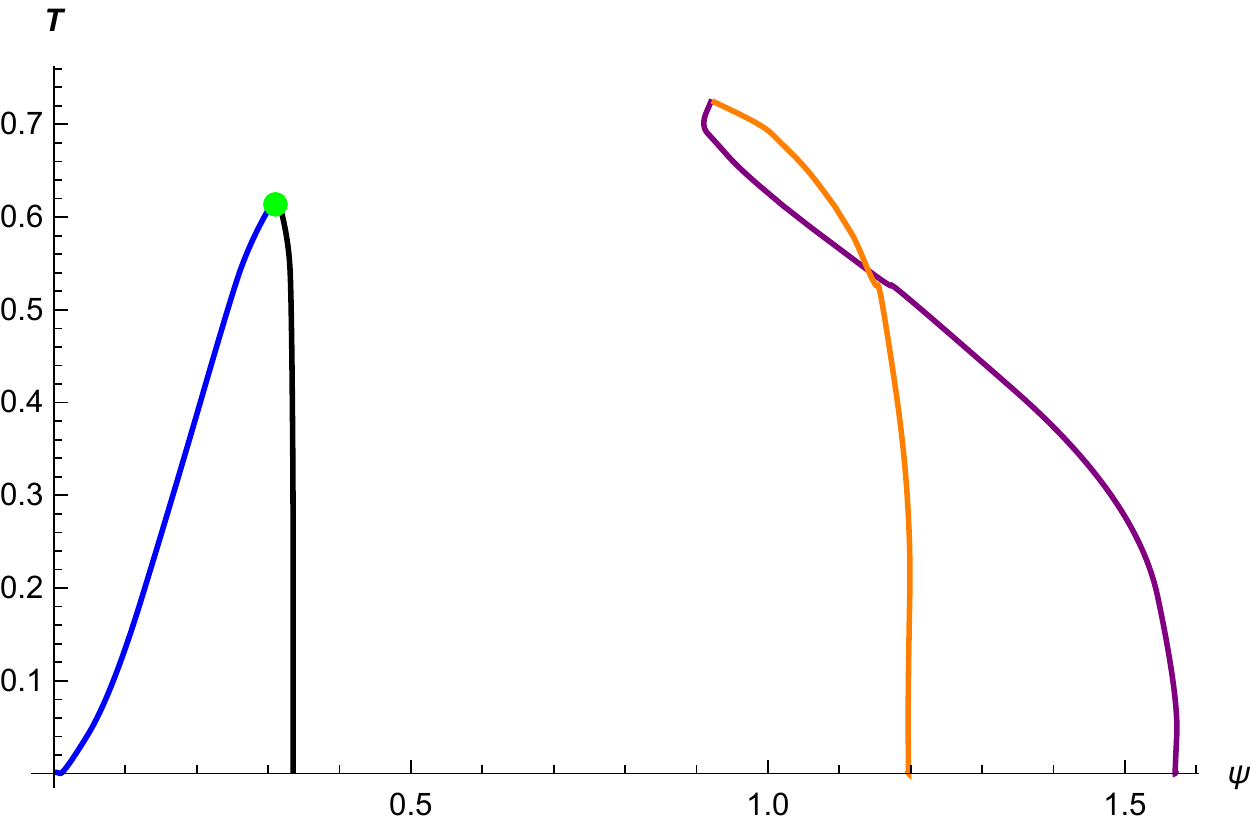}
\end{subfigure}
\begin{subfigure}{0.5\textwidth}
\includegraphics[width=0.95\linewidth]{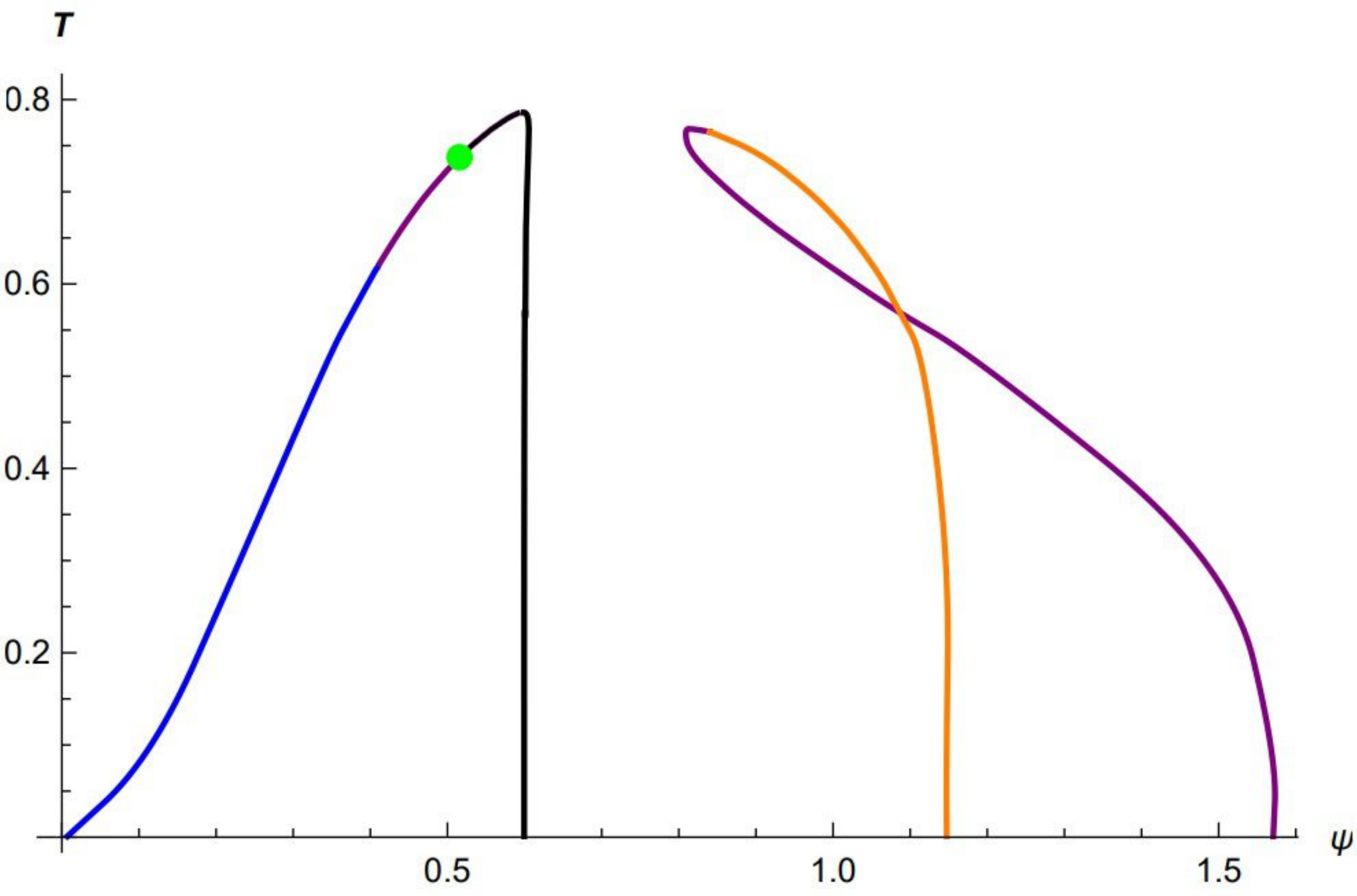}
\end{subfigure}
\caption{On the left, we plot the temperature of all the static M5-brane configurations as a function of their position $\psi$ on the four-sphere at fixed $p/\tilde M=0.01$. The colour conventions are explained in the main text. There is a single fat-thin merger in this regime represented by the green dot. On the right, we plot the temperature of all the static M5-brane configurations as a function of their position $\psi$ at fixed $p/\tilde M = 0.03$. This is still a phase diagram in regime I.}
\label{SolTemp_1_3}
\end{figure}

\begin{figure}
\begin{subfigure}{0.5\textwidth}
\includegraphics[width=0.95\linewidth]{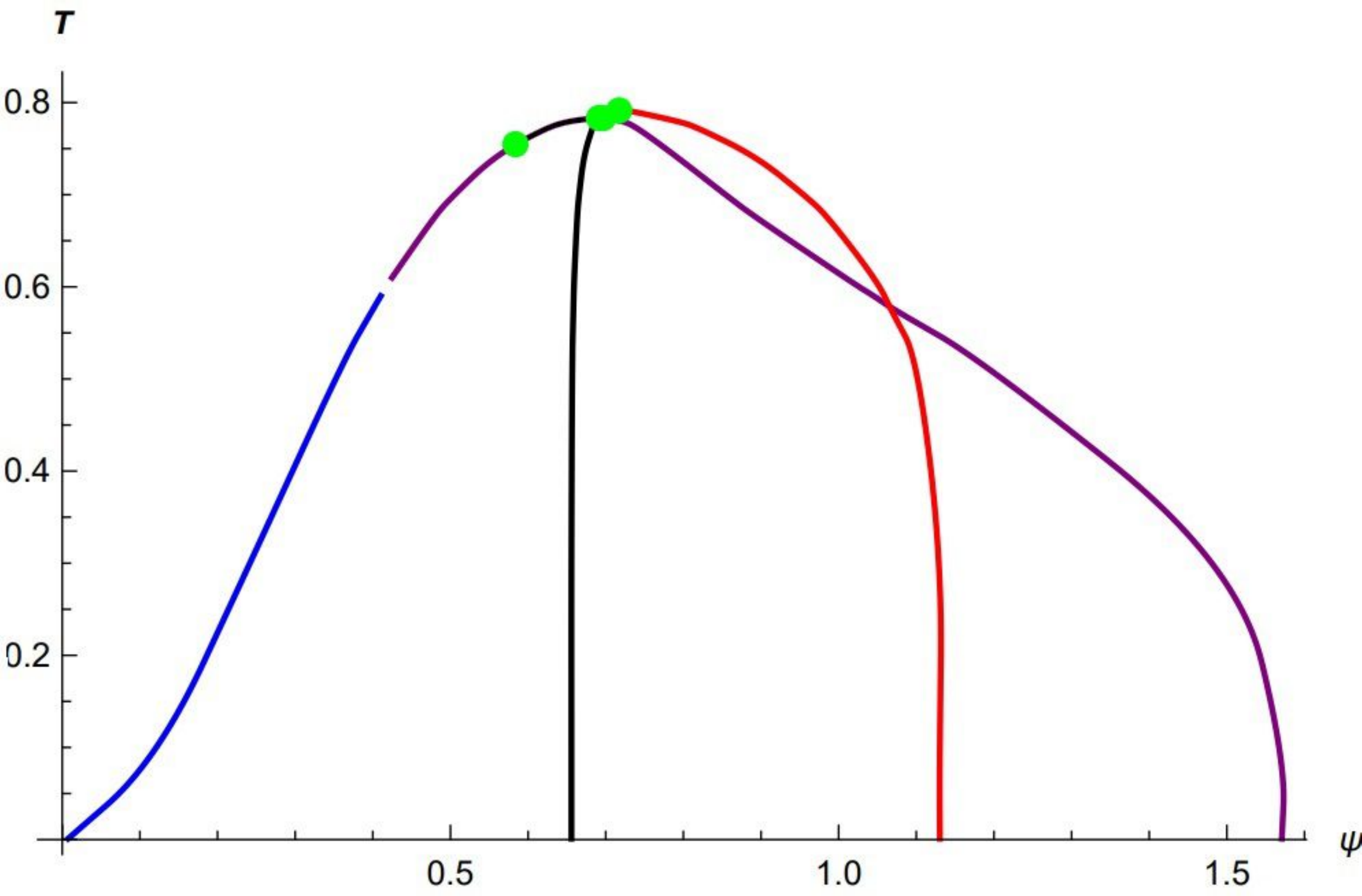}
\end{subfigure}
\begin{subfigure}{0.5\textwidth}
\includegraphics[width=0.95\linewidth]{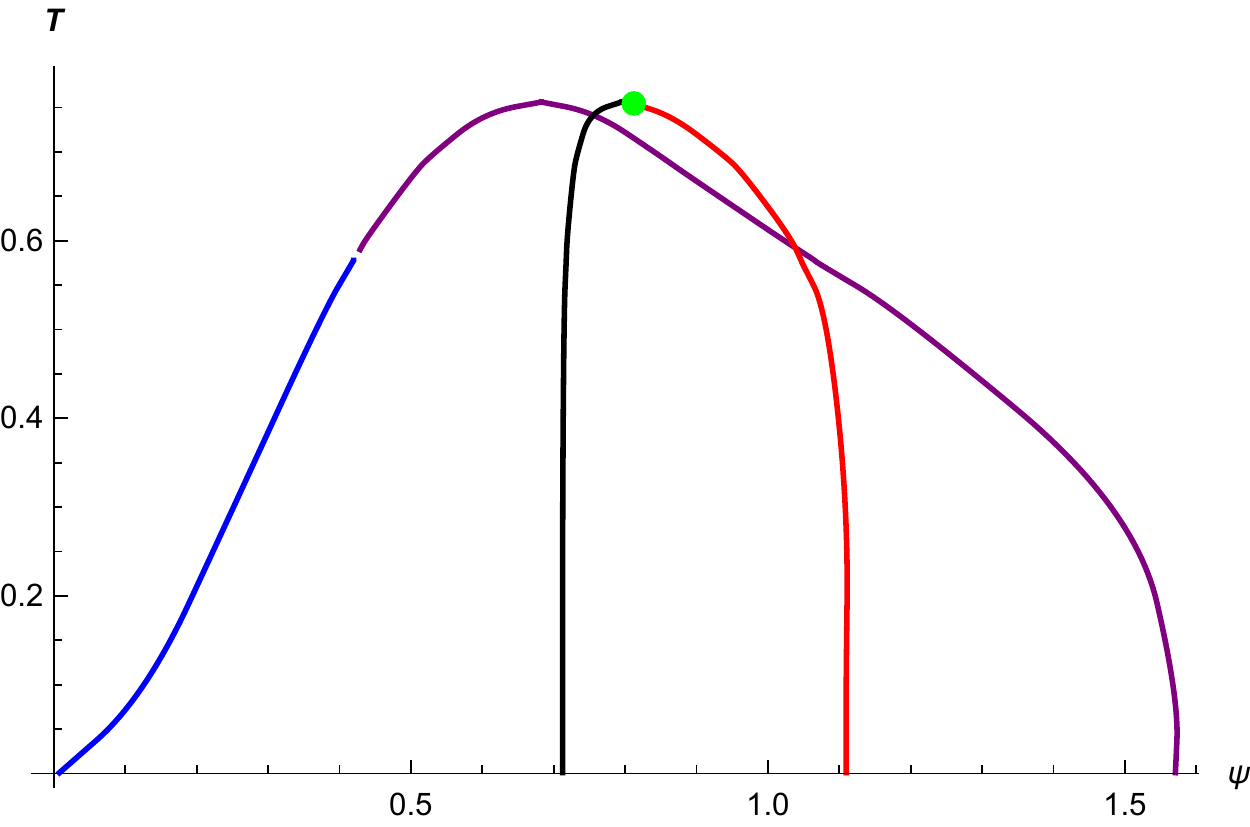}
\end{subfigure}
\caption{On the left, we plot the temperature of all the static M5-brane configurations as a function of their position $\psi$ at fixed $p/\tilde M = 0.035$. This is a representative phase diagram in regime II. There are three merger points in this diagram represented by green dots and two separate branches of metastable states represented by black segments. On the right, we plot the temperature of all the static M5-brane configurations as a function of their position $\psi$ at fixed $p/\tilde M = 0.04$. This is a representative phase diagram in regime III.}
\label{SolTemp_35_4}
\end{figure}

An interesting alternative perspective to the thermal properties of the wrapped M5 branes arises from the analysis of the blackfold equations at fixed temperature $T$. Since the fixed-$T$ potential $V_T$ is not defined for all angles $\psi$, it is more informative to plot $\T$ as a function of $\psi$ for  the extrema of $V_T$ at each value of $p/\tilde M$. In Figs.\ \ref{SolTemp_1_3} and \ref{SolTemp_35_4} we present these plots for $p/\tilde M=0.01,0.03,0.035,0.04$. The colour conventions for these plots are as follows:
\begin{itemize}
\item The blue curves represent unstable configurations of fat M5 black branes with values of $\alpha$ in the $+$ branch in \eqref{potTac}.
\item The purple curves in Figs.\ \ref{SolTemp_1_3} and Figs.\ \ref{SolTemp_35_4} represent unstable configurations in the $-$ branch in \eqref{potTac}.
\item The black curves represent metastable states. They belong to both the $+$ branch and the $-$ branch.
\item The orange curves in Figs.\ \ref{SolTemp_1_3} and the red curves in Figs.\ \ref{SolTemp_35_4} represent unstable thin M5-brane configurations. They belong to the $+$ branch.
\item The green dots represent mergers of a metastable with an unstable black hole phase. These dots are in direct correspondence with the green dots in the previous plots. Other points where different curves intersect are not merger points. At these intersections there are two black hole states with the same $T$ and $\psi$ but different values of $\alpha$.
\end{itemize}

The salient features of the temperature diagrams in Figs.\ \ref{SolTemp_1_3} and \ref{SolTemp_35_4} are the following. There are again three separate regimes I, II, III, which are the fixed-$T$ versions of the corresponding regimes in the fixed-$\alpha$ and fixed-$S$ analyses.

\vspace{0.2cm}
\noindent
{\bf Regime I: small $p/\tilde M$.}
At small enough values of $p/\tilde M$ the characteristic behaviour of the phase diagram is represented by Figs.\ \ref{SolTemp_1_3}. Let us first consider the features of the left plot of Figure\ \ref{SolTemp_1_3}. At small values of the temperature, there are four black hole phases: the blue unstable fat M5 state, the black metastable state, the orange unstable thin M5 state and the purple unstable thin M5 state. The combined branch of the orange and purple states is in one-to-one correspondence with the unstable states represented by red dots in e.g.\ Figure\ \ref{Valpha_3}. There is no merger in this branch. In this regime, there is only a single merger which is located on the blue-black branch. The blue-black merger is the fat-metastable merger that we noted also in Figure\ \ref{Valpha_3} in the context of the fixed-$\alpha$ and fixed-$S$ analyses. This merger involves two states in the $+$ branch. 

As we increase $p/\tilde M$ we observe two effects, which are clearly visible in the right plot of Figure\ \ref{SolTemp_1_3}. The first one is the appearance of intermediate purple states in the blue-black pair. Unlike the case of the left plot of Figure\ \ref{SolTemp_1_3}, in this case the merger happens in the $-$ branch, involving a metastable state (which close to the merger is in the $-$ branch) and a state in the purple branch (which is, by default, also a state in the $-$ branch). This merger occurs now at higher values of $\T$ compared to those in regime I. The second observation is that the blue-black and red-purple branches have moved closer.

\vspace{0.2cm}
\noindent
{\bf Regime II: intermediate $p/\tilde M$.}
In the second regime, where $p/\tilde M \in (\mathfrak p_1,\mathfrak p_2)$, the phase diagram has clearly rearranged. A representative case is depicted in the left plot of Figure\ \ref{SolTemp_35_4}. In this diagram, the main metastable branch has joined to the thin unstable (red) branch. The previous blue-purple branch has joined with the remaining purple branch at large $\psi$ through a new intermediate metastable set of states. These metastable states are represented by the black segment between the two green dots at $\psi\simeq 0.6$ and 0.7 in the left plot of Figure\ \ref{SolTemp_35_4}. There are three merger points in this diagram in direct correspondence to the mergers observed in the left plot of Figure\ \ref{Valpha_35_4}.

\vspace{0.2cm}
\noindent
{\bf Regime III: high $p/\tilde M$.}
Above the second critical value $p/\tilde M = \mathfrak p_2$, the phase diagram exhibits the behaviour 
represented by the right plot of Figure\ \ref{SolTemp_35_4}. In this regime there is a single thin-thin merger represented by the green dot at the joining point of the black (metastable) and red (unstable) branches. The combined blue-purple branch has no merger points and all its states are unstable. This phase diagram is directly related to the features observed in the right plot of Figure\ \ref{Valpha_35_4} and the right plot of Figure\ \ref{Ventro}.

\subsection{Nature of the merger points}
\label{mergergeometry}

In the previous subsections we distinguished between different branches of solutions by characterising them as thin or fat depending on the relative size of the Schwarzschild radius and the radius of the $S^3$ that the black M5 wraps. This characterisation can be made quantitatively more specific by introducing the dimensionless ratio
\be
\label{mergeaa}
d \equiv  \frac{ p^{1/3} \hat r_0}{\hat R_{S^3}} 
= \left( \frac{p}{\tilde M} \right)^{\frac{1}{3}} \frac{1}{\sin\psi (\cos\theta \sinh\alpha \cosh\alpha)^{\frac{1}{3}}}
~,
\ee
where 
\be
\label{mergeab}
\hat r_0 =r_0 \left( \frac{\CC}{\mathcal{Q}_5} \right)^{\frac{1}{3}} 
\ee
is a dimensionless quantity proportional to the local Schwarzschild radius of the black hole and
\be
\label{mergeac}
\hat R_{S^3} = \frac{(18\pi^2)^{\frac{1}{3}}}{2\pi \ell_P} R_{S^3}
\ee
is a dimensionless quantity proportional to the radius $R_{S^3} = m^{\frac{1}{3}}b_0 \sin\psi$ of the $S^3$ that the M5 black brane wraps at an angle $\psi$. A similar measure of black hole `fatness' was introduced in chapter \ref{EX} to describe wrapped NS5 black holes in the Klebanov-Strassler background. An analogous quantity, called $\nu$, that distinguishes between thin and fat neutral black ring solutions was introduced in \cite{Emparan:2006mm}.

\begin{figure}
\begin{subfigure}{0.5\textwidth}
\includegraphics[width=0.95\linewidth]{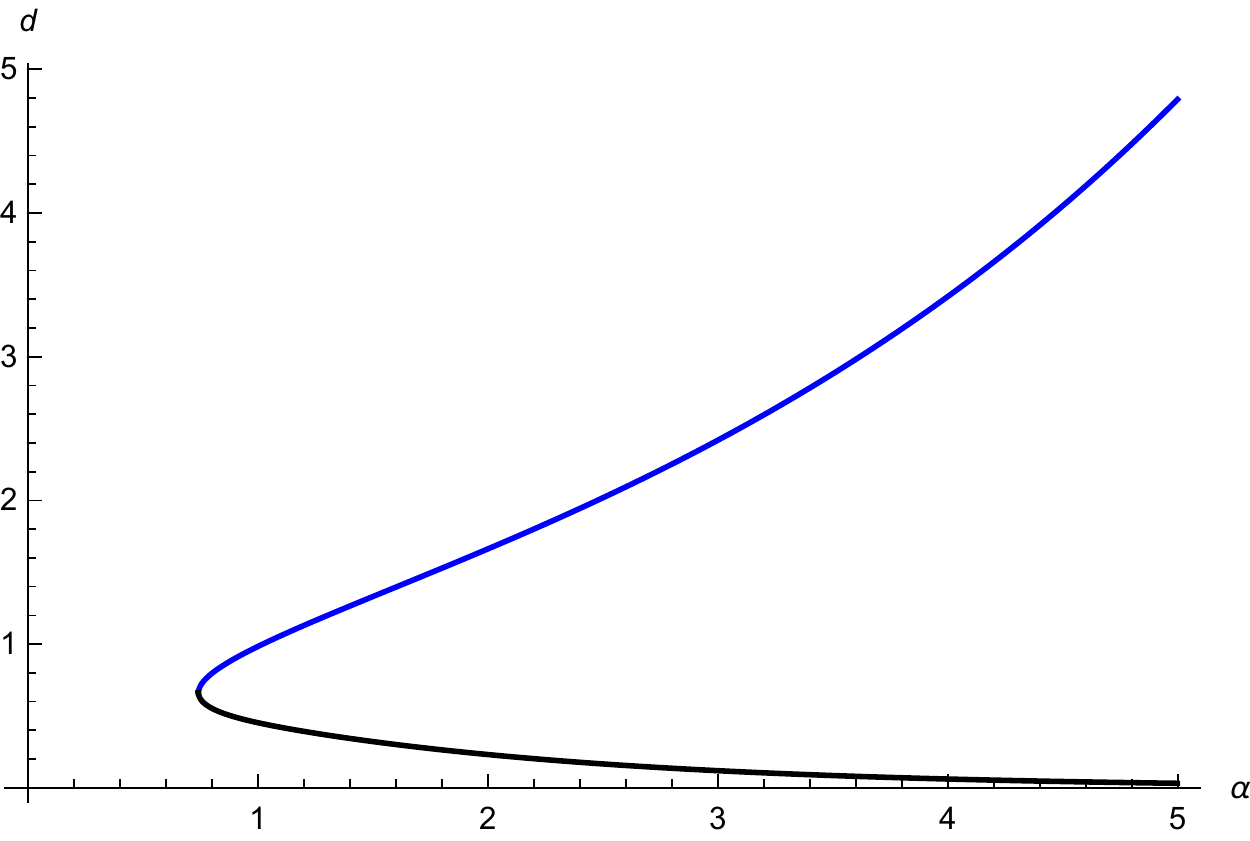} 
\end{subfigure}
\begin{subfigure}{0.5\textwidth}
\includegraphics[width=0.95\linewidth]{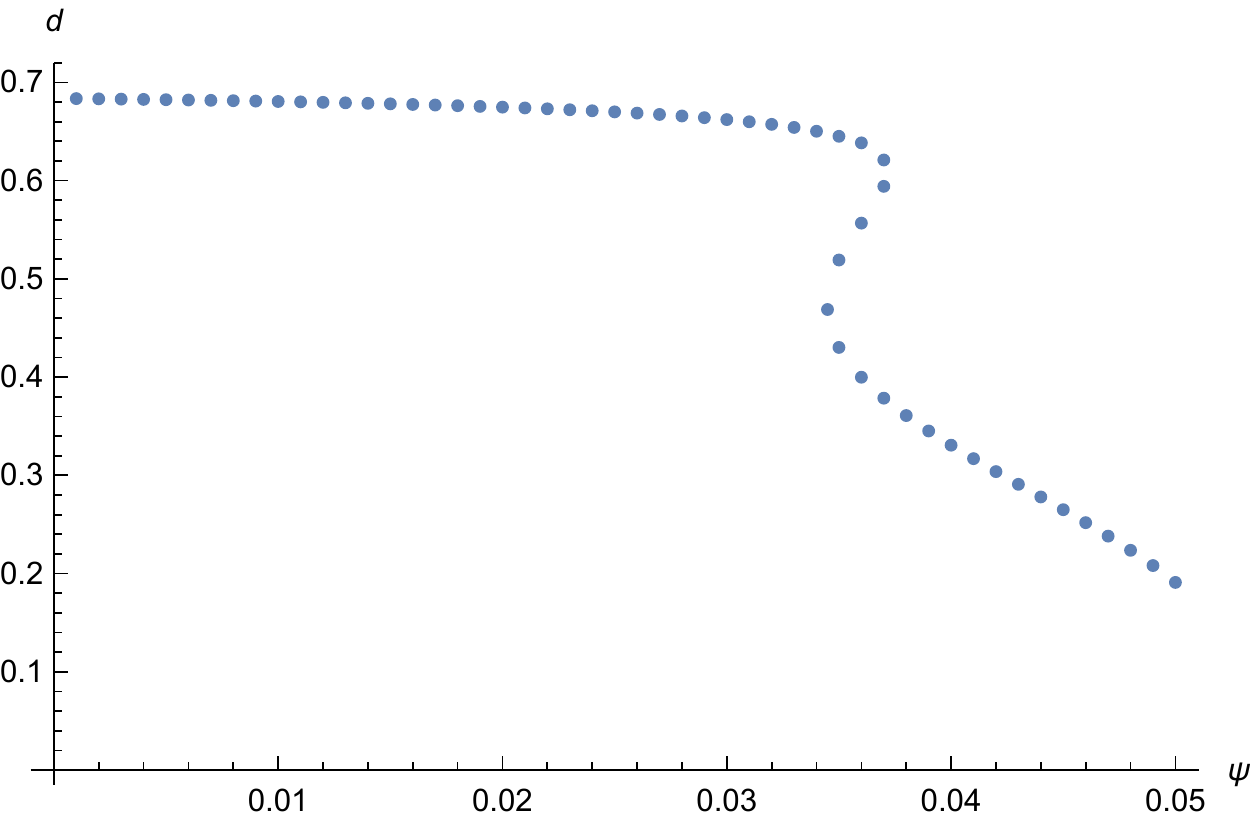}
\end{subfigure}
\caption{On the left, we plot the fatness ratio $d$ for blue unstable and black metastable states at $p/\tilde M =0.03$. The merger of the two branches occurs at $d\simeq 1$. The near-extremal metastable configurations are thin (small values of $d$), whereas the near-extremal unstable ones are fat (large values of $d$). On the right, we plot the fatness ratio $d$ (defined in equation\ \eqref{mergeaa}) at the merger points of the metastable states as a function of $p/\tilde M$. In region I, that involves a thin-fat merger, $d$ is almost constant. In region III, that involves a thin-thin merger, $d$ depends strongly on $p/\tilde M$. The intermediate regime II involves a multiplet of merger points.}
\label{fatthin_d1}
\end{figure}

Black hole states with $d\ll 1$ are by definition thin states where the Schwarzschild radius is comparatively smaller to the $S^3$ radius. States on the opposite part of the spectrum with $d\gg 1$ are fat states. For example, in the left plot of Figure\ \ref{fatthin_d1} we present the value of the ratio $d$ for the blue unstable and black metastable branches at $p/\tilde M=0.03$ that are depicted by blue and black dots in Figure\ \ref{Valpha_3}. We note that the near-extremal (i.e.\ large $\alpha$) metastable states have very low value of $d$ and are therefore thin states, whereas the corresponding unstable blue states have a very large value of $d$ and are therefore fat states. The merger occurs at a value of $d$ close to one.

According to the validity analysis \eqref{the4}, one finds
\be
d\ll\left(\frac{p}{N_5}\right)^{1/3}\sinh\alpha^{-1}~~.
\ee 
Thus, by appropriately tuning $p/N_5\gg1$, it is always possible to keep the merger points within our regimes of validity. On the other hand, as $\alpha$ increases, the validity of the unstable branch (blue curve in the left plot of Figure~\ref{fatthin_d1}) becomes more and more restricted.

A particularly interesting part of our discussion in this paper concerns the physics of the mergers where the metastable state is lost. The exact analysis of \cite{Cohen-Maldonado:2016cjh} shows that the existence of metastable anti-M2 states relies on the topology of the horizon. Supergravity configurations that describe point-like anti-M2s are not allowed by no-go theorems at zero temperature. At finite temperature anti-M2 black holes with spherical horizon topology are in principle allowed but require special boundary conditions for the fluxes on the horizon. Configurations of M5 black branes with non-spherical horizon topology evade these restrictions and are allowed by no-go theorems at zero temperature. As we thermalise the state, its horizon grows (namely, $d$ increases). At sufficiently high temperature one expects a transition where the horizon geometry can play a role. In that case, a scenario where the metastable state is lost would be consistent with the existing no-go theorems. 

We have already seen that the blackfold analysis verifies the expectation that the metastable state is lost at sufficiently high temperatures. The idea that the loss of the metastable state occurs because of a horizon-driven transition can be tested quantitatively by evaluating the fatness ratio $d$ at the merger points of the metastable states. Two things should happen if the above expectations are correct. Firstly, if the mergers are fat-thin mergers the transitions should occur at some value of $d$ of order 1. Secondly, some feature, e.g.\ a weak dependence of $d$ on $p/\tilde M$, should signal the dominant role of the horizon geometry. This is a highly non-trivial expectation. 

In the right plot of Figure\ \ref{fatthin_d1} we present numerical data on $d$ evaluated at the merger points of the metastable states as a function of $p/\tilde M$. We observe that the above expectation is verified extremely well in regime I, where the metastable state is lost via a fat-thin merger. Exactly the same type of physics was observed also in the case of polarised anti-D3 branes in the Klebanov-Strassler background in chapter \ref{EX}. These results are very suggestive about the validity of the overall picture that emerges from the use of the leading order blackfold analysis.

A new feature of the M2-M5 system in CGLP compared to the D3-NS5 system in Klebanov-Strassler is the existence of regimes II and III. In Regime III the metastable state is lost in a thin-thin merger via a completely different mechanism. In this case, the metastable state does not disappear because of the horizon-related effects, but because it develops a classical instability at some critical temperature. Consistently with this picture, in the right plot of Figure\ \ref{fatthin_d1} we observe that the values of $d$ at those mergers in regime III are much smaller and exhibit strong dependence on $p/\tilde M$. In the intermediate regime II, where multiple metastable branches occur, we observe an interesting feature of multi-valuedness in the dependence of $d$ on $p/\tilde M$. 

The co-existence of these patterns of mergers in the M2-M5 system is an interesting new prediction of the blackfold formalism. It would be very interesting to uncover further evidence for these transitions in supergravity (perhaps with numerical methods) and to understand the implications of these features in the three-dimensional QFT dual.

\chapter{Outlook}
\label{IN6}
The works in this thesis serve as a foundation for further (on-going) investigations on metastable antibranes in warped throats. We present here a short summary of these investigations and some of the relevant open problems.

It would be very interesting to complete the construction of the full leading-order backreacted solution of wrapped NS5 and M5 (black) branes in the appropriate scheme of matched asymptotic expansions. The benefits from such construction are threefold: First, it would serve as highly non-trivial evidence for the blackfold conjecture in the case antibranes backreaction, which claims that satisfying the blackfold equations guarantees a regular perturbative solution of all the supergravity equations. Second, it would be conclusive evidence for the existence of metastable state of antibranes in warped throats. Third, it would allow us to construct a blackfold set-up that enables the detection of possible brane fragmentation instabilities in the antibranes state. If the fragmentation instability is observed for extremal antibranes states, then it would be interesting to study the effects of non-extremality to see if such instability is resolved.

For unstable vacua, as well as semiclassically for the metastable ones, it is interesting to ask how the corresponding instabilities evolve dynamically and what is the end-state of the instabilities. When the solutions are extremal, there is an obvious end-point of the instability: the supersymmetric vacuum at the south pole. For example, the metastable vacuum is expected to evolve through vacuum tunnelling to the supersymmetric state through a process that is known as brane/flux annihilation \cite{Kachru:2002gs}. In the blackfold effective description of the thermal physics, we have seen that the north and south poles are strictly outside the regime of validity. However, even with this issue set aside for a moment, we notice that unlike the extremal case, the potential $V_S$ does not have any naive extrema at the two poles. Because of these features it is rather unclear what happens at the end-point of these instabilities. What kind of thermal solution, or black hole lies at the end of the evolution process of classical and semiclassical instabilities for the wrapped NS5 and M5 black branes that we described?

Finally, it would be interesting to elaborate further on the dual QFT interpretation of our results. We have uncovered an intricate pattern of thermal transitions of black hole phases in the bulk. What is the interpretation of these transitions in QFT? In this context, it would be useful to gain a more complete understanding of the nature of the finite-temperature merger points already in the gravity description. For the fat-metastable mergers, besides the features of the ratio $d$ that we reported above, we have not detected any other characteristic feature of the merger. For the thin-metastable mergers, one may note the similarity with the zero-temperature case at the maximum value of the antibrane charge where the metastable vacuum is lost.

\begin{appendices}

\chapter{Embedding geometry}
\label{EG}
The purpose of this appendix is to provide the preliminaries for discussions of embedding geometry. In section \ref{EG1}, we provide the definitions and properties of common geometric objects. In section \ref{EG2}, we show explicitly some useful properties of worldvolume conservation equations.

\section{Preliminaries} 
\label{EG1}

Given a manifold $\mathcal{M}$ with metric $g_{\mu \nu} d x^\mu d x^\nu$ and a submanifold $\mathcal{W}$ defined by the embedding $X^{\mu} (\sigma^a)$, we define the \textit{induced metric} $\gamma_{ab}$ by projecting the background metric $g_{\mu \nu}$ onto the worldvolume:
\be
\label{27}
\gamma_{ab} \equiv \p_a X^\mu \p_b X^\nu g_{\mu \nu} ~ .
\ee
Conversely, we can project any worldvolume tensors onto the background spacetime using the projector $\p_a X^\mu$. For example, we can project the induced metric $\gamma^{ab}$ onto the background to get a spacetime tensor $h^{\mu \nu}$ with
\be
h^{\mu \nu} \equiv \gamma^{a b} \p_a X^\mu \p_b X^\nu ~ .
\ee
$h^{\mu \nu}$ is the \textit{first fundamental form}, or equivalently, the \textit{tangential metric}. From the tangential metric, we can easily define the \textit{orthogonal metric} as
\be 
\perp_{\mu \nu} \equiv g_{\mu\nu} - h_{\mu\nu} ~ .
\ee
With the above definitions, we can easily show the following properties
\begin{align}
&h^\mu_{\nu}  \p_a X^\nu =  \p_a X^\mu ~ ,  &  &h^\mu_\nu h^\nu_\gamma = h^\mu_\gamma ~ , & &\perp^\mu_\nu h^\nu_\gamma = 0 ~,\\
&\perp_{\mu \nu} \p_a X^\mu = 0 ~, &  &\perp^{\mu}_\nu \perp^\nu_\gamma = \perp^\mu_\gamma ~.
\end{align}
The definitions of tangential metric and orthogonal metric here are equivalent to the definitions given by non-coordinates vectors where
\begin{align}
h^\mu_\nu &\equiv l^\mu_A l^A_\nu = \eta^{AB} l^\mu_A l_{\nu B} = - l^\mu_1 l_{\nu 1} + l^\mu_2 l_{\nu 2} + ...  ~ ,\\
\perp^\mu_\nu &\equiv n_X^\mu n^X_{\nu} = \delta^{X Y} n_X^\mu n_{Y \nu} = n_1^\mu n_{1 \nu} + ... 
\end{align}
with $l_A^\mu$ being vectors tangential to the worldvolume, $n_X^\mu$ being vectors orthogonal to the worldvolume, and together $l,n$ form an orthogonal basis.

Defining the object $\p^a X_\mu$ as
\be
\p^a X_\mu \equiv g_{\mu \nu} \gamma^{a b} \p_b X^\nu ~ ,
\ee
the pullback of a general tensor from the spacetime manifold $\mathcal{M}$ to the worldvolume $\mathcal{W}$ is given by
\be
T^{a_1 a_2 ... a_n}_{\,\,\, \,\,\, \,\,\, \,\,\, \,\,\, \,\,\, \,\,\, \,\,\, b_1 b_2 ... b_m} \equiv \p^{a_1} X_{\mu_1} ... \ \p_{b_1} X^{\nu_1} ... \ T^{\mu_1 ... \mu_n}_{\,\,\, \,\,\, \,\,\, \,\,\, \,\,\, \,\, \nu_1 ... \nu_m} ~ .
\ee
Conversely, one can also write down a general formula for projecting worldvolume tensors onto the background. However, it is important to note that for the projected worldvolume tensors, it is not well defined to take derivatives along arbitrary directions. Instead, it is only sensible to take derivatives in directions tangential to the worldvolume. As such, we define the \textit{tangential derivative operator}:
\be
\overline{\nabla}_\mu \equiv h^\mu_\nu \nabla_\nu ~ .
\ee
We define the \textit{extrinsic curvature tensor} as
\be
K_{\mu \nu}^{\,\,\, \,\,\, \rho} \equiv h^\sigma_\nu \overline{\nabla}_\mu  h^\rho_\sigma = - h^\sigma_\nu \overline{\nabla}_\mu \perp^\rho_\sigma ~ .
\ee
By contracting $K_{\mu \nu}^{\,\,\,\,\,\, \rho}$ with appropriate tangential and orthogonal metric, we can show that the extrinsic curvature tensor $K_{\mu \nu}^{\,\,\,\,\,\, \rho}$ is tangential in the first two indices and orthogonal in the third. It is also convenient to define the \textit{pullback extrinsic curvature tensor}
\be
\label{28}
K_{ab}^{\,\,\, \,\,\,\rho} \equiv \p_a X^\mu \p_b X^\nu K_{\mu \nu}^{\,\,\, \,\,\ \rho} =  \nabla_a \left( \p_b X^\rho \right) + \Gamma^\rho_{\mu \nu} \p_a X^\mu \p_b X^\nu
\ee
where $\nabla_a$ acts only on the $b$ index of $\p_b X^\rho$: $\nabla_a (\p_b X^\rho) = \p_a (\p_b X^\rho) - \Theta_{a b}^c \p_c X^\rho$ with $\Theta_{a b}^c$ the Christoffel symbols of the induced metric $\gamma_{ab}$.

\section{Worldvolume conservation equations}
\label{EG2}

As it is important for our discussions in the main text, let us consider conservation equations of projected worldvolume tensors, e.g. $\overline{\nabla}_\mu t^{\mu \nu}$ with $t^{\mu \nu} = t^{a b} \p_a X^\mu \p_b X^\nu$. In the case that $t^{\mu \nu}$ is 
antisymmetric, let us call this tensor $J^{\mu \nu}$, we note that
\be
\perp_\nu^\rho \overline{\nabla}_\mu J^{\mu \nu} = \overline{\nabla}_\mu \left( \perp_\nu^\rho J^{\mu \nu} \right) - J^{\mu \nu} \overline{\nabla}_\mu \perp^{\rho}_\nu = J^{\mu \nu} K_{\mu \nu}^{\,\,\,\,\,\, \rho} = 0
\ee
so
\be
\label{a1}
\overline{\nabla}_\mu J^{\mu \nu} = h_\rho^\nu \overline{\nabla}_\mu J^{\mu \rho} =  \p_b X^\nu \nabla_a J^{a b} 
\ee
where the last equality can be shown though direct substitution of background Christoffel symbols $\Gamma$ and worldvolume Christoffel symbols $\Theta$.
Thus, for antisymmetric worldvolume current, $\overline{\nabla}_\mu J^{\mu \nu} = 0$ is equivalent to $\nabla_a J^{a b}=0$. Of course, it can easily be shown that this statement holds for antisymmetric currents of any rank.

In the case that $t^{\mu \nu}$ is symmetric, let us imagine the energy-stress tensor $T^{\mu \nu}$, we can decompose the conservation equation $\overline{\nabla}_\mu T^{\mu \nu} = 0$ into equations along the worldvolume directions and equations orthogonal to the worldvolume directions, which we call \textit{intrinsic} and \textit{extrinsic} equations respectively. In particular, we note that
\be
\label{a2}
\overline{\nabla}_\mu T^{\mu \nu} = \overline{\nabla}_\mu \left( T^{\mu \nu} \, h_\nu^\rho \right) = T^{a b} K_{a b}^{\ \ \, \rho} + \p_b X^\rho \nabla_a T^{ab}
\ee
so, contracting $\overline{\nabla}_\mu T^{\mu \nu} = 0$ with $h^{\sigma}_\rho$, we obtain the intrinsic equation 
\be
\nabla_a T^{ab} = 0
\ee
and, contracting with $n^{(i)}_{\rho}$, we obtain the extrinsic equation
\be
T^{a b} K_{a b}^{\ \ \, (i)} = 0
\ee
where $K_{a b}^{\ \ \, (i)} \equiv K_{a b}^{\ \ \, \rho} n^{(i)}_{\rho}$. By using \eqref{28}, we can express $K_{a b}^{\ \ \, (i)}$ as 
\be
\label{a3}
K_{a b}^{\ \ \, (i)} = \p_a \p_b X^\rho + \Gamma^{(i)}_{a b}
\ee
where
\be
\Gamma^{(i)}_{a b} \equiv \Gamma^{\rho}_{\mu \nu} \p_{a} X^{\mu} \p_{b} X^{\nu} n_{\rho}^{(i)}  ~ .
\ee

\chapter{Warped deformed conifolds}
\label{CON}
In this appendix, we provide discussions on the Klebanov-Strassler (KS) \cite{Klebanov:2000hb} and Cvetic-Gibbons-Lu-Pope (CGLP) \cite{ Cvetic:2000db} warped deformed conifolds, which are also commonly referred to as KS and CGLP throats, focusing on aspects immediately relevant for their roles in the main text as background geometries of metastable antibranes. In particular, in our discussion of the KS throat in section \ref{KS}, we write down the explicit description of the throat close to the tip. Subsequently, we derive an expression for such description in an adapted coordinate system. We discuss the CGLP throat in section \ref{CON2}. However, as there are obvious similarities, such discussion is brief. 

\section{The Klebanov-Strassler throat}\label{KS}
The KS throat is a 10-dimensional type IIB supergravity solution. The throat involves a 6 dimensional deformed conifold, a 4 dimensional Minkowskian space, and non-trial $F_3, F_5, H_3$ fluxes which in turn induce warping effects on the flat space and the conifold. In this section, we shall discuss aspects of this geometry that are relevant for our blackfold analysis. For further information on the KS throat, we refer readers to the review \cite{Herzog:2001xk}.

\subsection{The Klebanov-Strassler from the Klebanov-Tseytlin}
\paragraph{A 6-dimensional conifold} Let us start with a 6-dimensional conifold defined by the equation
\be
\label{con1}
\sum_{a = 1}^{4} z_a^2 = 0 
\ee
where $z_a$ are complex numbers. For illustrative purposes, we can check that for the case that $a$ max is 2, this function reproduces the familiar description of a 2-dimensional cone: $x^2 + y^2 = z^2$. In such case, the defining equation \eqref{con1} reduces to the system
\be
\begin{cases}
&x_1^2 + x_2^2 = y_1^2 + y_2^2 \\
&x_1 y_1 + x_2 y_2 = 0
\end{cases}
\ee
where $z_a = x_a + i y_a$. Eliminating $y_1$ in the first equation using the second equation, reparametrising the RHS as $z^2 = y_2^2 \left(1 + x_2^2/x_1^2 \right)$, we see that the system indeed becomes  $x^2 + y^2 = z^2$. 

Back to the task at hand, we note that, as derived in \cite{Candelas:1989js}, the metric of the 6-dimensional conifold defined by equation \eqref{con1} can be written as
\be
\label{con2}
ds_6^2 = dr^2 + r^2 ds^2_{T^{1,1}}
\ee
with
\be
ds^2_{T^{1,1}} = \frac{1}{9} \Big( d\psi + \sum^2_{i = 1} \cos \theta_i d \phi_i \Big)^2 + \frac{1}{6} \sum_{i = 1}^{2} (d \theta_i^2 + \sin^2 \theta_i d \phi_i^2)
\ee
with $\psi \in (0, 4 \pi)$ and $\theta_i, \phi_i$ parametrise two $S^2$ in a standard way. Here, $T^{1,1}$ refers to the 5-dimensional base of our 6-dimensional conifold, and $r$ denotes the radial direction away from the tip of the cone. The base geometry $T^{1,1}$ is an Einstein manifold with topology $S^2 \times S^3$. For more details on the base geometry $T^{1,1}$ and its parametrisation, we refer readers to \cite{Candelas:1989js}. At this point, our 6-dimensional conifold is Ricci flat as one can easily check. 

For later convenience, let us note that the metric can also be described using the following 1-forms 
\begin{align}
\label{con3}
g^1 &=  \frac{- \sin \theta_1 d \phi_1 - \cos \psi \sin \theta_2 d \phi_2 + \sin \psi d \theta_2}{\sqrt{2}} ~ ,\\
g^2 &=  \frac{d \theta_1 - \sin \psi \sin \theta_2 d \phi_2 - \cos \psi d \theta_2}{\sqrt{2}} ~ ,\\
g^{3} &= \frac{- \sin \theta_1 d \phi_1 + \cos \psi \sin \theta_2 d \phi_2 - \sin \psi d \theta_2}{\sqrt{2}} ~ ,\\
g^{4} &= \frac{d \theta_1 + \sin \psi \sin \theta_2 d \phi_2 + \cos \psi d \theta_2}{\sqrt{2}} ~ , \\
\label{con4}
g^5 &= d \psi + \cos \theta_1 d \phi_1 + \cos \theta_2 d \phi_2
\end{align} 
as
\be
ds_6^2 = dr^2 + r^2 \left( \frac{1}{9} (g^5)^2 + \frac{1}{6} \sum^4_{i = 1} (g^i)^2 \right) ~ .
\ee

\paragraph{Adding charges to the cone, the Klebanov-Tseytlin throat} Let us begin with a Ricci flat 10-dimensional spacetime given by the 4-dimensional Minkowski space $\times$ the 6-dimensional conifold. The metric of such geometry is given by
\be
ds^2_{10} = - dt^2 + dx_1^2 + dx_2^2 + dx_3^2 + ds_6^2
\ee
where $ds_6^2$ is the conifold metric \eqref{con2}. We would like to place some D3 branes and some wrapped D5 branes at the tip of the conifold. The D3 branes will lie along the 4 flat directions and the D5 branes will further wrap around an $S^2$ of the base $T^{1,1}$ of the conifold. Such introduction means our spacetime will have non-zero $F_3, F_5$ fluxes obeying the following requirements:
\begin{align}
\label{ks1}
&\frac{1}{4 \pi^2 \alpha'} \int_{S^3} F_3 = M ~ , & &\frac{1}{(4 \pi^2 \alpha')^2} \int_{T^{1,1}} F_5 = N
\end{align}
where $M, N$ denote the number of D5 branes and D3 branes respectively. Note further that, not unlike the familiar cases of D-brane bound state such as the D3-D5 configuration \eqref{3}-\eqref{4}, because of the Chern-Simon terms in the supergravity action, our introduction will also induce an $H_3$ flux as a by product. The introduction of branes to the tip of the conifold will also mean that our conifold now becomes warped and no long Ricci flat. The solution describing such warped, fluxed conifold is called the Klebanov-Tseytlin throat.

\paragraph{The Klebanov-Strassler from the Klebanov-Tseytlin}
The motivation behind the construction of the Klebanov-Strassler throat was originally because people do not like the pointy tip of the 6-dimensional conifold in the Klebanov-Tseytlin throat. To get rid of the pointy tip in the 6-dimensional conifold, we can deform the defining equation \eqref{con1} by introducing an $\varepsilon^2$:
\be
\sum^4_{i = 1} z_i^2 = \varepsilon^2 ~ .
\ee
The effect is that the pointy tip is replaced by a finite size $S^3$. The solution describing the configuration resulted from blowing the point tip of the Klebanov-Tseytlin into a finite size $S^3$ is the Klebanov-Strassler solution.

By deforming the conifold, we have certainly changed some aspects of the geometry. Nevertheless, the intuitive understanding for the Klebanov-Strassler remains similar to that of the Klebanov-Tseytlin. In particular, we should think of the Klebanov-Strassler as the solution resulting from placing D3 and D5 branes at the tip of a Ricci-flat (deformed) conifold. The D3 and D5 branes will induce non-trivial $F_3, F_5, H_3$ fluxes as well as warping effects to the conifold, giving us a warped, fluxed throat geometry.

\subsection{The 6-dimensional deformed conifold}
Let us now take a closer look at the description of the KS throat. In particular, in this section, we discuss the parametrisation of the 6-dimensional deformed conifold. Even though there are already detailed discussion of the Euler angles parametrisation of the conifold in the literature, we find it necessary to remind our readers here as we will make use of this procedure later on to derive the description of the KS throat in adapted coordinates.      

The 6 dimensional deformed conifold of the KS solution is given by the equation
\be
\sum^4_{i = 1} z_i^2 = \varepsilon^2
\ee
where $z_i$ are complex numbers and $\varepsilon$ characterises the degree of deformation, i.e. if $\varepsilon = 0$, we have a normal cone. In order to obtain a parametrisation of the space, a clever trick one can do is to define the matrix
\be
W = \left( \begin{matrix}
z_3 + i z_4 & z_1 - i z_2 \\
z_1 + i z_2 & - z_3 + i z_4 
\end{matrix}\right) ~.
\ee
Then, the defining equation becomes
\be
\det W = - \varepsilon^2 ~.
\ee
It is easy to see that 
\be
W_0 = \left( \begin{matrix}
0 & \varepsilon e^{\tau/2} \\
\varepsilon e^{- \tau/2} & 0
\end{matrix}\right)
\ee
is one possible solution. Furthermore, if we define two $SU(2)$ matrices $L_j$ with $j = 1,2$ then 
\be
W = L_1 . W_0 . L_2^{\dagger}
\ee
also satisfies the equation $\det W = - \varepsilon^2$. As argued in \cite{Minasian:1999tt}, the metric of the deformed conifold is then given by 
\be
\label{a}
ds^2 = \mathcal{F} tr \left( d W^{\dagger} d W \right) + \mathcal{G} | tr (W^\dagger d W) |^2
\ee
where
\begin{gather}
\mathcal{F} (\tau) =  \frac{(\sinh 2 \tau  - 2 \tau)^{1/3}}{2 \times 2^{1/3} \times   \varepsilon^{2/3} \sinh \tau} ~,\\
\mathcal{G}(\tau) = \frac{2 - 3 \coth^2 \tau + 3 \tau (\cosh \tau/ \sinh^3 \tau) }{12  \times   \varepsilon^{8/3} (\cosh \tau \sinh \tau - \tau)^{2/3}} ~.
\end{gather}

\paragraph{Euler angles parametrisation of the deformed conifold}
One can parametrise the $L_j$ matrices using Euler angles as
\be
L_j = \left( \begin{matrix}
\cos \frac{\theta_j}{2} e^{i (\psi_j + \phi_j)/2} & - \sin \frac{\theta_j}{2} e^{-i (\psi_j - \phi_j)/2} \\
\sin \frac{\theta_j}{2} e^{i (\psi_j - \phi_j)/2} & \cos \frac{\theta_j}{2} e^{-i (\psi_j + \phi_j)/2} 
\end{matrix}\right)
\ee
with $(\psi_j, \phi_j)$ range from $0$ to $ 2 \pi$ and $\theta$ ranges from $0$ to $\pi$. Plugging the parametrised expression of $W = L_1 . W_0 . L_2^{\dagger}$ into (\ref{a}) yields the metric of the deformed conifold written in angular coordinates $\psi_j, \theta_j, \phi_j$. As the coordinates $\psi_1$ and $\psi_2$ only appear in $W$ as $\psi_1 + \psi_2$, we can define a new coordinate $\psi = \psi_1 + \psi_2$. The deformed conifold metric in these coordinates is then given by
\begin{equation}
ds_6^2 = \frac{1}{2} \varepsilon^{4/3} K(\tau) \Bigg[ \frac{1}{3 K^3 (\tau) } (d \tau^2 + (g^5)^2 ) + \cosh^2 \left ( \frac{\tau}{2} \right) [(g^3)^2 + (g^4)^2]  + \sinh^2 \left( \frac{\tau}{2} \right) [(g^1)^2 + (g^2)^2  ] 
\Bigg]
\end{equation}
where the function $K(\tau)$ is given by
\be
K (\tau) = \frac{(\sinh 2 \tau  - 2 \tau)^{1/3}}{2^{1/3} \sinh \tau}  ~,
\ee
and the $g^i$ forms are given in \eqref{con3}-\eqref{con4}.

\subsection{The Klebanov-Strassler throat near the apex in Euler angles}
For our purposes, we are only interested in the description of the KS throat near the apex. From the full description of the throat, we expand the metric and gauge fields in $\tau$ and keep only the relevant terms. To be more specific, we keep in the metric and gauge fields terms of the required order such that the profile of metric and fields solve the Supergravity equations to first order in $\tau$. For convenience, let us also set\footnote{Setting $g_s = 1$ is possible because the KS solution a has constant dilaton.} $g_s = 1$ and $\alpha' = 1$ in all further discussions of the KS throat. 

The KS metric near the apex is approximated by
\begin{multline}
ds_{10}^2 = A_1(\tau) \left( - (dx^0)^2 + ( d x^1)^2+ (d x^2)^2 + (dx^3)^2 \right) + A_2 (\tau) \left( d(\tau )^2 + (g^5)^2 \right) \\
+ A_3 (\tau) \left( (g^3)^2+ (g^4)^2 \right) + A_4 (\tau) \left( (g^1)^2 + (g^2)^2 \right)
\end{multline}
where
\be
A_1 (\tau) = \frac{\epsilon^{4/3}}{2^{1/3} (a_0)^{1/2} M} - \frac{ a_2 \, \tau^2 \, \epsilon^{4/3}}{2 \times 2^{1/3} (a_0)^{3/2} M} + \frac{3 \, (a_2)^2 \, \tau^4 \, \epsilon^{4/3}}{8 \times 2^{1/3} (a_0)^{5/2} M} -\frac{a_4 \, \tau^4 \, \epsilon^{4/3}}{2 \times 2^{1/3} (a_0)^{3/2} M}  ~,
\ee
\begin{multline}
A_2 (\tau) = \frac{(a_0)^{1/2} M}{2 \times 6^{1/3}} + \frac{(a_0)^{1/2} M \, \tau^2}{10 \times 6^{1/3}} + \frac{a_2 \, M \, \tau^2}{4 \times 6^{1/3} (a_0)^{1/2}} - \frac{ (a_2)^2 M \, \tau^4}{16 \times 6^{1/3} (a_0)^{3/2}} \\ 
+\frac{(a_0)^{1/2} M \, \tau^4}{210 \times 6^{1/3}}  +\frac{ a_4 \, M \, \tau^4}{4 \times 6^{1/3} (a_0)^{1/2}} + \frac{a_2 \, M \, \tau^4}{20 \times 6^{1/3} (a_0)^{1/2}} ~,
\end{multline}
\begin{multline}
A_3 (\tau) = \frac{(a_0)^{1/2} M}{6^{1/3}} + \frac{3^{2/3} (a_0)^{1/2} M \, \tau^2}{20 \times 2^{1/3}} +  \frac{a_2 \, M \, \tau^2}{2 \times 6^{1/3} (a_0)^{1/2}} + \frac{a_4 \, M \, \tau^4}{2 \times 6^{1/3} (a_0)^{1/2}}  \\
+ \frac{17 \, (a_0)^{1/2} M \, \tau^4}{2800 \times 6^{1/3}}   - \frac{(a_2)^2 M \, \tau ^4}{8 \times 6^{1/3} (a_0)^{3/2}} + \frac{3^{2/3} a_2 \, M \tau^4}{40 \times 2^{1/3} (a_0)^{1/2}}  ~,
\end{multline}
\begin{multline}
A_4 (\tau)  = \frac{(a_0)^{1/2} M \, \tau ^2}{4 \times 6^{1/3}}  - \frac{(a_0)^{1/2} M \, \tau^4}{240 \times 6^{1/3}} + \frac{a_2 \, M \, \tau ^4}{8 \times 6^{1/3} (a_0)^{1/2}} + \frac{a_4 \, M \, \tau^6}{8 \times 6^{1/3} (a_0)^{1/2}} \\
- \frac{(a_2)^2 M \, \tau^6}{32 \times 6^{1/3} (a_0)^{3/2}} - \frac{ a_2 \, M \, \tau ^6}{480 \times 6^{1/3} (a_0)^{1/2}} +\frac{59 \, (a_0)^{1/2} \, M \,\tau ^6}{50400 \times 6^{1/3}}
\end{multline}
with the constants $a_0 \approx 0.71805$, $a_2 = - (3 \times 6^{1/3})^{-1} $, and $a_4 = (18 \times 6^{1/3})^{-1}$.

The KS fluxes near the apex are approximated by\footnote{As our convention of the Hodge star operator is different from that of \cite{Klebanov:2000hb}, our description of $H_3$ and $\tilde{F}_5$ have different signs from those of \cite{Klebanov:2000hb}.}
\begin{multline}
H_3 = - \frac{M}{2} \Bigg( \left( \frac{\tau^2}{4} - \frac{\tau^4}{16} \right) d \tau \wedge g^1 \wedge g^2 + \left( \frac{1}{3} + \frac{\tau^2}{60} + \frac{\tau^4 }{1008} \right) d \tau \wedge g^3 \wedge g^4  \\
+ \left( \frac{\tau}{6} - \frac{7}{180} \tau^3 \right) g^5  \wedge (g^1 \wedge g^3 + g^2 \wedge g^4) \Bigg) ~,
\end{multline}
\begin{multline}
H_7 = - \frac{\epsilon^{8/3}}{2 \times 2^{2/3} a_0 M} \Bigg( \left(1 - \frac{\tau^2}{12} - \frac{a_2 \tau^2 }{a_0}\right) d x^0 \wedge ... \wedge dx^3 \wedge g^3 \wedge g^4 \wedge g^5 \\
+ \frac{\tau}{6} d x^0 \wedge ... d x^3 \wedge d \tau \wedge \left( g^1 \wedge g^3 + g^2 \wedge g^4 \right) + \frac{\tau^2}{12} d x^0 \wedge ... \wedge d x^3 \wedge g^1 \wedge g^2 \wedge g^5 \Bigg) ~,
\end{multline}

\begin{multline}
F_3 = \frac{M}{2} \Bigg( \left( 1 - \frac{\tau^2}{12} + \frac{7 \, \tau^4}{720} \right) g^5 \wedge g^3 \wedge g^4 + \left( \frac{\tau^2}{12}  - \frac{7 \, \tau^4}{720} \right)g^5 \wedge g^1 \wedge g^2 \\
+ \left ( \frac{\tau}{6} - \frac{7 \, \tau^3 }{180} \right) d \tau \wedge (g^1 \wedge g^3 + g^2 \wedge g^4 ) \Bigg) ~,
\end{multline}
\be
F_5 = \frac{\epsilon^{8/3}}{M^2} \left( \frac{\tau}{3 \times 3^{1/3} a_0^2 } - \frac{\tau^3}{9 \times 3^{1/3} \, a_0^2} - \frac{2 \, a_2 \, \tau^3}{3 \times 3^{1/3} \, a_0^3} \right) dx^0 \wedge dx^1 \wedge dx^2 \wedge dx^3 \wedge d \tau ~,
\ee
\begin{multline}
\label{201}
\tilde{F}_5 = \frac{\epsilon^{8/3}}{M^2} \left( \frac{\tau}{3 \times 3^{1/3} a_0^2 } - \frac{\tau^3}{9 \times 3^{1/3} \, a_0^2} - \frac{2 \, a_2 \, \tau^3}{3 \times 3^{1/3} \, a_0^3} \right) dx^0 \wedge dx^1 \wedge dx^2 \wedge dx^3 \wedge d \tau\\
-  \left( \frac{M^2 \, \tau^3}{36} \right) g^1 \wedge g^2 \wedge g^3 \wedge g^4 \wedge g^5   ~.
\end{multline}

\subsection{The Klebanov-Strassler metric near the apex in adapted coordinates}
The description of the KS throat near the apex above is in the angular coordinates $x^0$, $x^1$, $x^2$, $x^3$,
$\tau$, $\psi$, $\theta_1$, $\phi_1$, $\theta_2$, $\phi_2$ as presented in the original paper of Klebanov and Strassler. However, for our purpose, it proves useful to express the KS metric near the apex in adapted coordinates $t$, $x^1$, $x^2$, $x^3$, $r$, $\psi$, $\omega$, $\varphi$, $\theta_2$, $\phi_2$ as used in the main text\footnote{Note that the duplicate coordinates $x^1,x^2,x^3$, and $\psi$ of the two coordinates system are different. We decided not to change them to be consistent with the literature.}.

One might also wish to write the fluxes in term of the adapted coordinates. But, as the fluxes enter the blackfold equations only when coupled to the anti-D3-NS5 currents, just some components are relevant. As a result, we shall not attempt to transform the full description of the fluxes to the adapted coordinates but only the relevant components when needed.  

The Minkowskian coordinates $x^0$, $x^1$, $x^2$, $x^3$ and the radial coordinates $\tau$ of the angular coordinate system are respectively, up to some scaling, equivalent to the coordinates $t$, $x^1$, $x^2$, $x^3$, and $r$ of the adapted coordinates system. In particular, one can transform from one to the other as
\begin{align}
\label{1002}
x_0 &\rightarrow \frac{\sqrt{2} \sqrt{a_0} M}{3^{1/6} \times  \epsilon^{2/3}} \ t ~,\\
x_i &\rightarrow \frac{\sqrt{2} \sqrt{a_0} M}{3^{1/6} \times  \epsilon^{2/3}} \ x_i ~,\\
\label{1003}
\tau &\rightarrow 2 \ r  ~ .
\end{align}
Let us turn to the base of the conifold, which originally was expressed using Euler angles  $(\psi$, $\theta_1$, $\phi_1$, $\theta_2$, $\phi_2 )$, and attempt to parametrise it using  the spherical coordinates $( \psi, \omega, \varphi, \theta_2, \phi_2)$.

\paragraph{Spherical parametrisation of the deformed conifold}
For our analysis, it is most convenient to parametrise both the $S^3$ at the tip and the transverse $S^2$ using spherical coordinates, i.e. $(\psi, \omega, \varphi)$ and $(\theta_2, \phi_2)$ respectively. To do this, we shall apply the same parametrisation process as before but with an emphasis on identifying the 3 parameters of the tip $S^3$ and incorporate the remaining 2 parameters as we go up the throat. Recall from (\ref{a}) that the metric of the deformed conifold is given by
\be
\label{G}
ds^2 = \mathcal{F} tr \left( d W^{\dagger} d W \right) + \mathcal{G} | tr (W^\dagger d W) |^2
\ee
where 
\be
W = L_1 . W_0 . L_2^{\dagger}
\ee
with
\be
W_0 = \left( \begin{matrix}
0 & \varepsilon e^{\tau/2} \\
\varepsilon e^{- \tau/2} & 0
\end{matrix}\right)
\ee
and $L_j$ with $j=1,2$ are two $SU(2)$ matrices. As noted before that the coordinates $\psi_1$ and $\psi_2$ only appear in $W$ as $\psi_1 + \psi_2$, so instead of relabelling the final result, we parametrise $L_2$ with only two variables $(\theta_2, \phi_2)$
\be
L_2 = \left( \begin{matrix}
\cos \frac{\theta_2}{2} e^{i \phi_2/2} & - \sin \frac{\theta_2}{2} e^{i \phi_2/2} \\
\sin \frac{\theta_2}{2} e^{- i \phi_2/2} & \cos \frac{\theta_2}{2} e^{-i \phi_2/2} 
\end{matrix}\right) ~.
\ee 

Expanding $W_0$ in $\tau$, we have 
\be
\label{1000}
W_0 = \varepsilon f(\tau) \sigma_1 + \varepsilon  g(\tau) \sigma_2
\ee
where
\begin{align}
&\sigma_{1} = \left( \begin{matrix}
0 & 1 \\
1 & 0 
\end{matrix}\right) ~, & & \sigma_2 = \left( \begin{matrix}
0 & 1 \\
- 1 & 0 
\end{matrix}\right) ~,
\end{align}
and
\begin{align}
f(\tau) &=  1 + \frac{\tau^2}{8} + \frac{\tau^4}{384} + \mathcal{O} \left( \tau^6 \right) ~, &
g(\tau) &=  \frac{\tau}{2} + \frac{\tau^3}{48} + \mathcal{O} \left( \tau^5 \right) ~.
\end{align}
Thus, we have
\begin{align}
W &= L_1 . \big( \varepsilon f(\tau) \sigma_1 + \varepsilon  g(\tau) \sigma_2 \big) . L_2^\dagger \\
&= \varepsilon f(\tau) L + \varepsilon g(\tau) L . \hat{L}   
\end{align}
where $L \equiv L_1 . \sigma_1 . L_2^\dagger$ and $\hat{L} \equiv L_2 . (\sigma_1)^{-1} . \sigma_2 . L_2^\dagger$ .

As $L$ is an unitary complex matrix with $\det L = - 1$, we can parametrise $L$ using spherical coordinates as\footnote{To obtain the deformed conifold metric, it is algebraically simpler to write the matrix $L$ in Hopf coordinates first, carry out the necessary computations, then transform Hopf to spherical. Nevertheless, the final answers are the same.}
\be
L = \left( \begin{matrix}
- \sin \psi \sin \omega \cos \varphi + i \sin \psi \sin \omega \sin \varphi & \cos \psi - i \sin \psi \cos \omega \\
\cos \psi + i \sin \psi \cos \omega & \sin \psi \sin \omega \cos \varphi + i \sin \psi \sin \omega \sin \varphi 
\end{matrix}\right) ~.
\ee
On the other hand, the parametrisation of $\hat{L}$ comes directly from the parametrisation of $L_2$. We have
\be
\hat{L} = \left( \begin{matrix}
- \cos \theta_2 & - e^{i \phi_2} \sin \theta_2 \\
- e^{- i \phi_2} \sin \theta_2 & \cos \theta_2
\end{matrix}\right) ~.
\ee
Plugging the spherically parametrised $W$ into (\ref{G}), we obtain the metric of the deformed conifold in spherical coordinates.

\paragraph{Klebanov-Strassler metric near the apex in adapted coordinates} Recall from \cite{Klebanov:2000hb}, the KS metric is given by
\be
\label{BB49}
ds_{10}^2 = h^{-1/2} (\tau) \left( - d x_0^2 + d x_1^2 + dx_2^2 + dx_3^2 \right) + h^{1/2} (\tau) ds_6^2
\ee
where $ds_6^2$ is the metric of the deformed conifold and the $h(\tau)$ is the warping effects induced by the non-trivial fluxes:
\begin{align}
\label{A44}
h (\tau) &= M^2 \, 2^{2/3} \epsilon^{-8/3} \int_\tau^\infty dx \frac{x \coth x - 1}{\sinh^2 x} (\sinh 2 x - 2 x)^{1/3}\\
&= M^2 2^{2/3} \epsilon^{-8/3} \ ( a_0 + a_2 \tau^2  + a_4 \tau^4)  + \mathcal{O} (\tau^6)
\end{align}
where, as written down earlier, $a_0 \approx 0.71805$, $a_2 = - (3 \times 6^{1/3})^{-1} $, and $a_4 = (18 \times 6^{1/3})^{-1}$.

Substituting in \eqref{BB49} the spherically parametrised deformed conifold metric, applying the coordinate transformations (\ref{1002})-(\ref{1003}), and restricting our attention to some leading orders of $r$, we obtain the expression of the KS metric near the apex in our desired adapted coordinates. However, as the expression is long and ugly, we shall not write it explicitly here. Instead, we shall only write down components/properties that are immediately relevant for us.

Firstly, as you would expect, if we subdue terms of order $r^2$ or higher in all but the $(\theta_2, \phi_2)$ directions, we recover the metric in (\ref{ex3}):
\begin{multline}
\label{bb51}
g_{\mu \nu} d x^\mu d x^\nu = M b_0^2  \Big ( - dt^2 + (dx^1)^2 + (dx^2)^2 + (dx^3)^2  + dr^2 \\
+ d \psi ^2 + \sin^2 \psi \left( d \omega^2 + \sin^2 \omega d \varphi^2 \right)  + r^2 (d \theta_2^2 +  \sin^2 \theta_2 d \phi_2^2) \Big)
\end{multline}
where $b_0^2= \frac{2^{2/3} \sqrt{a_0}}{3^{1/3}} \approx 0.93266$.

Secondly, as they will be relevant for our stability analysis, we note the following derivatives
\begin{align}
&\p_r^2 g_{tt}\Big|_{r = \theta_2 = \phi_2 = 0}  = \frac{4 \times 2^{2/3} a_2 M}{3^{1/3} \sqrt{a_0}}  ~, &  &\p_r^2 g_{x^i x^i}\Big|_{r = \theta_2 = \phi_2 = 0} =  - \frac{4 \times 2^{2/3} a_2 M}{3^{1/3} \sqrt{a_0}} ~,
\end{align}
\be
\p_r^2 g_{\omega \omega} \Big|_{r = \theta_2 = \phi_2 = 0}  = \frac{ 4 \times 2^{2/3} M }{5 \times 3^{1/3} \sqrt{a_0}} \sin ^2 \psi  \Big(  4 a_0 + 5 a_2 - 2 a_0 \cos^2 \psi \sin^2 \omega \Big) ~,
\ee
\be
\p_r^2 g_{\varphi \varphi} \Big|_{r = \theta_2 = \phi_2 = 0}  =
\frac{4 \times 2^{2/3} M }{5 \times 3^{1/3} \sqrt{a_0}} \sin ^2\psi \sin ^2\omega  \left(4 a_0+5 a_2 - 2 a_0 \sin^2 \psi \sin ^2 \omega \right)
\ee
with $a_0 \approx 0.71805$ and $a_2 = - (3 \times 6^{1/3})^{-1} $.



\section{The Cvetic-Gibbons-Lu-Pope throat}
\label{CON2}
Analogous to the KS throat in type IIB supergravity, the CGLP throat is a warped, fluxed, deformed conifold solution of the eleven-dimensional supergravity theory. As the discussion of the CGLP throat has many overlaps with that of the KS throat, we include only the minimal details here. The eleven-dimensional background metric has the form
\be
\label{setupaa}
ds_{11}^2 = g_{\mu\nu} dx^\mu dx^\nu = H^{-\frac{2}{3}} \left( - (dx^0)^2 +  (dx^1)^2 +  (dx^2)^2 \right)
+ H^{\frac{1}{3}} ds_8^2
~,
\ee
where $ds_8^2$ is the metric element of the Stenzel space. Details on the full structure of this metric and the function $H$ can be found in \cite{Cvetic:2000db,Klebanov:2010qs}. The background also involves a non-trivial profile for the four-form field strength $G_4$ and its Hodge dual $G_7 = \star_{11} G_4$. It is convenient to collect the following constants that appear in this background (we follow closely the notation in \cite{Klebanov:2010qs}) 
\begin{itemize}

\item A constant $m$ appears in the expressions of $G_4$ and $G_7$ and is related to the $\tilde{M}$ units of $G_4$ flux through an $S^4$ of the background via the relation
\be
\label{setupab}
\tilde{M} = \frac{18\pi^2 m}{(2\pi \ell_P)^3}
~,
\ee
where $\ell_P$ is the eleven-dimensional Planck length.

\item The complex structure deformation of the four-complex dimensional conifold that gives rise to the Stenzel space is expressed in terms of the constant $\varepsilon$. In complex coordinates $z_i$ in $\IC^5$ the Calabi-Yau space $\sum_{i=1}^5 z_i^2 = 0$ is deformed to  $\sum_{i=1}^5 z_i^2 = \varepsilon^2$.

\item At the tip of the cone the value of the function $H$ is 
\be
\label{setupac}
 \hat H_0 \frac{m^2}{\epsilon^{\frac{9}{2}}} \simeq 1.0898\, \frac{m^2}{\epsilon^{\frac{9}{2}}}
~.
\ee

\item For simplicity, it is helpful to define and use the related constants:
\be
\label{setupad}
a_0^2 = \left( \frac{m^2}{\epsilon^{\frac{9}{2}}} \hat H_0 \right)^{-\frac{2}{3}}~, ~~
b_0^2 = \frac{3}{2} \hat H_0^{\frac{1}{3}}
~.
\ee

\end{itemize}

Anti-M2 branes placed in the CGLP background are attracted towards and eventually stabilise at the tip of the eight-dimensional cone. Hence, for our purposes, it is enough to focus on the tip of the conifold ($\tau = 0$ in the appropriate radial coordinate $\tau$ \cite{Klebanov:2010qs}). After a trivial rescaling of the Minkowski coordinates $x^0, x^1, x^2$ by the constant factor $a_0 /({b_0 m^{1/3}})$ one obtains the metric\footnote{As can be analogously seen from setting $r=0$ in the description of the KS metric \eqref{bb51}, at the tip of the CGLP throat, only the $S^4$ of the conical base space remains.}
\be
\label{setupae}
ds^2 = m^{\frac{2}{3}} b_0^2 \left( - (dx^0)^2 +  (dx^1)^2 +  (dx^2)^2 + d\psi^2 + \sin^2\psi \, d\Omega_3^2 \right)
~.
\ee
   We will use spherical coordinates $\vartheta, \omega, \varphi$ to express the metric element of the unit round $S^3$ as $d\Omega_3^2 = d\vartheta^2 + \sin^2\vartheta \left( d\omega^2 + \sin^2\omega \,d\varphi^2 \right)$. The four-form flux $G_4=d A_3$ is given in terms of the gauge field
\be
\label{setupaf}
A_3 = \frac{27}{4} m f(\psi) \sin^2 \vartheta \sin\omega \, d\vartheta \wedge d\omega \wedge d\varphi
\ee
where
\be
\label{setupag}
f(\psi) = \frac{1}{3} \cos^3 \psi - \cos\psi + \frac{2}{3}
~,
\ee
while the seven-form flux $G_7$ takes the form
\be
\label{setupai}
G_7 = - \frac{27}{4} m^2 b_0^3 \sin^3\psi  \sin^2\vartheta \sin\omega \, dx^0 \wedge dx^1 \wedge dx^2 \wedge d\psi \wedge d\vartheta \wedge d\omega \wedge d\varphi
~.
\ee

\chapter{Brane bound states}\label{BS}
In this appendix, we describe the D3-NS5 and the M2-M5 brane bound states, focusing on aspects relevant to their role in the main text as the seed solutions for the metastable antibranes. We begin, in section \ref{BS1}, by obtaining the supergravity description of the D3-NS5 bound state via S-dualising the D3-D5 solution. Subsequently, we collect its thermodynamics information, and construct its far-zone equivalent currents. As the construction of equivalent currents is an important component of the blackfold approach, we will take our time with such discussions. In section \ref{C2}, we describe the M2-M5 brane bound state and its equivalent currents. 
As there are obvious overlaps, the discussion of the M2-M5 bound state will be less detailed. However, it is still a worthwhile read because the M2-M5 equivalent currents possess an unique and important feature that one might naively miss.

\section{D3-NS5 bound state \& its equivalent currents} \label{BS1}
\subsection{D3-NS5 from S-dualising D3-D5 }\label{D3-NS5Sugra}
\paragraph{D3-D5 bound state} The familiar D3-D5 brane bound state supergravity solution is given by \cite{Harmark:1999rb}\footnote{Let us note that our description is slightly different from that of \cite{Harmark:1999rb}. Nevertheless, one can easily check through direct substitution that this description is indeed the correct description with respect to our conventions stated in section \ref{IN4}.}
\be
ds^2 = H^{-1/2} \left( -f dt^2 + D \left( (dx^1)^2 + (dx^2)^2 \right) + \sum_{i=3}^5 (dx^i)^2 \right)
+ H^{1/2} \left( f^{-1} dr^2 +r^2 d\Omega_3^2 \right)
~ ,
\ee
\begin{align} 
e^{2 \phi} &= H^{-1} D ~ ,\\
B_2 &= \tan\theta (H^{-1} D -1) dx^1 \wedge dx^2 ~ ,\\
C_2 &=  - r_0^2 \sinh 2 \alpha \, \cos \theta  \ \varphi \sin^2 \psi \sin \omega \, d \psi \wedge d \omega ~ ,\\
C_4 &= (H^{-1}-1) \coth\alpha \, \sin\theta ~ dt \wedge dx^3 \wedge d x^4 \wedge dx^5 -  \left( \frac{r^2}{r_0^2 \sinh^2 \alpha \cos^2 \theta} + 1 \right) B_2 \wedge C_2
\end{align}
with
\begin{align}
f &= 1 - \frac{r^2_0}{r^2} ~ , & D = \left( \sin^2 \theta H^{-1} + \cos^2 \theta \right) ^{-1} ~ ,\\
 H &=1 + \frac{r_0^2 \sinh^2 \alpha}{r^2} 
\end{align}
where $d \Omega_3^2$ is the standard $S^3$ metric $d \Omega_3^2 = d \psi^2 + \sin^2 \psi \left( d \omega^2 + \sin^2 \omega d \varphi^2 \right)$.

\paragraph{D3-NS5 bound state via S-duality} 
In the scope of supergravity, S-duality can be described as a symmetry of the type IIB supergravity action. By S-dualising, we mean making use of such symmetry to transform one solution, here the D3-D5 bound state, to another solution, here the D3-NS5 bound state. For our purposes, in the string frame, the transformation can be described as 
\begin{align}
e^{\phi^{\text{NS5}}} &= \frac{1}{e^{\phi^{\text{D5}}}} ~ , & ds_2^{\text{NS5}} &=  e^{-\phi^{\text{D5}}} ds_2^{\text{D5}} ~ ,
\end{align}
\begin{align}
H_{3}^{\text{NS5}} &= F_3^{\text{D5}}  ~ ,& F_3^{\text{NS5}} &= - H_3^{\text{D5}} ~ ,& \tilde{F}_5^{\text{NS5}} &= \tilde{F}_5^{\text{D5}} ~ .
\end{align}

As a result, we have the D3-NS5 bound state given by
\begin{equation}\label{3}
ds^2 = D^{-1/2} \left( -f dt^2 + D \left( (dx^1)^2 + (dx^2)^2 \right)+ \sum_{i=3}^5 (dx^i)^2 \right)+ H D^{-1/2} \left( f^{-1} dr^2 + r^2 d\Omega_3^2 \right) ~ ,
\end{equation}
\begin{align} 
e^{2 \phi} &= H D^{-1} ~ ,\\
C_2 &= - \tan \theta (H^{-1} D - 1) \, d x^1 \wedge d x^2 ~ ,\\
B_2 &= - r_0^2 \sinh 2 \alpha \, \cos \theta  \ \varphi \sin^2 \psi \sin \omega \, d \psi \wedge d \omega ~ ,\\
\label{4}
C_4 &= (H^{-1}-1) \coth\alpha \, \sin\theta ~ dt \wedge dx^3 \wedge d x^4 \wedge dx^5 + \frac{r^2}{r_0^2 \sinh^2 \alpha \cos^2 \theta} B_2 \wedge C_2 ~ .
\end{align}
As one can easily check against the type IIB supergravity equations, this is indeed a solution. In this description, the extremal D3-NS5 solution can be obtained by taking the limit $r_0 \rightarrow 0, \alpha \rightarrow \infty$ in such a way that we can define a finite extremal horizon radius $r_h \equiv r_0 \sinh \alpha$. As such, $r_0$ and $\alpha$ can be thought of as nonextremal parameters. 

\paragraph{Thermodynamics} As computed in \cite{Harmark:1999rb}, the thermodynamics of this solution are
\begin{align}
\label{bs2}
&\varepsilon = \frac{\Omega_3}{16\pi G} r_0^2 \left( 3 + 2 \sinh^2 \alpha \right) ~ ,& &s = \frac{\Omega_3}{4G} r_0^3 \cosh\alpha ~ ,& &\mathcal{T} = \frac{1}{2\pi r_0 \cosh\alpha} ~ ,
\end{align}
\begin{align}
\label{0010}
&\Phi_3 = \sin\theta\, \tanh\alpha ~ ,& & \mathcal{Q}_3 = \frac{\Omega_3}{8\pi G} r_0^2\, \sin\theta \, \sinh\alpha \, \cosh\alpha ~ , \\
&\Phi_5 = \cos\theta\, \tanh\alpha ~ , & & \mathcal{Q}_5 = \frac{\Omega_3}{8\pi G} r_0^2\, \cos\theta \, \sinh\alpha \, \cosh\alpha
\end{align}
where $\Omega_3=2\pi^2$ is the volume of the unit radius round $S^3$. And, the effective energy stress tensor is given by \cite{Emparan:2011hg}
\be
T_{ab} = \mathcal{T} s \left( u_a u_b - \frac{1}{n} \gamma_{ab} \right) - \sum_{q \, = \, 3, 5} \Phi_q \mathcal{Q}_q h_{ab}^{(q)}  ~ .
\ee

\subsection{Far-zone equivalent currents}
As discussed in \cite{Marolf:2000cb}, there are at least three sensible notions of charges in a supergravity theory. For the purpose of constructing equivalent currents, we shall be interested in something called the Maxwell charge. The key idea for the Maxwell charges is that the Chern-Simons terms in the equation of motion can be thought of as a source for the gauge field. For example, let us look at the equation of motion for the $C_4$ gauge field in type IIB supergravity: 
\be
d \star \tilde{F}_5 - H_3 \wedge F_3 = - 16 \pi G \star J_4 ~ .
\ee
In this case, the Maxwell current is given by
\be
d \star \tilde{F}_5 = - 16 \pi G \star J_{4}^{Maxwell} = - 16 \pi G \star J_4 + H_3 \wedge F_3
\ee
where the sign and factors in front of $J_4^{Maxwell}$ is to make sure it is compatible with our conventions of $J_4$. 

The Maxwell current $J_4^{Maxwell}$ can be interpreted as modelling how the branes interact with the background flux $\tilde{F}_5$. The Maxwell charge can be computed easily from Gauss's law of the $\tilde{F}_5$ flux and, thus, can be interpreted as the monopole source that will reproduce the $\tilde{F}_5$ flux far away. As we shall see explicitly soon, this is what we need for the construction of equivalent currents.

Turning our attention to the case of D3-NS5 bound state, the relevant forced Maxwell equations are
\begin{align}
d \star \tilde{F}_3 &= - 16 \pi G \star J_2^{Maxwell} ~ ,\\
d \star \tilde{F}_5 &= -16 \pi G \star J_4^{Maxwell} ~ ,\\
d \star H_7 &= 16 \pi G \star j_6^{Maxwell}  ~ .
\end{align}
We do not know the exact expressions of these Maxwell currents, however, we can mimic their effects far away by using Maxwell charges to construct a set of equivalent currents. Adopting the convention that $Q = \int \star J$, using the description of the D3-NS5 bound state in (\ref{3})-(\ref{4}), we obtain the Maxwell charges
\begin{align}
Q_1^{Maxwell}&= - \frac{1}{16 \pi G} \lim_{r \rightarrow \infty} \int \star \tilde{F}_3  = Vol_4 \ C r_0^2 \sinh^2 \alpha \sin \theta \cos \theta ~ ,\\
\label{7}
Q_3^{Maxwell}&= - \frac{1}{16 \pi G} \lim_{r \rightarrow \infty} \int \star \tilde{F}_5   = Vol_2 \ C  r_0^2 \sinh \alpha \cosh \alpha \sin \theta ~ ,\\
Q_5^{Maxwell}&= \frac{1}{16 \pi G} \lim_{r \rightarrow \infty} \int  \star H_7 = - C r_0^2 \sinh \alpha \cosh \alpha \cos \theta 
\end{align}
where $C = \displaystyle \frac{\Omega_3}{8 \pi G} =  \frac{\pi}{4 G}$ and $Vol_n$ is the volume of the n-dimensional flat space. Requiring that our equivalent currents reproduce the same Maxwell charges at $r \rightarrow \infty$, they can now be easily constructed:\footnote{The equivalent currents are localised ($\delta$ function) currents in the full 10 dimensional picture.}
\begin{align}
J_2^{equiv} &= C r_0^2 \sinh^2 \alpha \sin \theta \cos \theta \ v \wedge w ~ , \\
J_4^{equiv} &=   C r_0^2 \sinh \alpha \cosh \alpha \sin \theta \ * ( - v \wedge w) ~ , \\
j_6^{equiv} &= - C r_0^2 \sinh \alpha \cosh \alpha \cos \theta \ * (- 1)
\end{align}
where $*$ is the 6-dimensional worldvolume Hodge star, and $v, w$ are orthogonal vectors used to describe the distribution of the dissolved D3 charge. 

In the description of D3-NS5 branes above, we have not restricted the range of $\theta \in (0, 2 \pi)$. For the construction of KPV state, we are interested in anti-D3-NS5 branes, which corresponds to the range $\theta \in (\pi, 3 \pi/4)$ of our description\footnote{The statement that anti-D3-NS5 branes are described by $\theta$ in the regime of  $(\pi, 3 \pi/4)$ is only strictly true for background where Maxwell charges and Page charges are the same. As we shall see in our study of the KPV state, the metastable state of anti-D3-NS5 branes can have $\theta$ outside of this regime.}. For convenience, we can do a reparametrisation $\theta \rightarrow \theta - \pi$ to bring it to the regime $\theta \in (0, \pi/2)$. In the new $\theta$, our currents are given by
\begin{align}
J_2 &= C r_0^2 \sinh^2 \alpha \sin \theta \cos \theta \ v \wedge w ~ , \\
J_4 &=  C r_0^2 \sinh \alpha \cosh \alpha \sin \theta \ * (  v \wedge w) ~ , \\
j_6 &= - C r_0^2 \sinh \alpha \cosh \alpha \cos \theta \ * ( 1) 
\end{align}
where we have drop the superscript $equiv$ for syntactical simplicity. 

\section{M2-M5 bound state and its equivalent currents}
\label{C2}
\subsection{M2-M5 bound state}
Following \cite{Harmark:1999rb}, we can easily read off the description for the M2-M5 brane bound state and its thermodynamics quantities. In flat space, the M2-M5 branes has the metric
\begin{multline}
ds^2 = (HD)^{-1/3} \left[ - f dt^2 + (dx^1)^2 + (dx^2)^2 + D \left( (dx^3)^2 + (dx^4)^2 + (dx^5)^2 \right) \right] \\
+ H \left(f^{-1} dr^2 + r^2 d\Omega^2_4 \right)
\end{multline}
where 
\begin{align}
f &= 1 - \frac{r_0^3}{r^3} ~ ,& H &= 1 + \frac{r_0^3 \sinh^2 \alpha}{r^3} ~ ,\\
D &= \left( \sin^2 \theta H^{-1} + \cos^2 \theta \right)^{-1} ~ .
\end{align}
The gauge fields are given by
\be
\label{bs5}
A_3 = - \sin \theta \coth \alpha (H^{-1} -1 ) dt \wedge dx^1 \wedge dx^2 + \tan \theta D H^{-1} dx^3 \wedge dx^4 \wedge dx^5  ~ , 
\ee
\be
A_6 = \cos \theta  \coth \alpha D (H^{-1} -1 ) dt \wedge dx^1 ... \wedge dx^5 ~ .
\ee
The thermodynamics of this solution are 
\begin{align}
\label{bs20}
\varepsilon &= \frac{\Omega_4}{16 \pi G} r_0^3 \left(4 + 3 \sinh^2 \alpha \right) ~ , & \mathcal{T} &= \frac{3}{4 \pi r_0 \cosh \alpha} ~ , &
\end{align}
\begin{align}
s &= \frac{\Omega_4}{4 G} r_0^4 \cosh \alpha ~ ,& \Phi_5 &= \cos \theta \tanh \alpha ~ ,& \Phi_2 &= - \sin \theta \tanh \alpha ~ ,
\end{align}
\begin{align}
\label{bs6}
\mathcal{Q}_5 &=  \frac{3 \Omega_4}{16 \pi G} \cos \theta  r_0^3 \sinh \alpha \cosh \alpha ~ , &
\mathcal{Q}_2 &= -  \frac{ 3 \Omega_4}{16 \pi G}  \sin \theta r_0^3\sinh \alpha \cosh \alpha 
\end{align}
where $\Omega_4$ is the volume of the unit 4-sphere $S^4$. A corresponding free energy $\mathcal{F}$ can be defined as
\be
\label{forceaed}
\mathcal{F} = \varepsilon - \mathcal{T} s = \frac{\Omega_4}{16 \pi G} r_0^3 (1 + 3 \sinh^2 \alpha)
\ee
and the worldvolume energy-stress tensor as
\be
T_{ab} = \mathcal{T} s \left( u_a u_b - \frac{1}{3} \eta_{ab} \right) - \sum_{q= 2,5} \Phi_q \mathcal{Q} h^{(q)}_{ab}
\ee
where $h_{ab}$ are the projector on to the M2 and M5 branes. 

\subsection{Far-zone equivalent currents}
Following the same procedure as before, we can easily write down an expression for the M2-M5 equivalent currents. The feature of interest is that, in this case, $J_3$ has two legs. This is due to the fact that $A_3$ \eqref{bs5} has two legs, one along the M2 branes and the other orthogonal to it. When trying to mimic the effects of the M2-M5 bound state on the $G_4$ flux\footnote{Think of the required form of $J_3$ for which the sourced eleven-dimensional supergravity gauge field equation $d \star G_4 = - 16 \pi G  \star J_3$ produces a field strength $G_4 = d A_3$ where $A_3$ has 2 legs.}, we need to take this into account. As a result, besides the leg along the familiar directions of the M2 branes, $J_3$ also has an extra leg in the orthogonal directions. In order to write down an expression for this extra leg, it is useful to compute a charge-like quantity $\tilde{\mathcal{Q}}_2$ from an integral of the four-form field strength $G_4$ over an $S^7$ that surrounds the ``orthogonal'' leg of the field strength:
\be
\label{bs7}
\tilde{\mathcal{Q}}_2 = -\frac{1}{16\pi G} \int_{S^7} \star G_4 = C r_0^3 \sin\theta \cos\theta \sinh^2\alpha ~ .
\ee
where $C= \frac{3 \Omega_4}{16 \pi G} = \frac{\pi}{2 G}$. With the expressions of the charges $\mathcal{Q}_2$, $\mathcal{Q}_5$ \eqref{bs6}, and the charge-like quantity  $\tilde{\mathcal{Q}}_2$ \eqref{bs7}, we have the far-zone equivalent currents of the M2-M5 bound state:
\begin{equation}
J_3 =  \CC r_0^3 \sin\theta \sinh\alpha \bigg[ \cosh\alpha * (v \wedge w \wedge z ) 
- \cos\theta \sinh\alpha \, v \wedge w \wedge z \bigg] ~ ,
\end{equation}
\begin{equation}
\mathcal{J}_6 = - C \cos \theta \,  r_0^3 \sinh \alpha \cosh \alpha \, ( * 1 )
\end{equation}
where $*$ is the 6-dimensional worldvolume Hodge star, and $v$, $w$, $z$ are orthogonal vectors used to describe the distribution of the dissolved M2 charge.

\chapter{Blackfold perturbation equations} \label{secB}

In this appendix, we derive the blackfold perturbation equations for deformations around the KPV state. We start with the computations of some useful variational expressions. Subsequently, we present the derivation of the blackfold perturbation equations used in the main text. For further discussions on variational properties of embedding geometry or blackfold perturbation equation, see e.g. \cite{Armas:2017pvj,Armas:2019iqs}.

\section{Useful variations}

\paragraph{Variation of induced metric} Hitting $\delta$ to the definition of $\gamma_{ab}$ in (\ref{27}), we obtain the expression
\be
\delta \gamma_{a b} = \p_a X^\mu \p_b X^\nu \Big( \nabla_\mu \left( \delta X^\alpha g_{\alpha \nu} \right) + \nabla_\nu \left( \delta X^\alpha g_{\alpha \mu} \right) \Big) ~ .
\ee
When we embed a surface without edges in a higher-dimensional background, the variations along the brane directions of the embedding functions $X^\mu (\sigma)$ can be cancelled by a reparametrisation of the worldvolume coordinates. As a result, we only have to worry about the variations of the transverse scalars $\delta X^\mu_{\perp} (\sigma)$ (i.e. $\p^a X_\mu \delta X^\mu_\perp = 0$). Making use of equation (\ref{28}), we have
\be
\label{d1}
\delta \gamma_{ab} = - 2 K_{ab}^{\,\,\, \,\,\, \rho} \left( \delta X^\alpha_{\perp} g_{\alpha \rho} \right) ~ .
\ee
Using the identity $\gamma_{ab} \gamma^{bc} = \delta_a^c$, we can easily deduce that
\be
\delta \gamma^{ab} = 2 K^{ab}_{\,\,\, \,\,\, \rho} \delta X^\rho_{\perp} ~ . 
\ee

\paragraph{Variation of normal vectors} We note that the normal vectors are implicitly defined by
\begin{align}
\p_a X^\rho n_\rho^{(i)} &= 0 ~ ,\\
n_\rho^{(i)} n^\rho_{(j)} &= \delta^{(i)}_{\ (j)} ~ .
\end{align}
Hitting $\delta$ to both equations yields respectively the variation of $n^{(i)}_\rho$ along the worldvolume directions and normal to the worldvolume directions\footnote{As normal vectors are used collectively to specify the position of the branes inside the background, it is obvious that we have a rotational gauge symmetry in defining these vectors. Therefore, we can safely ignore variations regarding rotations of the normal vectors among themselves.}: 
\begin{align}
h^\rho_\sigma \, \delta n_\rho^{(i)} &= - \p^a X_\sigma \p_a \delta X^\rho_\perp n_\rho^{(i)} ~ , \\
\perp^{\rho}_\sigma \delta n_{\rho}^{(i)} &=  \frac{1}{2} n^{\alpha \, (i)} n^{\beta \, (i)} \p_\gamma g_{\alpha \beta} \delta X^\gamma_\perp n_\sigma^{(i)} ~ .
\end{align}
All together, we have 
\be
\label{30}
\delta n^{(i)}_\rho = - \p^a X_\rho \p_a \delta X^\sigma_\perp n_\sigma^{(i)} + \frac{1}{2} n^{\alpha \, (i)} n^{\beta \, (i)} \p_\gamma g_{\alpha \beta} \delta X^\gamma_\perp n_\rho^{(i)} ~ .
\ee

\paragraph{Variation of extrinsic curvature} Hitting $\delta$ to the expression of $K_{ab}^{\ \ \rho}$ in (\ref{28}), we obtain
\begin{equation}
\label{31}
\delta K_{ab}^{\,\,\, \,\,\, \rho} = \nabla_a \left( \p_b \delta X^\rho_\perp  \right) - \delta \Theta^c_{ab} \p_c X^\rho + \delta \Gamma^{\rho}_{\mu \nu} \p_a X^\mu \p_b X^\nu + 2 \Gamma^\rho_{\mu \nu} \p_a \delta X^\mu_\perp \p_b X^\nu ~ .
\end{equation}
Considering the variation of the projected extrinsic curvature $K_{ab}^{\ \ \, (i)}$, we have
\be
\delta \left( K_{ab}^{\,\,\, \,\,\, (i)} \right) = \delta \left( K_{ab}^{\,\,\, \,\,\, \rho} n_\rho^{(i)} \right) = \delta \left( K_{ab}^{\,\,\,\,\,\, \rho}  \right) n_{\rho}^{(i)} + K_{ab}^{\,\,\,\,\,\, \rho} \delta \left( n_\rho^{(i)} \right) ~ .
\ee
Making use of results in (\ref{30}) and (\ref{31}), we can write 
\begin{multline}
\label{42}
\delta\left( K_{ab}^{\,\,\, \,\,\, (i)}  \right) = n_\rho^{(i)} \nabla_a \left( \p_b \delta X^\rho_\perp \right) + n^{(i)}_\rho \delta X^\alpha_\perp \p_\alpha \Gamma^\rho_{\mu \nu} \p_a X^\mu \p_b X^\nu + 2 \, n_\rho^{(i)}  \Gamma^\rho_{\mu \nu} \p_a \delta X^\mu_\perp \p_b X^\nu \\
+ \frac{1}{2} K_{ab}^{\,\,\,\,\,\, \rho} \left( n^{\alpha \, (i)} n^{\beta \, (i)} \p_\gamma g_{\alpha \beta} \delta X^\gamma_\perp n_\rho^{(i)} \right) ~ .
\end{multline}

\paragraph{Variation of anti-D3-NS5 blackfold energy-momentum tensor} Hitting $\delta$ to the expression of $T^{ab}$ in (\ref{ex5}), we obtain the expression
\begin{multline}
\label{43}
\delta T^{a b} = - \mathbb{Q}_5 \sin \theta \delta ( \tan \theta )  \gamma^{ab} - \mathbb{Q}_5 \frac{1}{\cos\theta} \left( 2 K^{ab}_{\ \ \, \rho} \delta X^\rho_\perp \right) \\
+ \mathbb{Q}_5  \Big( \delta (v^a) v^b + v^a \delta (v^b) + \delta(w^a) w^b + w^a \delta (w^b) \Big) \tan \theta \sin \theta \\
+ \mathbb{Q}_5 (v^a v^b + w^a w^b) \sin \theta \delta  ( \tan \theta ) +  \mathbb{Q}_5 (v^a v^b + w^a w^b) \sin \theta \cos^2 \theta \delta (\tan \theta) ~ .
\end{multline}
We can also provide the general expressions for the variations of the blackfold currents. However, as the blackfold currents either enter our equations with a Hodge dual or coupled to the background fluxes, let us write down only the needed components when we use them.

\section{Current conservation equations}
Recall from (\ref{34})-(\ref{35}) the blackfold current conservation equations
\begin{align}
d * j_6 &= 0 ~ ,\\
d * J_4 - * j_6 \wedge F_3 &= 0 ~ ,\\
d * J_2 + H_3 \wedge * J_4 &= 0  ~ .
\end{align}
\begin{enumerate}
\item Considering the $j_6$ conservation equation, we can easily show that it gives rise to the perturbation equation
\be 
\label{A}
\p_a \delta \mathbb{Q}_5 = 0
\ee
where we have used $ * j_6 = \mathbb{Q}_5$.

\item Considering the $J_4$ conservation equation, firstly, we note that it can be rewritten as
\be
d * \tilde{J}_4 = 0
\ee 
where
\begin{align}
* \tilde{J}_4 &= * J_4 - *j_6 \wedge C_2\\
&=  - C r_h^2 \sin \theta \, v \wedge w - C r_h^2 \cos \theta \, C_2 ~ .
\end{align}
From the unitary condition $v^a v_a = w^a w_a = 1$, it can be easily shown that
\begin{align}
&\delta v_\omega = \sqrt{M} b_0 \cos \psi ~ , & &\delta w_\varphi =  \sqrt{M} b_0 \cos \psi \sin \omega ~ .
\end{align}
Therefore, we have
\begin{multline}
\delta \left( * \tilde{J}_4 \right) = - \mathbb{Q}_5 \delta \tan \theta \, v \wedge w - \mathbb{Q}_5 \tan \theta \left( \delta v \wedge w + v \wedge \delta w \right) - \mathbb{Q}_5 \, \delta C_2 \\
= -\Bigg( \mathbb{Q}_5 M b_0^2 \sin^2 \psi \delta \tan \theta + 2 \mathbb{Q}_5 M b_0^2 \tan \theta \cos \psi \sin \psi \delta \psi + 2  \mathbb{Q}_5 M \sin^2 \psi \delta \psi \Bigg)  \sin \omega d \omega \wedge d \varphi \\
- \Big(\mathbb{Q}_5 \tan \theta \sqrt{M} b_0 \sin \psi \delta w_t\Big) d \omega \wedge dt  - \Big( \mathbb{Q}_5 \tan \theta \sqrt{M} b_0 \sin \psi  \sin \omega \delta v_t \Big) dt \wedge d \varphi 
\end{multline}
where we have used that $C_2$ at the tip is given by $C_2 = M (\psi - \frac{1}{2} \sin 2 \psi) \sin \omega d \omega \wedge d \varphi$ and corrections away from the tip start at order $\mathcal{O} \left( r^2 \right)$. Thus, the $J_4$ perturbation equation is given by
\begin{multline}
\label{B}
- \mathbb{Q}_5 M b_0^2 \sin^2 \psi \sin \omega
\Bigg(  \p_t \delta \tan \theta + 2 \tan \theta \cot \psi \p_t \delta \psi + \frac{2}{b_0^2}  \p_t \delta \psi \Bigg) \\
=  \mathbb{Q}_5  M^{3/2} b^3_0 \tan \theta  \sin \psi \Big( \p_\varphi \delta w^t + \p_\omega  \left( \sin \omega \delta v^t \right) \Big)
\end{multline}
where we have used $\delta v_t = - M b_0^2 \, \delta v^t$ and $\delta w_t = - M b_0^2 \, \delta w^t$.

\item Considering the $J_2$ conservation equation, we have the variation of $* J_2$ is given by 
\begin{multline}
\delta \big( * J_2 \big) =  \mathbb{Q}_5 \left( \delta \sin \theta \right) * (v \wedge w)+ \mathbb{Q}_5 \sin \theta \delta \left( * (v \wedge w) \right) \\
= \mathbb{Q}_5 \left( \cos^3 \theta \delta \tan \theta \right) \sqrt{-\gamma} \Big(v^\omega w^\varphi d t \wedge ... \wedge dx_3  \Big) - 2 \mathbb{Q}_5 \sin \theta ( \sqrt{-\gamma} \gamma_{\omega \omega} K^{\omega \omega}_{\ \ \, \psi} \delta \psi) \Big(v^\omega w^\varphi d t \wedge ... \wedge dx_3  \Big)\\
+ \mathbb{Q}_5 \sin \theta \sqrt{-\gamma} \Big( \delta v^t w^\varphi d x_1 \wedge ... \wedge dx_3 \wedge d \omega + v^\omega \delta w^t d x_1 \wedge ... d x_3 \wedge d \varphi\\
+ \delta v^\omega w^\varphi d t \wedge ... \wedge dx_3 + v^\omega \delta w^\varphi d t \wedge ... \wedge dx_3   \Big) ~ .
\end{multline}

As $\delta \left( H_3 \wedge * J_4 \right) = \delta H_3 \wedge * J_4 + H_3 \wedge \delta (* J_4) = 0$, the $J_2$ perturbation equation is equivalent to the set of equations
\begin{align}
\label{C}
\cot \theta \cos^2 \theta \p_\omega \delta \tan \theta + \sqrt{M} b_0 \sin \psi \p_t \delta v^t  &= 0 ~ ,\\
\cot \theta \cos^2 \theta  \p_\varphi \delta \tan \theta +  \sqrt{M} b_0 \sin \psi \sin \omega \p_t \delta w^t &= 0 ~ , \\
\p_\varphi \delta v^t - \p_\omega (\sin \omega  \delta w^t ) &= 0
\end{align}
where we have used (\ref{5})-(\ref{6}).

\end{enumerate}

\section{Energy-momentum conservation equations}\label{energy-momentum}
Recall from (\ref{3000})-(\ref{3001}), the intrinsic and extrinsic blackfold equations 
\begin{align}
\nabla_a T^{a b} &= \p^b X_\mu \, \mathcal{F}^\mu ~ , \\
T^{ab} K_{ab}^{\,\,\, \,\,\, (i)}  &= \mathcal{F}^\mu \, n^{(i)}_\mu
\end{align}
where $\mathcal{F}^\mu$ denotes the force terms coming from the coupling of the currents to the fluxes (\ref{90}).

\subsection{Intrinsic perturbation equation} \label{intrinsic}
The blackfold intrinsic perturbation equation is given by
\be
\delta \left( \nabla_a T^{a b} \right) = \delta \left( \p^b X_\mu \, \mathcal{F}^\mu \right) ~ .
\ee
Considering the LHS, we have
\be
\delta \left( \nabla_a T^{ab} \right) = \nabla_a \delta T^{ab} - T^{b c} \nabla_c \left( K_\rho \delta X^\rho_\perp \right) - 2 T^{ac} \nabla_c \left( K_{a \ \rho}^{\ b} \delta X^\rho_\perp \right) + T^{ac} \nabla^b \left( K_{ac \rho} \delta X^\rho_\perp \right)
\ee
where $K^\rho = \gamma^{ab} K_{ab}^{\ \ \rho}$ and we have used the identity
\be
\delta \, \Theta^{b}_{ac} = \frac{1}{2} \gamma^{b d} (\nabla_a  \delta \gamma_{c d} + \nabla_c  \delta \gamma_{a d} - \nabla_d  \delta \gamma_{ac})  ~ .
\ee
Considering the RHS, we have
\begin{align}
\delta \left( \p^b X_\mu \mathcal{F}^\mu \right) &= \delta \left(  \p^b X_\mu \right) \mathcal{F}^\mu +  \p^b X_\mu \delta \left(\mathcal{F}^\mu \right) \\
&=  \gamma^{t b} g_{\psi \psi} \p_t \delta \psi  \left( F_3^{\psi \omega \varphi} J_{2 \omega \varphi} \right)
\end{align}
where we have made use of the explicit expression of $\mathcal{F}^\mu$ in (\ref{90}). Altogether, we have the intrinsic perturbation equation
\begin{multline}
\nabla_a \delta T^{ab} - T^{b c} \nabla_c \left( K_\rho \delta X^\rho_\perp \right) - 2 T^{ac} \nabla_c \left( K_{a \ \rho}^{\ b} \delta X^\rho_\perp \right) + T^{ac} \nabla^b \left( K_{ac \rho} \delta X^\rho_\perp \right)\\
=  \gamma^{t b} g_{\psi \psi} \p_t \delta \psi \left( F_3^{\psi \omega \varphi} J_{2 \omega \varphi} \right) ~ .
\end{multline}
Substituting in appropriate expressions, we obtain for $b = t,\omega, \varphi$ respectively 

\begin{enumerate}
\item The $t$ intrinsic perturbation equation  
\begin{multline}
\label{D}
\p_t \delta \tan \theta +  \frac{\sqrt{M} b_0}{\sin \psi} \tan \theta \left( \p_\omega \delta v^t + \frac{1}{\sin \omega} \p_\varphi \delta w^t + \cot \omega \delta v^t \right) 
 \\
 + 2 \left( \cot \psi \tan \theta + \frac{1}{b_0^2} \right) \p_t \delta \psi = 0 ~ ,
\end{multline} 

\item The $\omega$ intrinsic perturbation equation 
\be
\sqrt{M} b_0 \sin \psi  \tan^2 \theta  \p_t \delta v^t +   \sin \theta \cos \theta  \p_\omega \delta \tan \theta    = 0 ~ ,
\ee

\item The $\varphi$ intrinsic perturbation equation 
\be
\sqrt{M} b_0 \sin \psi \sin \omega \tan^2 \theta \p_t \delta w^t  + \sin \theta \cos \theta \p_\varphi \delta \tan \theta  = 0 ~ .
\ee
\end{enumerate}

\subsection{Extrinsic equation} \label{extrinsic}
The extrinsic blackfold perturbation equation is given by
\be
\delta \left( T^{ab} K_{ab}^{\,\,\, \,\,\, (i)} \right) = \delta \left( \mathcal{F}^\mu \, n^{(i)}_\mu \right) ~ .
\ee
Making use of the results in (\ref{42}), we can easily write the LHS as
\begin{multline}
\delta \left( T^{ab} K_{ab}^{\,\,\, \,\,\, (i)} \right) = \delta T^{ab} K_{ab}^{\ \ \, (i)} + T^{ab} n_\rho^{(i)} \nabla_a \left( \p_b \delta X^\rho_\perp \right) + T^{ab} n^{(i)}_\rho \delta X^\alpha_\perp \p_\alpha \Gamma^\rho_{\mu \nu} \p_a X^\mu \p_b X^\nu\\
+ 2 \,T^{ab} n_\rho^{(i)}  \Gamma^\rho_{\mu \nu} \p_a \delta X^\mu_\perp \p_b X^\nu + \frac{1}{2} T^{ab} K_{ab}^{\,\,\,\,\,\, \rho} \left( n^{\alpha \, (i)} n^{\beta \, (i)} \p_\gamma g_{\alpha \beta} \delta X^\gamma_\perp n_\rho^{(i)} \right) ~ .
\end{multline}
For our purpose, we are interested in the orthogonal directions $\psi$ and $r$. The unitary normal vectors specifying these directions are respectively
\begin{align}
&n^{(1)} = \sqrt{M} b_0 d \psi ~ ,& &n^{(2)} = \sqrt{M} b_0 d r  ~ .
\end{align}
For the $\psi$ direction, the RHS is given by 
\be
\delta \left( \mathcal{F}^\mu n_\mu^{(1)} \right) = \delta  \mathcal{F}^\mu n_\mu^{(1)} + \mathcal{F}^\mu \delta n_\mu^{(1)} = \delta \mathcal{F}^\psi n_\psi^{(1)} ~ .
\ee
The expression of $\delta \mathcal{F}^\psi$ can be easily obtained by hitting $\delta$ to the force term $\mathcal{F}^\mu$ (\ref{90}). As the computation is tedious but straightforward, we shall not include all the details here. Nevertheless, for the convenience of the readers, let us note down the final results along with some useful (non-vanishing) intermediate steps. We have
\begin{align}
\delta F_3^{\psi \omega \varphi} &= \delta \left( g^{\psi \mu} \gamma^{\omega a_1} \p_{a_1} X^{\alpha_1} \gamma^{\varphi a_2} \p_{a_2} X^{\alpha_2} F_{3 \mu \alpha_1 \alpha_2} \right) \\
&= g^{\psi \psi} \left( \delta  \gamma^{\omega \omega} \right) \gamma^{\varphi \varphi} F_{3 \psi \omega \varphi} + g^{\psi \psi} \gamma^{\omega \omega} \left( \delta  \gamma^{\varphi \varphi} \right) F_{3 \psi \omega \varphi} + g^{\psi \psi} \gamma^{\omega \omega} \gamma_{\varphi \varphi} \left( \delta  F_{3 \psi \omega \varphi}  \right) \\
&= \left( 4 g^{\psi \psi} K^{\omega \omega}_{\ \ \ \psi} \gamma^{\varphi \varphi} F_{3 \psi \omega \varphi} + g^{\psi \psi} \gamma^{\omega \omega} \gamma^{\varphi \varphi}  \p_\psi F_{3 \psi \omega \varphi} \right) \delta \psi ~ .
\end{align}
Similarly, we have 
\be
\delta H_{7}^{\psi t ... \varphi} = \left( 4 g^{\psi \psi} \gamma^{tt} ... \gamma^{x^3 x^3} K^{\omega \omega}_{\ \ \ \psi} \gamma^{\varphi \varphi} H_{7 \psi t ... \varphi} + g^{\psi \psi} \gamma^{tt}... \gamma^{\varphi \varphi} \p_\psi H_{7 \psi t ... \varphi} \right) \delta \psi ~ .
\ee
Let us note also that 
\begin{align}
\delta J_{2 \omega \varphi} &= \mathbb{Q}_5 \left( \delta \sin \theta \right) v_\omega w_\varphi + \mathbb{Q}_5 \sin \theta \left( \delta v_\omega w_\varphi + v_\omega \delta w_\varphi \right) \\
&= \left( M b_0^2 \mathbb{Q}_5 \cos^3 \theta  \sin^2 \psi \sin \omega \right) \delta \tan \theta + \left( 2  M b_0^2 \mathbb{Q}_5 \sin \theta \cos \psi \sin \psi \sin \omega \right) \delta \psi
\end{align}
and
\begin{align}
\delta j_{6 t ... \varphi} &= - \mathbb{Q}_5 \left( \delta \sqrt{-\gamma} \right) = - \frac{1}{2} \mathbb{Q}_5 \sqrt{-\gamma} \gamma^{\alpha \beta} \delta \gamma_{\alpha \beta} \\
&=  \left( 2 \, \mathbb{Q}_5 \sqrt{-\gamma} \gamma^{\omega \omega} K_{\omega \omega}^{\ \ \ \psi} g_{\psi \psi} \right) \delta \psi  ~ .
\end{align}
Altogether, we have the variation of the force term $\delta \mathcal{F}^\psi$ is given by
\be
\delta \mathcal{F}^\psi = - \left( \delta H_7^{\psi t ... \varphi} \right) j_{6 t ... \varphi} - H_7^{\psi t ... \varphi} \left( \delta j_{6 t ... \varphi} \right) + \left( \delta F_3^{\psi \omega \varphi} \right) J_{2 \omega \varphi} + F_3^{\psi \omega \varphi} \left( \delta J_{2 \omega \varphi} \right) ~ .
\ee
For the $r$ direction, the RHS is given by
\be
\delta \left( \mathcal{F}^\mu n_\mu^{(2)} \right) = \delta  \mathcal{F}^\mu n_\mu^{(2)} + \mathcal{F}^\mu \delta n_\mu^{(2)} = \delta \mathcal{F}^r n_r^{(2)} ~ .
\ee
Similar to our treatment of $\delta \mathcal{F}^\psi$, we shall not present here the full computation of $\delta \mathcal{F}^r$ but only the final results along with some useful (non-vanishing) intermediate steps. We have
\begin{align}
\delta \tilde{F}_5^{r t ... x^3} &= \delta \left( g^{r \nu} \gamma^{t a_1} ... \gamma^{x^3 a_4} \p_{a_1} X^{\alpha_1} ... \p_{a_4} X^{\alpha_4} \tilde{F}_{5 \nu \alpha_1 ... \alpha_4} \right) \\
&= \left(  g^{rr} \gamma^{tt} ... \gamma^{x^3 x^3}  \p_r \tilde{F}_{5 r t ... x^3} \right) \delta r ~ .
\end{align}
The variation of the force term $\delta \mathcal{F}^r$ is given by
\be
\delta \mathcal{F}^r = \left( \delta \tilde{F}_5^{r t ... x^3} \right) J_{4 t ... x^3} ~ .
\ee
Substituting in appropriate expressions and simplify where possible, we obtain respectively 
\begin{enumerate}
\item The $\psi$ extrinsic perturbation equation
\be
\label{E}
(\p_t)^2 \delta \psi -  \frac{ \cos^2 \theta}{ \sin^2 \psi}   \nabla^2 \delta \psi  = \frac{2  \cos^2 \theta}{ \sin^2 \psi}  \delta \psi + \frac{2}{b_0^2}  \cos^2 \theta \left( 1 + \sin \theta \right) \delta \tan \theta ~ ,
\ee

\item The $r$ extrinsic perturbation equation 
\begin{multline}
 (\p_t)^2 \delta r - \frac{\cos^2 \theta}{ \sin^2 \psi}  \nabla^2 \delta r   = \frac{8 a_2}{a_0} \sin \theta \delta r +  \frac{8 a_2 }{ a_0}  \delta r   -  \frac{  16 a_0 + 20 a_2  }{5  a_0} \cos^2 \theta  \delta r \\
 +  \frac{ 4  }{5} \cos^2 \theta   \sin^2 \omega  \delta r 
\end{multline}
where $\nabla^2$ is the normalised Laplacian, i.e. $\nabla^2 = (\p_\omega)^2 + 1/\sin^2 \omega (\p_\varphi)^2 + \cot \omega \p_\omega$. 
\end{enumerate}

\end{appendices}

%% file: content/back_matter.tex
%

\cleardoublepage

\input{content/colophon}
